\documentclass[journal, draftclsnofoot, onecolumn]{IEEEtran}
\IEEEoverridecommandlockouts
\usepackage[cmex10]{amsmath}
\pdfoutput=1
\usepackage{graphicx,amsthm,amsbsy,amssymb,cite,algorithm,algorithmic}
\usepackage{enumitem}
\ifCLASSOPTIONcompsoc
\usepackage[caption=false,font=normalsize,labelfont=sf,textfont=sf]{subfig}
\else
\usepackage[caption=false,font=footnotesize]{subfig}
\fi

\graphicspath{ {./figures/} }

\def\Ex{\mathop{\bf E\/}}
\DeclareMathOperator{\vecmat}{vec}

\DeclareMathOperator{\real}{Re}
\DeclareMathOperator{\imag}{Im}

\DeclareMathOperator{\trace}{tr}
\DeclareMathOperator{\Tr}{Tr}
\DeclareMathOperator{\rank}{rank}
\DeclareMathOperator{\Span}{span}

\def\abf{\mathbf{a}}
\def\xbf{\mathbf{x}}
\def\wbf{\mathbf{w}}
\def\ybf{\mathbf{y}}

\def\vbf{\mathbf{v}}

\def\sbf{\mathbf{s}}
\def\zbf{\mathbf{z}}
\def\ubf{\mathbf{u}}
\def\Abf{\mathbf{A}}
\def\Bbf{\mathbf{B}}
\def\Cbf{\mathbf{C}}
\def\Dbf{\mathbf{D}}
\def\Ebf{\mathbf{E}}
\def\Fbf{\mathbf{F}}

\def\Rbf{\mathbf{R}}
\def\Pbf{\mathbf{P}}

\def\Tbf{\mathbf{T}}
\def\Xbf{\mathbf{X}}
\def\Ybf{\mathbf{Y}}
\def\Zbf{\mathbf{Z}}
\def\Vbf{\mathbf{V}}
\def\Ubf{\mathbf{U}}

\def\Sigbf{\boldsymbol{\Sigma}}

\def\Ex{\mathop{\bf E\/}}

\def\PsiS{\boldsymbol{\Psi}_{S}}
\def\PsiW{\boldsymbol{\Psi}_{W}}
\def\Pss{\Pbf_{SS}}
\def\Pww{\Pbf_{WW}}
\def\Pws{\Pbf_{WS}}
\def\Psw{\Pbf_{SW}}
\def\Bww{\Bbf_1}
\def\Bws{\Bbf_2}
\def\Bss{\Bbf_3}
\def\Sigws{\Sigbf_2}
\def\Rni{\mathbf{R_{ni}}}
\def\Rniest{\mathbf{\hat{R}_{ni}}}
\def\Rniinv{\mathbf{R}_{\mathbf{ni}}^{-1}}
\def\Rnipinv{\mathbf{R}_{\mathbf{ni}}^{\dagger}}
\def\Rnihalf{\mathbf{R}_{\mathbf{ni}}^{1/2}}
\def\vecBws{\boldsymbol{\beta}_2}
\def\vecSigws{\boldsymbol{\sigma}_2}
\def\vecTarg{\boldsymbol{\tau}}
\def\Rnirep{\breve{\mathbf{R}}_{\mathbf{ni}}}

\def\Clutmat{\Cbf_V}
\def\ClutProj{\Pbf_{\Clutmat}}
\def\ClutProjPerp{\ClutProj^{\perp}}
\def\NIProj{\Pbf_{\Rni}} 
\def\NIProjPerp{\NIProj^{\perp}}
\def\NIBreveProj{\Pbf_{\Rnirep}}
\def\NIBreveProjPerp{\NIBreveProj^{\perp}}

\def\Complex{\mathbb{C}}
\def\Reals{\mathbb{R}}

\newcommand{\normsq}[1]{{\norm{#1}^{2}}}
\newcommand{\frobnorm}[2]{{\norm{#1}^{#2}_{F}}}
\newcommand{\kronecker}{\raisebox{1pt}{\ensuremath{\:\otimes\:}}} 
\newcommand{\Identity}[1]{\ensuremath{\mathbf{I}_{#1}}}
\newcommand{\Commut}[2]{\ensuremath{\mathbf{K}_{#1,#2}}}
\newcommand{\CommutO}[1]{\ensuremath{\mathbf{K}_{#1}}}
\newcommand{\ZeroMat}[2]{\ensuremath{\mathbf{0}_{#1\times#2}}}

\newcommand{\Range}[1]{\mathcal{R}(#1)}
\newcommand{\Nullspace}[1]{\mathcal{N}(#1)}

\DeclareMathOperator{\relint}{relint}
\DeclareMathOperator{\dom}{dom}

\newcommand{\dif}{\mathop{}\!\mathrm{d}}

\makeatletter
\newcommand{\spx}[1]{%
	\if\relax\detokenize{#1}\relax
	\expandafter\@gobble
	\else
	\expandafter\@firstofone
	\fi
	{^{#1}}%
}
\makeatother

\newcommand\dpd[3][]{\dfrac{\partial\spx{#1}#2}{\partial#3\spx{#1}}}

\newcommand{\genericdel}[4]{%
	\ifcase#3\relax
	\ifx#1.\else#1\fi#4\ifx#2.\else#2\fi\or
	\bigl#1#4\bigr#2\or
	\Bigl#1#4\Bigr#2\or
	\biggl#1#4\biggr#2\or
	\Biggl#1#4\Biggr#2\else
	\left#1#4\right#2\fi
}

\newcommand{\sVert}[1][0]{%
	\ifcase#1\relax
	\rvert\or\bigr|\or\Bigr|\or\biggr|\or\Biggr
	\fi
}

\let\norm\enVert
\newcommand{\fullfunction}[5]{%
	\begin{array}{@{}r@{}r@{}c@{}l@{}}
		#1 \colon & #2 & {}\longrightarrow{} & #3 \\
		& #4 & {}\longmapsto{}     & #5
	\end{array}
}
\newtheorem{theorem}{\bf{Theorem}}
\newtheorem{lemma}{\bf{Lemma}}
\newtheorem{remark}{\bf{Remark}}
\newtheorem{prop}{Proposition}

\title{Relaxed Bi-quadratic Optimization for Joint Filter-Signal Design in Signal-Dependent STAP}
\author{Sean~M.~O'Rourke,~\IEEEmembership{Member, IEEE,} Pawan~Setlur,~\IEEEmembership{Member, IEEE,} \\
	Muralidhar Rangaswamy,~\IEEEmembership{Fellow, IEEE,} A. Lee Swindlehurst,~\IEEEmembership{Fellow, IEEE}%
\thanks{S.~M.~O'Rourke is with the Sensors Directorate, US AFRL, WPAFB, OH and the University of California, Irvine. Email: sean.orourke.3@us.af.mil}%
\thanks{M. Rangaswamy is with the Sensors Directorate, US AFRL, WPAFB, OH. Email: muralidhar.rangaswamy@us.af.mil}%
\thanks{P. Setlur is affiliated with the Wright State Research Inst. as a research contractor for the US AFRL, WPAFB, OH. Email: pawan.setlur@wright.edu}%
\thanks{A. L. Swindlehurst is with the University of California, Irvine. Email: swindle@uci.edu}%
\thanks{Approved for Public Release, No.: 88ABW-2016-5820}}
\begin{document}

\maketitle
\begin{abstract}

We investigate an alterative solution method to the joint signal-beamformer optimization problem considered by Setlur and Rangaswamy\cite{SetlurRangaswamy2016}. First, we directly demonstrate that the problem, which minimizes the recieved noise, interference, and clutter power under a minimum variance distortionless response (MVDR) constraint, is generally non-convex and provide concrete insight into the nature of the nonconvexity. Second, we employ the theory of biquadratic optimization and semidefinite relaxations to produce a relaxed version of the problem, which we show to be \emph{convex}. The optimality conditions of this relaxed problem are examined and a variety of potential solutions are found, both analytically and numerically.
\end{abstract}

\section{Introduction}\label{sec:intro}
In 1972, at a NATO conference in the East Midlands of England, two luminaries of signal processing agreed that an adaptive system was incapable of being simultaneously spatially and temporally optimal. Agreeing with M. Mermoz's proposition, the Naval Underwater System Center's Norman Owsley asserted that since ``the spectrum of the signal must be known \emph{a priori}" for temporal optimality, ``the post beamformer temporal processor cannot be fully adaptive and fully optimum (sic) simultaneously"\cite{Owsley1972}. In many ways, this fundamental philosophy has carried forward in multichannel signal processing and control theory unabated in the last 40-plus years since that conference in Loughborough. Under the fully-adaptive radar paradigm, however, it is necessary to find transmit and receive resources that \emph{are} simultaneously optimal, or at least optimal under prior knowledge assumptions. With these additional degrees of freedom, we can in fact develop a result that defies this wisdom. The purpose of this report is to document such an attempt, which is a significant extension of the work of \cite{SetlurRangaswamy2016} in joint waveform-filter design for radar space-time adaptive processing (STAP), that challenges the pre-existing status quo. 

In the general scenario, we assume that an airborne radar equipped with an array of sensing elements observes a moving target on the ground. Furthermore, this radar can change its transmitted waveform every coherent processing interval (CPI), instead of on a per-pulse basis. In order to develop a strategy for waveform design, we consider a STAP model that includes fast-time (range) samples as well as the usual slow-time and spatial samples. This is a departure from traditional STAP, which operates on spatiodoppler responses from the radar after matched filtering \cite{Klemm2002, Ward1994}, but recent advances in the literature have considered incorporating fast-time data for more accurate clutter modeling, both transmitter- and jammer-induced \cite{Madurasinghe2006, Seliktar2006}. 

An obvious result of including the fast-time data in the model is that the clutter representation is now signal-dependent, since in airborne STAP, the dominant clutter source is non-target ground reflections that persist over all range bins. We assume that, if modeled as a random process, the clutter is uncorrelated with any interference or noise (which, unlike the clutter, are assumed to have no signal dependence), and the related clutter correlation matrix is also signal dependent. If we formulate our waveform-filter design in the typical minimum variance distortionless-response (MVDR) framework \cite{Capon1969}, this dependence in the correlation structure leads to what multiple authors (\cite{Aubry2012, Aubry2013, ChenVaidyanathan2009}, among others) have empirically or intuitively identified as a non-convex optimization problem. However, the authors in \cite{SetlurRangaswamy2016} correctly identified that for a fixed transmit waveform, the problem is convex in the receive filter, and vice versa. This led to a collection of design algorithms based on alternating minimization, a reasonable heuristic. 

While the alternating minimization method is useful, it has no claims of optimality, nor do other methods in the literature that dealt with signal-dependent interference in other contexts, like the single sensor radar and reverberant channel of \cite{StoicaHeLi2012}. Furthermore, none of these authors directly proved or cited anything that demonstrated the non-convexity of the problem or its engineering consequences. We will ameliorate some of these concerns in this paper, first by directly proving the non-convexity of the objective, then relating it to the existing literature on \emph{biquadratic programming} (BQP). While the biquadratic program is demonstrably non-convex, it is possible to relax it to a convex quadratic program using semidefinite programming (SDP) (see \cite{BoydVandenberghe} for details on SDP and \cite{Ling2009, Ling2011} on relaxing the BQP). This relaxation will permit us to efficiently solve the problem computationally and analytically (up to a matrix completion), as well as reveal important structural information about the solution that matches with engineering intuition. 

Before continuing, we outline some mathematical notation that will be used throughout the paper. The symbol $\mathbf{x} \in \Reals^{N\times1} (\mathbf{x} \in \Complex^{N\times1})$ indicates a column vector of real (complex) values, while $\mathbf{X} \in \Reals^{M\times N} (\mathbf{X} \in \Complex^{M\times N})$ indicates an $N$ row, $M$ column matrix of real (complex) values. The superscripts $T$, $*$, and $H$ indicate the transpose, conjugate, and Hermitian (conjugate) transpose of a matrix or vector. The symbol $\kronecker$  indicates a Kronecker product. The operator $\vecmat$ turns a matrix into a vector with the matrix stacked columnwise -- that is, for $\mathbf{A} \in \Reals^{p\times q}, \vecmat(\mathbf{A}) \in \Reals^{pq\times 1}$. Special matrices that will recur often include:
\begin{itemize}
	\item the $n\times n$ identity matrix \Identity{n};
	\item the $n \times 1$ vector of all ones $\mathbf{1}_{n}$;
	\item the $n\times m$ all-zero matrix \ZeroMat{n}{m};
	\item and, via Magnus \& Neudecker\cite{Magnus1979}, the \emph{$(p,q)$ commutation matrix} $\Commut{p}{q} \in \Reals^{pq \times pq}$ that is non-uniquely determined by the relation $\vecmat(\mathbf{A}^{T}) = \Commut{p}{q}\vecmat(\mathbf{A})$  for $\mathbf{A}$ as above. $\Commut{p}{q}$ is an orthogonal permutation matrix (\emph{i.e.}, $\Commut{p}{q}^{T}  = \Commut{p}{q}^{-1}$ ). Transposing the matrix swaps the indices as well ($\Commut{p}{q}^{T} = \Commut{q}{p}$), and clearly then $\Commut{p}{q}\Commut{q}{p} = \Identity{pq}$. Additionally, $\Commut{1}{q} = \Identity{q}$ and $\Commut{p}{1} = \Identity{p}$. We will also frequently use the one-index form $\CommutO{p} = \Commut{p}{p}$ for the square commutation matrix. 
\end{itemize}
We will often use calligraphic letters like $\mathcal{C}$ to indicate tensors or multilinear operators (like the Hessian matrix). Additionally, the vectorizations of various matrices will be indicated by the equivalent lowercase Greek symbol in bold -- for example, a matrix $\Bbf$ will have the vectorization $\boldsymbol{\beta} = \vecmat(\Bbf)$. 

The rest of the paper continues as follows: First, we will outline the general signal model and relate it to the work on transfer functions by \cite{Guerci2016}. Next, we will define the joint design problem and demonstrate, after some transformation, that the problem is non-convex unless all clutter patches are known to be nulled a priori. Section IV will describe the relationship with the biquadratic problem and methods to relax the joint design into a convex quadratic semidefinite program. We then attempt to find analytic solutions and insights into this relaxed problem in Section V. These solutions are generally verified in Section VI through simulation, as well as demonstrate that the resulting rank-one solution performs similarly to the alternating minimization. Finally, we summarize our conclusions and highlight future research paths in Section VII. 

\section{STAP Model}\label{sec:STAPmodel}
Consider a radar that consists of a calibrated airborne linear array of $M$ identical sensor elements, where the first element is the phase center and also acts as the transmitter. During the transmission period, the radar probes the environment with a number of pulses $s(t)$ of width $T$ seconds and bandwidth $B$ hertz at a carrier frequency $f_o$. Within each burst, we transmit $L$ pulses at a rate of $f_p$ (i.e. the pulses are transmitted every $T_p = 1/f_p$ seconds) and collect them in a coherent processing interval (CPI). We also assume that the phase center is located at $\xbf_{r}$ and the platform moves at a rate that is relatively constant over the CPI. 

Assume the probed environment contains a target that lies at an azimuth $\theta_t$ and an elevation $\phi_t$ relative to the array phase center, moving with a relative velocity vector $\boldsymbol{\delta}v = \begin{bmatrix}{\delta v}_x & {\delta v}_y & {\delta v}_z\end{bmatrix}^{T}.$ If we assume that the array interelement spacing $d$ is small relative to the distance between the platform and the target, then the target's doppler shift is independent of the element index and given by
\begin{equation}
f_d = 2f_o\frac{\boldsymbol{\delta}v^{T}[\sin(\phi_t)\sin(\theta_t) \, \sin(\phi_t)\cos(\theta_t) \, \cos(\phi_t)]}{c}
\end{equation}
where $c$ is the usual speed of light. 

Let us now assume that we discretely sample the pulse $s(t)$ into $N$ samples, resulting in the sample vector $\sbf = [s_1, s_2, \cdots, s_N] \in \Complex^{N}$. Assuming the data is aligned to a common reference and given other assumptions from \cite{SetlurDevroyeRangaswamy2014}, we can say that at the target range gate $\tau_t$, the combined target response over the entire CPI can be represented by a vector $\ybf_t \in \Complex^{NML}$ given by
\begin{equation}
\ybf_\mathtt{t} = \rho_t \vbf_t(f_d) \kronecker \sbf \kronecker \abf_t(\theta_t,\phi_t)
\end{equation}
where $\rho_t$ is the complex backscattering coefficent from the target, the vector $\vbf_t(f_d) \in \Complex^{L}$ is the doppler steering vector whose $i$th element is given by $e^{-\jmath2\pi f_d (i-1) T_p }$, and the vector $\abf_t(\theta_t,\phi_t) \in \Complex^{M}$ is the spatial steering vector whose $i$th element is given by $e^{-\jmath2\pi(i-1)\vartheta}$ where $\vartheta = d\sin(\theta_t)\sin(\phi_t) f_o/c$. In this form, our departure from traditional STAP is clear, given the dependence on the waveform $\sbf$. 

Since no radar operates in an ideal environment, the target return $\ybf_t$ is corrupted by a variety of undesired returns from the environments -- noise, signal-independent interference, and clutter. We can consider the overall return in the target range gate as an additive model:
\begin{equation}
\tilde{\ybf} = \ybf_\mathtt{t} + \ybf_\mathtt{n} + \ybf_\mathtt{i} +\ybf_\mathtt{c} = \ybf_\mathtt{t} +\ybf_\mathtt{u} ,
\end{equation}
where the subscripts n, i, and c stand for noise, interference, and clutter, respectively, and these are all collected in the Undesired signal term. We assume that these undesired energy sources are statistically uncorrelated from each other, with unique distributions. We shall subsequently describe each of these sources, starting with the noise. 

We assume the noise is zero mean and identically distributed across all sensors, pulses, and fast time samples. The covariance matrix of  $\ybf_n$ is given by $\Rbf_\mathtt{n} \in \Complex^{NML \times NML}$. 

The interference term consists of jammers and other spurious emitters that are intentional or unintentional and are ground-based, airborne, or both. We assume that there are $K$ known interference sources, but otherwise we have no knowledge of what they transmit into the surveillance region, and (we hope) they are not dependent on our transmitted signal. Thus, we model their contributions as a zero-mean random process spread over all pulses and fast time samples. Assume that the $k$th interferer is located at the azimuth-elevation pair $(\theta_k,\phi_k)$. In the $l$th PRI, we assume that the waveform is a complex continuous-time signal $\alpha_{kl}(t)$. Under the same sampling scheme, this transforms into a vector $\boldsymbol{\alpha}_{kl} \in \Complex^{N}$, similar in form to $\sbf$. Stacked across all PRIs, we obtain another random vector $\boldsymbol{\alpha}_{k} = [\boldsymbol{\alpha}_{k0}^{T} \, \boldsymbol{\alpha}_{k1}^{T} \, \cdots \, \boldsymbol{\alpha}_{k(L-1)}^{T} ]^{T} \in \Complex^{NL}$, whose covariance matrix we define as $\Ex\{\boldsymbol{\alpha}_k\boldsymbol{\alpha}_k^{H}\} = \Rbf_{\boldsymbol{\alpha},k}$. Then, the response from the $k$th interferer can be modeled as 
\begin{equation}
\ybf_{\mathtt{i},k} = \boldsymbol{\alpha}_{k} \kronecker \abf_\mathtt{i}(\theta_k,\phi_k)
\end{equation}
where, as with the target, $\abf_i(\theta_k,\phi_k)$ is the array response to the interferer. 
The covariance of $\ybf$ is then $\Rbf_{\boldsymbol{\alpha},k} \kronecker \abf_\mathtt{i}(\theta_k,\phi_k)\abf_i(\theta_k,\phi_k)^{H}$, since expectations follow through Kronecker products that don't have a random dependence. If we assume that each of the interferers is uncorrelated with each other, then the overall covariance of the combined signal-independent interference $\ybf_\mathtt{i}$ is
\begin{equation}
\Rbf_\mathtt{i} = \sum_{k = 1}^{K} \Rbf_{\boldsymbol{\alpha},k} \kronecker \abf_\mathtt{i}(\theta_k,\phi_k)\abf_i(\theta_k,\phi_k)^{H}.
\end{equation}
In future sections, we will use the notation $\Rni = \Rbf_\mathtt{n} + \Rbf_\mathtt{i}$ to denote the combined noise and interference covariance matrix, describing the second order effects of the signal-independent corruption. 

Finally, we come to the clutter. In airborne radar applications, the most significant clutter source is the ground, which produces returns persistent throughout all range gates up to the horizon. Though other clutter sources exist, like large discrete objects, vegetation, and targets not currently being surveilled, the specific stochastic model we apply here only concerns ground clutter. However, we note that a later formulation in this paper \emph{may} be amenable to considering those other sources in a manner that recalls efforts in the literature on channel estimation (see, for example, \cite{Guerci2016}). 
For now, let us assume we have a number of clutter patches (say, $Q$) each comprising $P$ distinct scatterers. As with the target, the return from the $p$th scatterer in the $q$th patch, located spatially at the azimuth-elevation pair $(\theta_{pq},\phi_{pq})$, maintains a Kronecker structure given by
\begin{align*}
\gamma_{pq}\vbf(f_{\mathtt{c},pq}) \kronecker \sbf \kronecker \abf(\theta_{pq},\phi_{pq})
\end{align*}
where the returned complex reflectivity is $\gamma_{pq}$ and the Doppler shift observed by the platform is $f_{\mathtt{c},pq}$. This Doppler shift is solely induced by the platform motion, characterized by the aforementioned velocity vector $\dot{\xbf}_{\textmd{r}}$, and is given by 
\begin{equation}
f_{\mathtt{c},pq} = 2f_o\frac{\dot{\xbf}_{\textmd{r}}^{T}[\sin(\phi_{pq})\sin(\theta_{pq}) \, \sin(\phi_{pq})\cos(\theta_{pq}) \, \cos(\phi_{pq})]}{c}.
\end{equation}
Thus, the overall response from the $q$th clutter patch is 
\begin{equation}
\ybf_{\mathtt{c},q} = \sum_{p = 1}^{P} \gamma_{pq}\vbf(f_{\mathtt{c},pq}) \kronecker \sbf \kronecker \abf(\theta_{pq},\phi_{pq}).
\end{equation}
In order to define the covariance matrix of this response, we require the $q$th combining matrix $\Bbf_q \in \Complex^{NML \times P}$ as
\begin{equation*}
\mathbf{B_q} = [\mathbf{v}(f_{C,1q})\otimes\mathbf{s} \otimes \mathbf{a} (\theta_{1q},\phi_{1q}) \, \cdots \,\mathbf{v}(f_{C,Pq})\otimes\mathbf{s} \otimes \mathbf{a}(\theta_{Pq},\phi_{Pq})].
\end{equation*}
and the covariance of the reflectivity vector $[\gamma_{1q} \, \gamma_{2q} \, \cdots \, \gamma_{Pq}]^{T}$ given by $\mathbf{R}^{pq}_{\gamma} \in \Complex^{P \times P}$. Then, the overall covariance matrix for the patch is
\begin{equation}
\mathbf{R}^{q}_{\boldsymbol{\gamma}} = \mathbf{B}_{q}\mathbf{R}^{pq}_{\gamma}\mathbf{B}^{H}_{q}
\end{equation}
If we assume that the scatterers in one patch are uncorrelated with the scatterers in any other patch, then the total clutter response is $\ybf_{\mathtt{c}} = \sum_{q = 1}^{Q}\ybf_{\mathtt{c},q}$ and its covariance is given by
\begin{equation}
\Rbf_{\mathtt{c}} = \sum\limits_{q = 1}^{Q}\mathbf{R}^{q}_{\boldsymbol{\gamma}}
\end{equation}
As above, we will denote the overall undesired response covariance matrix as $\Rbf_{\mathtt{u}} = \Rni + \Rbf_{\mathtt{c}}(\sbf)$. 

Since this is a rather cumbersome model, we can simplify our description of the clutter as follows. Let us assume that our range resolution is large enough that we cannot resolve individual scatterers in each patch -- as mentioned in \cite{SetlurRangaswamy2016}, this is typical in STAP applications. Thus, we can regard each scatterer in the patch as lying within the same range gate and having approximately equal Doppler shifts, hence $f_{\mathtt{c},pq} \approx f_{\mathtt{c},q}$. Similarly, if we assume far-field operation, the scatterers will lie in approximately the same angular resolution cell centered at $(\theta_q,\phi_q)$, which means $\theta_{pq} \approx \theta_{q}, \phi_{pq} \approx \phi_q$. Given this simplification, we can modify our representation of the patch response and its covariance. 

Under this assumption, the per-patch clutter response is
\begin{align*}
\ybf_{\mathtt{c},q} = \gamma_q \vbf(f_{\mathtt{c},q}) \kronecker \sbf \kronecker \abf(\theta_{q},\phi_{q})
\end{align*}
where $\gamma_q = \sum_{p = 1}^{P}\gamma_{pq}$ is the combined reflectivity of all scatterers within the patch.  For the covariance, the combining matrix for the $q$-th clutter patch is given by $\mathbf{B_q} = [\mathbf{v}(f_{\mathtt{c},q})\otimes\mathbf{s} \otimes \mathbf{a} (\theta_q,\phi_q),\ldots,\mathbf{v}(f_{\mathtt{c},q})\otimes\mathbf{s} \otimes \mathbf{a}(\theta_q,\phi_q)]$. Since the deterministic patch response is repeated $P$ times, via standard Kronecker product properties, this is equivalent to 
\begin{equation}
\mathbf{B_q} = \mathbf{1}_{P}^{T} \otimes \mathbf{v}(f_{\mathtt{c},q})\otimes\mathbf{s} \otimes \mathbf{a}(\theta_q,\phi_q).
\end{equation}
More importantly, since $\mathbf{R}^{q}_{\boldsymbol{\gamma}} = \mathbf{B}_{q}\mathbf{R}^{pq}_{\gamma}\mathbf{B}^{H}_{q}$, we have
\begin{subequations}
	\begin{align}
	\mathbf{R}^{q}_{\gamma} &=  (\mathbf{1}_{P}^{T} \otimes \mathbf{v}(f_{\mathtt{c},q})\otimes\mathbf{s} \otimes \mathbf{a}(\theta_q,\phi_q))\mathbf{R}^{pq}_{\gamma}(\mathbf{1}_{P}^{T} \otimes \mathbf{v}(f_{\mathtt{c},q})\otimes\mathbf{s} \otimes \mathbf{a}(\theta_q,\phi_q))^{H}\\
	& =  \overline{R}^{pq}_{\gamma}(\mathbf{v}_q \otimes \mathbf{s}\otimes \mathbf{a}_q)(\mathbf{v}_q \otimes \mathbf{s} \otimes \mathbf{a}_q)^{H}
	\end{align}
\end{subequations}
where $\overline{R}^{q}_{\gamma} = \mathbf{1}_{P}^{T}\mathbf{R}^{pq}_{\gamma}\mathbf{1}^{}_{P}$. 

\paragraph*{Relationship to the Channel Model construct}
In \cite{Guerci2016}, the authors considered a transfer function/matrix approach for simultaneous transmit and receive resource design in MIMO radar, similar to the typical literature on control theory and digital communications. Using a rather obvious mathematical fact, we can immediately reframe our model in this form. Observe the following Kronecker mixed product property: for conformable matrices/vectors/scalars $\Abf$ through $\Fbf$,
\begin{align*}
(\Abf \kronecker \Bbf \kronecker \Cbf)(\Dbf \kronecker \Ebf \kronecker \Fbf) = \Abf\Dbf \kronecker \Bbf\Ebf \kronecker \Cbf\Fbf.
\end{align*}

Clearly, we can apply this to a deterministic space-time-doppler response vector. For example, the per-patch clutter response under the simplification is $\vbf(f_{\mathtt{c},q}) \kronecker \sbf \kronecker \abf(\theta_{q},\phi_{q})$. Since we can regard the scalar 1 as a conformable matrix, we have
\begin{align*}
\vbf(f_{\mathtt{c},q}) \kronecker \sbf \kronecker \abf(\theta_{q},\phi_{q}) & = (\vbf(f_{\mathtt{c},q}) \kronecker \Identity{N} \kronecker \abf(\theta_{q},\phi_{q})) \sbf \\
& = \boldsymbol{\Gamma}_{q}\sbf
\end{align*}
where $\boldsymbol{\Gamma}_{q}$ is obviously defined. By the same token, the target response is equivalent to $\Tbf\sbf$ where $\Tbf = \vbf_t(f_d) \kronecker \Identity{N} \kronecker \abf_t(\theta_t,\phi_t)$. With this form, the overall received vector in the range gate of interest is
\begin{align*}
\tilde{\ybf} = \rho_t\Tbf\sbf + \sum_{q = 1}^{Q}\gamma_q\boldsymbol{\Gamma}_{q}\sbf+\ybf_\mathtt{n} + \ybf_\mathtt{i}.
\end{align*}
This is something considerably easier on the eyes and more comprehensibly relates to the transfer function approach. We will continue to use this notation in subsequent analysis, as it reveals the structure of the design problem in a much more direct fashion. 

\section{Joint Waveform-Filter Design}\label{sec:initialprob}
With the signal model in place, we now turn to the initial purpose of \cite{SetlurRangaswamy2016}. Our goal is to find a STAP beamformer vector $\wbf \in \Complex^{NML}$ and a transmit signal $\sbf \in \Complex^{N}$ that minimizes the combined effect of the noise and interference represented by the covariance matrix $\mathbf{R_{ni}} \in \Complex^{NML \times NML}$) and the signal-dependent clutter. However, we also want to ensure reasonable radar operation. We describe this process below.

At the range gate we interrogate (which we assume contains the target), the return $\tilde{\ybf}$ is processed by a filter characterized by the weight vector $\wbf$, forming the output return $\wbf^{H}\tilde{\ybf}$. As mentioned above, we want to design this vector and the signal to minimize the expected undesired power $\Ex\{|\wbf^{H}\ybf_\mathtt{u}|^2\} = \wbf^{H}\Rbf_{\mathtt{u}}(\sbf)\wbf$. Additionally, we would like to constrain this minimization to ensure reasonable radar operation. First, for a given target space-time-doppler bin, we want a particular filter output, say, $\kappa \in \Complex$. This filter output can be represented by the Capon beamformer equation $\mathbf{w}^H\mathbf{T}\mathbf{s}$, where $\mathbf{T} \in \Complex^{NML \times N}$ is the target channel response above. Second, we place an upper bound on the total signal power, say, $P_{o}$. Mathematically, we can represent this optimization problem as
\begin{equation}
\begin{aligned}
& \underset{\mathbf{w}, \mathbf{s}}{\text{min}}
& & \mathbf{w}^H\Rbf_{\mathtt{u}}(\sbf)\mathbf{w}\\
& \text{s.t.}
& & \mathbf{w}^H\mathbf{T}\mathbf{s} = \kappa \\
& & & \mathbf{s}^H \mathbf{s}\leq P_o
\end{aligned} \label{eq:OriginalProblem}.
\end{equation}	

In \cite{SetlurRangaswamy2016}, this problem was computationally shown to be non-convex, and therefore used a variety of alternating minimization algorithms to solve it. Our purpose will first be to analytically prove that the problem as formulated is non-convex from an engineering standpoint. However,determining convexity, or lack thereof, will be difficult in the current configuration, so we will make a notational change. Let $\mathbf{b}$ be the \emph{combined beamformer-signal} vector defined by $\mathbf{b}= [\mathbf{w}^{T}\; \mathbf{s}^{T}]^{T}$. 

In order to recover the individual elements $\mathbf{w}, \mathbf{s}$ from $\mathbf{b}$, we define two matrices:
\begin{align}
	\mathbf{w} &= \boldsymbol{\Psi}_W\mathbf{b}, \, \boldsymbol{\Psi}_W = \begin{bmatrix}\Identity{NML} & \ZeroMat{NML}{N} \end{bmatrix} \nonumber\\
	\mathbf{s} &= \boldsymbol{\Psi}_S\mathbf{b}, \, \boldsymbol{\Psi}_S = \begin{bmatrix} \ZeroMat{N}{NML} & \Identity{N}\end{bmatrix}.\label{eq:SplittingMatrices}
\end{align}
For further notational simplicity, let us also define the complete objective function $f_o$ as a sum of the noise-interference cost $f_{NI}$ \& the total clutter cost $f_{C}$ (which is itself a sum of the $Q$ per-patch clutter costs):
\begin{align*}
	\mathrm{f}_o & =  \mathrm{f}_{NI} + \mathrm{f}_{C} = \wbf^{H}\Rbf_{\mathtt{u}}(\sbf)\mathbf{w}\\
	\mathrm{f}_{NI} & =   \mathbf{w}^H(\mathbf{R_n}+\mathbf{R_i})\mathbf{w} = \mathbf{w}^{H} \mathbf{R_{ni}} \mathbf{w}\\
	& =   \mathbf{b}^{H}\boldsymbol{\Psi}_W^{T}\mathbf{R_{ni}}\boldsymbol{\Psi}_W\mathbf{b} = \mathbf{b}^{H}\mathbf{\widetilde{R}_{ni}}\mathbf{b} \\
	\mathrm{f}_{C} & =  \sum\limits_{q=1}^Q \mathbf{w}^{H}\mathbf{R}_{\boldsymbol{\gamma}}^q(\mathbf{s})\mathbf{w} = \sum\limits_{q=1}^Q \mathrm{f}_{q}
\end{align*}

\subsection{Massaging the clutter objective}
We start with the per-patch clutter objective function. Using the channel representation above, $\mathbf{R}^{q}_{\boldsymbol{\gamma}}$ can be distilled to $\mathbf{R}^{q}_{\boldsymbol{\gamma}} = \overline{R}^{q}_{\gamma} \boldsymbol{\Gamma}_{q}\mathbf{s}\mathbf{s}^{H}\boldsymbol{\Gamma}_{q}^{H}$.
Therefore, the cost functional for the $q$-th clutter patch is
\begin{subequations}
	\begin{eqnarray}
	\textsf{f}_{q}(\mathbf{w},\mathbf{s})  & = & \overline{R}^{pq}_{\gamma} \mathbf{w}^{H} \boldsymbol{\Gamma}_{q}\mathbf{s}\mathbf{s}^{H}\boldsymbol{\Gamma}_{q}^{H}\mathbf{w} \label{eq:simp_cost_fxn1} \\
	& = & \overline{R}^{q}_{\gamma} |\mathbf{w}^{H}\boldsymbol{\Gamma}_{q}\mathbf{s}|^2 \label{eq:simp_cost_fxn2}
	\end{eqnarray}
\end{subequations}
Substituting  $\mathbf{w} = \PsiW\mathbf{b}$ and $\mathbf{s} = \PsiS\mathbf{b}$ into (\ref{eq:simp_cost_fxn2}) above, the final form of this cost function in the joint vector is 
\begin{equation}
\textsf{f}_{q}(\mathbf{w},\mathbf{s})  = \overline{R}^{q}_{\gamma} |\mathbf{b}^{H}\PsiW^{T}\boldsymbol{\Gamma}_{q}\PsiS\mathbf{b}|^2 = \overline{R}^{q}_{\gamma} |\mathbf{b}^{H}\overline{\boldsymbol{\Gamma}}_{q}\mathbf{b}|^2 \label{eq:simp_cost_fxn3}
\end{equation}
where $\overline{\boldsymbol{\Gamma}}_{q} = \PsiW^{T}\boldsymbol{\Gamma}_{q}\PsiS$. 

\subsection{An alternative form of the per-patch clutter cost}
Although the form derived above is quite compact \& informative, we develop another equivalent formation of the per-patch clutter cost that will allow us to quickly identify the Hessian of each of the cost functions considered and, ultimately, determine the convexity of the problem. 

An alternative form of the per-patch clutter cost function comes from Equation (\ref{eq:simp_cost_fxn2}). First, for any complex number $z$, its squared magnitude is $|z|^2 = \real(z)^{2} + \imag(z)^{2}$. 
Applying this to Equation~(\ref{eq:simp_cost_fxn2}), we obtain:
\begin{subequations}
	\begin{eqnarray}
	\textsf{f}_{q}(\mathbf{w},\mathbf{s})  & = & \overline{R}^{q}_{\gamma} |\mathbf{w}^{H}\boldsymbol{\Gamma}_{q}\mathbf{s}|^2 \\
	& = &  \overline{R}^{q}_{\gamma} \real(\mathbf{w}^{H}\boldsymbol{\Gamma}_{q}\mathbf{s})^{2} +  \overline{R}^{q}_{\gamma}\imag(\mathbf{w}^{H}\boldsymbol{\Gamma}_{q}\mathbf{s})^{2}\label{eq:real_imag_form_ws}
	\end{eqnarray}
\end{subequations}
Using elementary algebra, the components of Equation (\ref{eq:real_imag_form_ws}) are given by:
\begin{align*}
	\real(\mathbf{w}^{H}\boldsymbol{\Gamma}_{q}\mathbf{s}) & =   \frac{1}{2}(\mathbf{w}^{H}\boldsymbol{\Gamma}_{q}\mathbf{s} + \mathbf{s}^{H}\boldsymbol{\Gamma}_{q}^{H}\mathbf{w})\\
	& =  \frac{1}{2}\mathbf{b}^{H}\begin{bmatrix}
		\mathbf{0}_{NML\times NML} & \boldsymbol{\Gamma}_{q} \\
		\boldsymbol{\Gamma}_{q}^{H} & \mathbf{0}_{N \times N}
	\end{bmatrix}\mathbf{b}\\
	\imag(\mathbf{w}^{H}\boldsymbol{\Gamma}_{q}\mathbf{s}) & =  \frac{1}{2j}(\mathbf{w}^{H}\boldsymbol{\Gamma}_{q}\mathbf{s} - \mathbf{s}^{H}\boldsymbol{\Gamma}_{q}^{H}\mathbf{w}) \\
	& =  \frac{1}{2}\mathbf{b}^{H}\begin{bmatrix}
		\mathbf{0}_{NML\times NML} & \frac{1}{j}\boldsymbol{\Gamma}_{q} \\
		-\frac{1}{j}\boldsymbol{\Gamma}_{q}^{H} & \mathbf{0}_{N \times N}
	\end{bmatrix}\mathbf{b}
\end{align*}
Before we continue, we require a particular property of complex matrices. Any complex matrix $\mathbf{M}$ can be decomposed into the matrix sum $\mathbf{M} = \mathbf{M}_{H} + \mathbf{M}_{AH}$ where $\mathbf{M}_{H} = \frac{1}{2}(\mathbf{M}+\mathbf{M}^{H}), \mathbf{M}_{AH} = \frac{1}{2}(\mathbf{M}-\mathbf{M}^{H})$ are the \emph{H}ermitian and \emph{A}nti-Hermitian parts of $\mathbf{M}$. We can see by inspection that the block matrices in the real \& imaginary parts listed above contain the Hermitian and anti-Hermitian parts of $\overline{\boldsymbol{\Gamma}}_{q}$. Therefore, these forms are 
\begin{eqnarray*}
	\real(\mathbf{w}^{H}\boldsymbol{\Gamma}_{q}\mathbf{s}) & = &  \frac{1}{2}\mathbf{b}^{H}\begin{bmatrix}
		\mathbf{0}_{NML\times NML} & \boldsymbol{\Gamma}_{q} \\
		\boldsymbol{\Gamma}_{q}^{H} & \mathbf{0}_{N \times N}
	\end{bmatrix}\mathbf{b}=  \mathbf{b}^{H}\overline{\boldsymbol{\Gamma}}_{q,H}\mathbf{b} \\
	\imag(\mathbf{w}^{H}\boldsymbol{\Gamma}_{q}\mathbf{s})& = & \frac{1}{2}\mathbf{b}^{H}\begin{bmatrix}
		\mathbf{0}_{NML\times NML} & \frac{1}{j}\boldsymbol{\Gamma}_{q} \\
		-\frac{1}{j}\boldsymbol{\Gamma}_{q}^{H} & \mathbf{0}_{N \times N}
	\end{bmatrix}\mathbf{b} = \frac{1}{j} \mathbf{b}^{H}\overline{\boldsymbol{\Gamma}}_{q,A}\mathbf{b}
\end{eqnarray*}
Substituting these into the forms above and simplifying results in
\begin{eqnarray}
\textsf{f}_{q}(\mathbf{w},\mathbf{s}) & = &  \overline{R}^{q}_{\gamma} \real(\mathbf{w}^{H}\boldsymbol{\Gamma}_{q}\mathbf{s})^{2} +  \overline{R}^{q}_{\gamma}\imag(\mathbf{w}^{H}\boldsymbol{\Gamma}_{q}\mathbf{s})^{2} \nonumber\\
& = &\overline{R}^{q}_{\gamma}\left((\mathbf{b}^{H}\overline{\boldsymbol{\Gamma}}_{q,H}\mathbf{b})^2 - (\mathbf{b}^{H}\overline{\boldsymbol{\Gamma}}_{q,A}\mathbf{b})^2\right)\nonumber \\ 
& = & \overline{R}^{q}_{\gamma}\begin{bmatrix}
\mathbf{b}^{H}\overline{\boldsymbol{\Gamma}}_{q,H}\mathbf{b} \\
\mathbf{b}^{H}\overline{\boldsymbol{\Gamma}}_{q,A}\mathbf{b}
\end{bmatrix}^{H} 
\begin{bmatrix}
1 & 0 \\
0 & 1 
\end{bmatrix}
\begin{bmatrix}
\mathbf{b}^{H}\overline{\boldsymbol{\Gamma}}_{q,H}\mathbf{b} \\
\mathbf{b}^{H}\overline{\boldsymbol{\Gamma}}_{q,A}\mathbf{b}
\end{bmatrix} \nonumber\\
& = & \begin{bmatrix}
\mathbf{b}^{H}\overline{\boldsymbol{\Gamma}}_{q,H}\mathbf{b} \\
\mathbf{b}^{H}\overline{\boldsymbol{\Gamma}}_{q,A}\mathbf{b}
\end{bmatrix}^{H} 
\begin{bmatrix}
\overline{R}^{q}_{\gamma} & 0 \\
0 & \overline{R}^{q}_{\gamma} 
\end{bmatrix}
\begin{bmatrix}
\mathbf{b}^{H}\overline{\boldsymbol{\Gamma}}_{q,H}\mathbf{b} \\
\mathbf{b}^{H}\overline{\boldsymbol{\Gamma}}_{q,A}\mathbf{b}
\end{bmatrix} \nonumber\\
& = & \mathbf{g}^{H}_{q}(\mathbf{b})\mathbf{M}_{q} \mathbf{g}_{q}(\mathbf{b}) \label{eq:fq_quad_quad_form}
\end{eqnarray} 
This is an even more compact form of $f_q$ that, as we will see in the next section, permits us to quickly and easily derive the Hessian of the per-patch clutter cost, the \emph{total} clutter cost, and the overall objective function.

\subsection{Summary of Forms}
Finally, we arrive at multiple equivalent final forms of the clutter function $f_C$, which is a sum of the per-patch clutter forms:
\begin{align*}
	\textsf{f}_{C}& =  \sum\limits_{q=1}^Q \overline{R}^{q}_{\gamma} |\mathbf{w}^{H}\boldsymbol{\Gamma}_{q}\mathbf{s}|^2 =  \sum\limits_{q=1}^Q \overline{R}^{q}_{\gamma} |\mathbf{b}^{H}\overline{\boldsymbol{\Gamma}}_{q}\mathbf{b}|^2\\
	& = \sum\limits_{q=1}^Q \overline{R}^{q}_{\gamma} (\real(\mathbf{w}^{H}\boldsymbol{\Gamma}_{q}\mathbf{s})^{2} +  \imag(\mathbf{w}^{H}\boldsymbol{\Gamma}_{q}\mathbf{s})^{2}) =  \sum\limits_{q=1}^Q \mathbf{g}^{H}_{q}(\mathbf{b})\mathbf{M}_{q} \mathbf{g}_{q}(\mathbf{b})
\end{align*}

In the joint variable $\mathbf{b}$, the overall optimization problem becomes
\begin{equation}
\begin{aligned}
& \underset{\mathbf{b}}{\text{min}}
& & \mathbf{b}^H\mathbf{\widetilde{R}_{ni}}\mathbf{b} + \sum\limits_{q=1}^Q \mathbf{g}^{H}_{q}(\mathbf{b})\mathbf{M}_{q} \mathbf{g}_{q}(\mathbf{b})\\
& \text{s.t.}
& & \mathbf{b}^H\PsiW^{T}\mathbf{T}\PsiS\mathbf{b}=\kappa \\
& & & \mathbf{b}^H\PsiS^{T}\PsiS\mathbf{b}\leq P_o
\end{aligned}\label{eq:JointNCProofOpt}
\end{equation}

Before we prove \emph{joint} non-convexity, the form in Equation (\ref{eq:simp_cost_fxn1}) provides an immediate proof of convexity of the per-patch cost in $\mathbf{w}$ for fixed $\mathbf{s}$ and vice versa. Since the cost function $f$ is real and the matrix $ \mathbf{R}^{q}_{\boldsymbol{\gamma}}$ is positive-semidefinite (by virtue of being a correlation/covariance matrix) for fixed $\mathbf{s}$, then for any non-zero $\mathbf{w}$, $f \geq 0$ and is thus convex in $\mathbf{w}$. 

\subsection{Proving Joint (Non-)Convexity}
\subsubsection{Methods of verifying convexity}
The traditional definition of convexity is technically only defined for real-valued functions of \emph{real} arguments (be they scalar, vector, or matrix).  However, newly developed theory in \cite{Hjorungnes2011} and the associated journal literature permits us to make similar claims for real-valued functions of \emph{complex} arguments.

Let us define a scalar function $f$ of the complex variable $\mathbf{b}$ as $$\fullfunction{f}{\Complex^{J\times1}}{\Reals}{\mathbf{b}}{\textsf{f}(\mathbf{b},\mathbf{b}^{*})}.$$
Under the traditional definition (see, for example, in \cite{BoydVandenberghe}), this function is convex if for any $\mathbf{b}_1, \mathbf{b}_2 \in \Omega$ (where $\Omega$ is some convex subset of $\Complex^{J\times1}$) and $\alpha \in \left[0,1\right] \subset \Reals$, 
\begin{equation}
f(\alpha\mathbf{b}_1 + (1-\alpha)\mathbf{b}_2) \leq \alpha f(\mathbf{b}_1) + (1-\alpha)f(\mathbf{b}_2). \label{eq:convexity}
\end{equation}
This is true because the function of interest is \emph{real-valued}, despite the fact that its arguments are complex-valued. Equivalently, the Hessian matrix $\mathcal{\widetilde{H}}_{\mathbf{v}}[f] = \nabla^{2}_{\mathbf{v}}f(\mathbf{v})$ must be positive semidefinite when evaluated at any stationary point $\mathbf{v}=\mathbf{b}_{0} \in \Omega$. We will use the Hessian method, because it reveals the interesting structure of the problem in addition to showing non-convexity.

\subsubsection{Determining complex Hessians}
Since our function is a real-valued function of complex-valued variables, we require the following from Hj{\o}rungnes' work on complex matrix derivatives \cite{Hjorungnes2011}. From \cite[Theorem~3.2]{Hjorungnes2011}, the stationary points of a real-valued function of complex variables are the points where 
\begin{align}
\mathcal{D}_{\mathbf{b}}f(\mathbf{b},\mathbf{b}^{*}) & = \ZeroMat{1}{N(ML+1)}\\
\intertext{or,}
\mathcal{D}_{\mathbf{b}^{*}}f(\mathbf{b},\mathbf{b}^{*}) & = \ZeroMat{1}{N(ML+1)}
\end{align}
where, for example, the form $\mathcal{D}_{\mathbf{b}}f(\mathbf{b},\mathbf{b}^{*})$ is the complex derivative with respect to $\mathbf{b}$. As is the standard, we treat the vector $\mathbf{b}$ and its conjugate $\mathbf{b}^{*}$ as separate variables for the purposes of differentiation. This fact derives from the Wirtinger calculus and is directly proven in the reference above. 

\paragraph{A Taylor series argument for Hessians of real-valued functions}
It is well known that for a twice-differentiable real-valued scalar function, positive semi-definiteness of its Hessian matrix over a convex set is an equivalent statement of convexity. This is true even if the arguments are complex-valued. See, for example, \cite[Lemma~5.2]{Hjorungnes2011} and the consequences thereof. Namely, if the Hessian matrix in \cite[Equation~5.44]{Hjorungnes2011} is positive semidefinite at all stationary points $\mathbf{b}_{o}$ within the convex set of interest, then the zeroth-order convexity requirement in Equation \ref{eq:convexity} is automatically satisfied if we set $\mathbf{b}_{1} = \mathbf{b}_{o} + \dif\mathbf{b}, \mathbf{b}_{2} = \mathbf{b}_{o}$.
Let us rearrange Equation \ref{eq:convexity} to directly prove this assertion from the reference. First, observe that
\begin{eqnarray*}
	f(\alpha\mathbf{b}_1 + (1-\alpha)\mathbf{b}_2) & = & f(\mathbf{b}_{2} + \alpha(\mathbf{b}_1 - \mathbf{b}_2)) \\
	\alpha f(\mathbf{b}_1) + (1-\alpha)f(\mathbf{b}_2)  & = &   f(\mathbf{b}_2) + \alpha (f(\mathbf{b}_1) - f(\mathbf{b}_2))
\end{eqnarray*}
Then, the convexity requirement becomes
\begin{eqnarray*}
	f(\mathbf{b}_{2} + \alpha(\mathbf{b}_1 - \mathbf{b}_2)) & \leq &  f(\mathbf{b}_2) + \alpha (f(\mathbf{b}_1) - f(\mathbf{b}_2)) \\
	f(\mathbf{b}_{2} + \alpha(\mathbf{b}_1 - \mathbf{b}_2)) -  f(\mathbf{b}_2)  & \leq & \alpha (f(\mathbf{b}_1) - f(\mathbf{b}_2)) \\
	\frac{f(\mathbf{b}_{2} + \alpha(\mathbf{b}_1 - \mathbf{b}_2)) -  f(\mathbf{b}_2)}{\alpha} & \leq & f(\mathbf{b}_1) - f(\mathbf{b}_2) \\
	\frac{f(\mathbf{b}_{2} + \alpha(\mathbf{b}_1 - \mathbf{b}_2)) -  f(\mathbf{b}_2)}{\alpha} + f(\mathbf{b}_2)  & \leq & f(\mathbf{b}_1)
\end{eqnarray*}
Let us set $\mathbf{b}_{1} = \mathbf{b}_{o} + \dif\mathbf{b}, \mathbf{b}_{2} = \mathbf{b}_{o}$, where $\mathbf{b}_{o}$ is a stationary point and  $\dif\mathbf{b}$ is a vector of  infintesimally small perturbations. Then, the convexity requirement is 
\begin{eqnarray*}
	\frac{f(\mathbf{b}_{o} + \alpha \dif\mathbf{b}) -  f(\mathbf{b}_o)}{\alpha} + f(\mathbf{b}_o)  & \leq & f(\mathbf{b}_o + \dif\mathbf{b}).
\end{eqnarray*} At a stationary point, the first term on the left hand side is zero. Why? We can clearly see that, as $\alpha \rightarrow 0$, it is essentially a derivative, which must vanish at a stationary point by definition. Thus, the condition again changes to:
\begin{eqnarray}
f(\mathbf{b}_o)  & \leq & f(\mathbf{b}_o + \dif\mathbf{b}) \label{eq:convexity_diff}
\end{eqnarray}
Using the Taylor series expansion from \cite[Lemma~5.2]{Hjorungnes2011}, we expand $f(\mathbf{b} + \dif\mathbf{b})$, 
\begin{equation*}
f(\mathbf{b} + \dif\mathbf{b})=   f(\mathbf{b}) + \begin{bmatrix}\mathcal{D}_{\mathbf{b}}f(\mathbf{b},\mathbf{b}^{*})^{T} \\ \mathcal{D}_{\mathbf{b}^{*}}f(\mathbf{b},\mathbf{b}^{*})^{T} \end{bmatrix}^{T} \begin{bmatrix}\dif\mathbf{b}^{\phantom*} \\ \dif\mathbf{b}^{*} \end{bmatrix}+ \frac{1}{2} \begin{bmatrix}\dif\mathbf{b}^{\phantom*} \\ \dif\mathbf{b}^{*} \end{bmatrix}^{H}\begin{bmatrix}
\mathcal{H}_{\mathbf{b}\mathbf{b}^{*} }& \mathcal{H}_{\mathbf{b}^{*}\mathbf{b}^{*}} \\
\mathcal{H}_{\mathbf{b}\mathbf{b}} & \mathcal{H}_{\mathbf{b}^{*}\mathbf{b}}
\end{bmatrix}  \begin{bmatrix}\dif\mathbf{b}^{\phantom*} \\ \dif\mathbf{b}^{*} \end{bmatrix}  + {HOT}(\mathbf{b})
\end{equation*}
where the higher order terms function $HOT$ converges to zero in the sense provided in the reference. Certainly then, regardless of the value of $\mathbf{b}$,
\begin{equation*}
f(\mathbf{b} + \dif\mathbf{b}) \geq f(\mathbf{b}) + \begin{bmatrix}\mathcal{D}_{\mathbf{b}}f(\mathbf{b},\mathbf{b}^{*})^{T} \\ \mathcal{D}_{\mathbf{b}^{*}}f(\mathbf{b},\mathbf{b}^{*})^{T} \end{bmatrix}^{T}\begin{bmatrix}\dif\mathbf{b}^{\phantom*} \\ \dif\mathbf{b}^{*} \end{bmatrix} + \frac{1}{2}\begin{bmatrix}\dif\mathbf{b}^{\phantom*} \\ \dif\mathbf{b}^{*} \end{bmatrix}^{H}\begin{bmatrix}
\mathcal{H}_{\mathbf{b}\mathbf{b}^{*} }& \mathcal{H}_{\mathbf{b}^{*}\mathbf{b}^{*}} \\
\mathcal{H}_{\mathbf{b}\mathbf{b}} & \mathcal{H}_{\mathbf{b}^{*}\mathbf{b}}
\end{bmatrix} \begin{bmatrix}\dif\mathbf{b}^{\phantom*} \\ \dif\mathbf{b}^{*} \end{bmatrix}
\end{equation*}
If we evaluate this expression at $\mathbf{b} = \mathbf{b}_{o}$, which is our aforementioned stationary point, the first order derivatives in the second term vanish by definition and therefore
\begin{equation*}
f(\mathbf{b}_{o} + \dif\mathbf{b}) \geq f(\mathbf{b}_{o}) + \frac{1}{2}\begin{bmatrix}\dif\mathbf{b}^{\phantom*} \\ \dif\mathbf{b}^{*} \end{bmatrix}^{H}\begin{bmatrix}
\mathcal{H}_{\mathbf{b}\mathbf{b}^{*} }& \mathcal{H}_{\mathbf{b}^{*}\mathbf{b}^{*}} \\
\mathcal{H}_{\mathbf{b}\mathbf{b}} & \mathcal{H}_{\mathbf{b}^{*}\mathbf{b}}
\end{bmatrix}\begin{bmatrix}\dif\mathbf{b}^{\phantom*} \\ \dif\mathbf{b}^{*} \end{bmatrix}.
\end{equation*}
In order for (\ref{eq:convexity_diff}) to be satisfied, we then have 
\begin{eqnarray*}
	f(\mathbf{b}_{o}) + \frac{1}{2}\begin{bmatrix}\dif\mathbf{b}^{\phantom*} \\ \dif\mathbf{b}^{*} \end{bmatrix}^{H}\begin{bmatrix}
		\mathcal{H}_{\mathbf{b}\mathbf{b}^{*} }& \mathcal{H}_{\mathbf{b}^{*}\mathbf{b}^{*}} \\
		\mathcal{H}_{\mathbf{b}\mathbf{b}} & \mathcal{H}_{\mathbf{b}^{*}\mathbf{b}}
	\end{bmatrix} \begin{bmatrix}\dif\mathbf{b}^{\phantom*}\\ \dif \mathbf{b}^{*} \end{bmatrix} & \geq &  f(\mathbf{b}_{o}) \\
	\frac{1}{2}\begin{bmatrix}\dif \mathbf{b}^{\phantom*} \\ \dif \mathbf{b}^{*} \end{bmatrix}^{H}\begin{bmatrix}
		\mathcal{H}_{\mathbf{b}\mathbf{b}^{*} }& \mathcal{H}_{\mathbf{b}^{*}\mathbf{b}^{*}} \\
		\mathcal{H}_{\mathbf{b}\mathbf{b}} & \mathcal{H}_{\mathbf{b}^{*}\mathbf{b}}
	\end{bmatrix} \begin{bmatrix}\dif\mathbf{b}^{\phantom*} \\ \dif\mathbf{b}^{*} \end{bmatrix} &\geq& 0,
\end{eqnarray*}
which is clearly the definition of positive semidefiniteness for a Hermitian complex matrix. Therefore, in order for the function $f$ to be convex, the matrix $\begin{bmatrix}
\mathcal{H}_{\mathbf{b}\mathbf{b}^{*} }& \mathcal{H}_{\mathbf{b}^{*}\mathbf{b}^{*}} \\
\mathcal{H}_{\mathbf{b}\mathbf{b}} & \mathcal{H}_{\mathbf{b}^{*}\mathbf{b}}
\end{bmatrix}$ \emph{must} be positive semidefinite at all possible $\mathbf{b}_{o}$ inside the convex domain of $f$. 
\begin{remark}
	Traditionally, convexity is only defined for real functions of \emph{real} variables, which may make the analysis above seem incomplete. However, as we will show below, the definition of convexity for a complex vector variable is equivalent to the definition of joint convexity for its real and imaginary parts. To show this, we will prove that the definiteness condition remains if one replaces the differentials with respect to the complex variables with those for the real variables. 
	
	First, let us define the real and imaginary parts of the complex vector as $\mathbf{x} = \real(\mathbf{b}), \mathbf{y} = \imag(\mathbf{b})$ (which we assume are linearly independent real vectors and will treat as such). Clearly, then, we have $\mathbf{b} = \mathbf{x} + j\mathbf{y}$ and $\mathbf{b}^{*} = \mathbf{x} - j\mathbf{y} $. Applying the differential rules of \cite{Hjorungnes2011} to these expressions, we obtain:
	\begin{eqnarray*}
		\dif \mathbf{b} & = & \dif \mathbf{x} + j \dif \mathbf{y} = \begin{bmatrix} \Identity{J} & j\Identity{J}\end{bmatrix} \begin{bmatrix}\dif \mathbf{x} \\  \dif \mathbf{y} \end{bmatrix}\\
		\dif \mathbf{b}^{*} & = &  \dif \mathbf{x} - j \dif \mathbf{y} = \begin{bmatrix} \Identity{J} & -j\Identity{J}\end{bmatrix} \begin{bmatrix}\dif \mathbf{x} \\  \dif \mathbf{y} \end{bmatrix}
	\end{eqnarray*}
	which clearly means that we can replace the vector of complex differentials we see throughout the above derivation with 
	\begin{equation*}
	\begin{bmatrix}\dif \mathbf{b}^{\phantom*} \\  \dif \mathbf{b}^{*} \end{bmatrix} =  \begin{bmatrix} \Identity{J} & j\Identity{J} \\  \Identity{J} & -j\Identity{J} \end{bmatrix}\begin{bmatrix}\dif \mathbf{x} \\  \dif \mathbf{y} \end{bmatrix} = \mathbf{T}_{RC}\begin{bmatrix}\dif \mathbf{x} \\  \dif \mathbf{y} \end{bmatrix},
	\end{equation*}
	where $\mathbf{T}_{RC}$ is a transformation matrix that maps the real components to the complex vectors. 
	Making this substitution in the Taylor series expansion from above results in
	\begin{eqnarray*}
		f(\mathbf{x} + \dif\mathbf{x}, \mathbf{y} + \dif\mathbf{y}) & = &   f(\mathbf{x},\mathbf{y}) + \begin{bmatrix}\mathcal{D}_{\mathbf{b}}f(\mathbf{b},\mathbf{b}^{*})^{T} \\ \mathcal{D}_{\mathbf{b}^{*}}f(\mathbf{b},\mathbf{b}^{*})^{T} \end{bmatrix}^{T}  \mathbf{T}_{RC}\begin{bmatrix}\dif \mathbf{x} \\  \dif \mathbf{y} \end{bmatrix} \\
		& & \, + \frac{1}{2} \begin{bmatrix}\dif \mathbf{x} \\  \dif \mathbf{y} \end{bmatrix}^{T}\mathbf{T}_{RC}^{H}\begin{bmatrix}
			\mathcal{H}_{\mathbf{b}\mathbf{b}^{*} }& \mathcal{H}_{\mathbf{b}^{*}\mathbf{b}^{*}} \\
			\mathcal{H}_{\mathbf{b}\mathbf{b}} & \mathcal{H}_{\mathbf{b}^{*}\mathbf{b}}
		\end{bmatrix}  \mathbf{T}_{RC}\begin{bmatrix}\dif \mathbf{x} \\  \dif \mathbf{y} \end{bmatrix}  + {HOT}(\mathbf{x},\mathbf{y})
	\end{eqnarray*}
	where ${HOT}(\mathbf{x},\mathbf{y})$ must still be positive \& converge to zero in the same sense as before.  As an aside, we note that the first derivatives are clearly those w.r.t. the components $\mathbf{x},\mathbf{y}$.
	
	Continuing on, it is still true that the following inequality holds, even if we substitute for the real and imaginary components:
	\begin{equation*}
	f(\mathbf{x} + \dif\mathbf{x},\mathbf{y} + \dif\mathbf{y}) \geq  f(\mathbf{x},\mathbf{y}) + \begin{bmatrix}\mathcal{D}_{\mathbf{b}}f(\mathbf{b},\mathbf{b}^{*})^{T} \\ \mathcal{D}_{\mathbf{b}^{*}}f(\mathbf{b},\mathbf{b}^{*})^{T} \end{bmatrix}^{T}  \mathbf{T}_{RC}\begin{bmatrix}\dif \mathbf{x} \\  \dif \mathbf{y} \end{bmatrix}  + \frac{1}{2} \begin{bmatrix}\dif \mathbf{x} \\  \dif \mathbf{y} \end{bmatrix}^{T}\mathbf{T}_{RC}^{H}\begin{bmatrix}
	\mathcal{H}_{\mathbf{b}\mathbf{b}^{*} }& \mathcal{H}_{\mathbf{b}^{*}\mathbf{b}^{*}} \\
	\mathcal{H}_{\mathbf{b}\mathbf{b}} & \mathcal{H}_{\mathbf{b}^{*}\mathbf{b}}
	\end{bmatrix}  \mathbf{T}_{RC}\begin{bmatrix}\dif \mathbf{x} \\  \dif \mathbf{y} \end{bmatrix}. 
	\end{equation*}
	Evaluating the inequality at the new stationary points $\mathbf{x}_{o}, \mathbf{y}_{o}$ eliminates the first derivatives, yielding
	\begin{equation*}
	f(\mathbf{x}_{o} + \dif\mathbf{x},\mathbf{y}_{o} + \dif\mathbf{y})  \geq f(\mathbf{x}_{o},\mathbf{y}_{o}) +\frac{1}{2} \begin{bmatrix}\dif \mathbf{x} \\  \dif \mathbf{y} \end{bmatrix}^{T}\mathbf{T}_{RC}^{H}\begin{bmatrix}
	\mathcal{H}_{\mathbf{b}\mathbf{b}^{*} }& \mathcal{H}_{\mathbf{b}^{*}\mathbf{b}^{*}} \\
	\mathcal{H}_{\mathbf{b}\mathbf{b}} & \mathcal{H}_{\mathbf{b}^{*}\mathbf{b}}
	\end{bmatrix}  \mathbf{T}_{RC}\begin{bmatrix}\dif \mathbf{x} \\  \dif \mathbf{y} \end{bmatrix}.
	\end{equation*}
	
	Again, (\ref{eq:convexity_diff}) is satisfied only if the right-hand side of the above expression is greater than the value of the function at the stationary point, ergo 
	\begin{eqnarray*}
		f(\mathbf{x}_{o},\mathbf{y}_{o}) +\frac{1}{2} \begin{bmatrix}\dif \mathbf{x} \\  \dif \mathbf{y} \end{bmatrix}^{T}\mathbf{T}_{RC}^{H}\begin{bmatrix}
			\mathcal{H}_{\mathbf{b}\mathbf{b}^{*} }& \mathcal{H}_{\mathbf{b}^{*}\mathbf{b}^{*}} \\
			\mathcal{H}_{\mathbf{b}\mathbf{b}} & \mathcal{H}_{\mathbf{b}^{*}\mathbf{b}}
		\end{bmatrix}  \mathbf{T}_{RC}\begin{bmatrix}\dif \mathbf{x} \\  \dif \mathbf{y} \end{bmatrix} & \geq &  f(\mathbf{x}_{o},\mathbf{y}_{o}) \\
		\frac{1}{2} \begin{bmatrix}\dif \mathbf{x} \\  \dif \mathbf{y} \end{bmatrix}^{T}\mathbf{T}_{RC}^{H}\begin{bmatrix}
			\mathcal{H}_{\mathbf{b}\mathbf{b}^{*} }& \mathcal{H}_{\mathbf{b}^{*}\mathbf{b}^{*}} \\
			\mathcal{H}_{\mathbf{b}\mathbf{b}} & \mathcal{H}_{\mathbf{b}^{*}\mathbf{b}}
		\end{bmatrix}  \mathbf{T}_{RC}\begin{bmatrix}\dif \mathbf{x} \\  \dif \mathbf{y} \end{bmatrix} &\geq& 0.
	\end{eqnarray*}
	
	Now, the convexity requirement is that $\mathbf{T}_{RC}^{H}\begin{bmatrix}
	\mathcal{H}_{\mathbf{b}\mathbf{b}^{*} }& \mathcal{H}_{\mathbf{b}^{*}\mathbf{b}^{*}} \\
	\mathcal{H}_{\mathbf{b}\mathbf{b}} & \mathcal{H}_{\mathbf{b}^{*}\mathbf{b}}
	\end{bmatrix}  \mathbf{T}_{RC}$ \emph{must} be positive semidefinite at all possible $\mathbf{x}_{o},\mathbf{y}_{o}$ inside the convex domain of $f$. However, this is \emph{exactly identical} to the complex condition above, since according to \cite[Lemma~7.63(b)]{Seber2007}, these are biconditional statements! For real-valued functions, therefore, convexity in the complex variable is identical to joint convexity in its real and imaginary parts. 
\end{remark}

For notational simplicity, let us define this matrix -- which we will call the \emph{Hessian} of the function $f$ with respect to the Taylor series argument above -- as:
\begin{equation*}
\mathcal{\widetilde{H}}[f] = \begin{bmatrix}
\mathcal{H}_{\mathbf{b}\mathbf{b}^{*}}(f) & \mathcal{H}_{\mathbf{b}^{*}\mathbf{b}^{*}}(f) \\
\mathcal{H}_{\mathbf{b}\mathbf{b}}(f) & \mathcal{H}_{\mathbf{b}^{*}\mathbf{b}}(f)
\end{bmatrix}.
\end{equation*}

\paragraph{The augmented Hessian}
While the Hessian $\mathcal{\widetilde{H}}[f]$ is useful for determining convexity, it can be quite difficult to find expressions in a simple and systematic way for a given function $f$. If we use a change of variables, however, we can quickly find a \emph{related} Hessian using a chain rule. As in \cite{Hjorungnes2011}, we first define the augmented variable $\mathcal{Z} = \begin{bmatrix}
\mathbf{b} & \mathbf{b}^{*}
\end{bmatrix}$. 
Derivatives taken with respect to $\mathcal{Z}$ are often just stackings of the derivatives with respect to $\mathbf{b}$ and $\mathbf{b}^{*}$. The matrix of second derivatives with respect to the augmented variable $\mathcal{H}_{\mathcal{Z}\mathcal{Z}}(f)$, henceforth called the \emph{augmented Hessian} of the function $f$, is given by:
\begin{equation*}
\mathcal{H}_{\mathcal{Z}\mathcal{Z}}(f) =  \begin{bmatrix}
\mathcal{H}_{\mathbf{b}\mathbf{b}}(f) & \mathcal{H}_{\mathbf{b}^{*}\mathbf{b}}(f) \\
\mathcal{H}_{\mathbf{b}\mathbf{b}^{*}}(f) & \mathcal{H}_{\mathbf{b}^{*}\mathbf{b}^{*}}(f)
\end{bmatrix}.
\end{equation*}

\paragraph{The Chain Rule for augmented Hessians}
A benefit of the augmented Hessian is that a \emph{true} chain rule  in complex arguments exists for them, as opposed to the desired Hessian form. While there is a more general form of the chain rule listed in the reference \cite{Hjorungnes2011}, we present below a stripped down version more appropriate for our purposes.

If $f = h \circ \mathbf{g}$ is a scalar function of a $J\times1$ vector $\mathbf{b}$ (a composition of the scalar function $h$ and the vector function $\mathbf{g}$), then the augmented Hessian of $f$ is given by 
\begin{equation*}
\mathcal{H}_{\mathcal{Z}\mathcal{Z}}(f) = ( \mathcal{D}_{\mathbf{g}}(h) \kronecker \Identity{2T}) \mathcal{H}_{\mathcal{Z}\mathcal{Z}}h + ( \mathcal{D}_{\mathcal{Z}} \mathbf{g})^{T} \mathcal{H}_{\mathbf{g}\mathbf{g}}(h)  (\mathcal{D}_{\mathcal{Z}} \mathbf{g}).
\end{equation*}
As usual, the operator $\mathcal{D}_{\mathbf{Z}}(\mathbf{F}) = \dpd{\vecmat(\mathbf{F})}{\vecmat^{T}(\mathbf{Z})}$ can be considered the derivative of the matrix function $\mathbf{F}$ with respect to the complex matrix variable $\mathbf{Z}$, and similarly for vector \& scalar functions. 
\paragraph{Mapping between augmented Hessian and the actual Hessian}
While the augmented Hessian is excellent for calculation purposes, it is the actual Hessian that is needed for determining convexity. Thankfully, there is a straightforward relationship between the two: 
\begin{equation}
\mathcal{\widetilde{H}}[f] = \begin{bmatrix}
\ZeroMat{J}{J} & \Identity{J}\\
\Identity{J} & \ZeroMat{J}{J}
\end{bmatrix}\mathcal{H}_{\mathcal{Z}\mathcal{Z}}(f)
\end{equation}
and vice versa. Functionally, this means that we exchange the (1,1) block with the (2,1) block and the (1,2) block with the (2,2) block. 

\subsection{STAP Step 0: Preliminaries}
\subsubsection{Procedural outline}
Given the various functional forms and preliminary proofs, we will now outline the rest of the procedure to determine convexity of the objective. First, we will determine the derivatives and Hessians of each component of the per-patch clutter cost $\textsf{f}_q$ and use the chain rule to establish its Hessian. Using elementary techniques, we will demonstrate that this matrix is always indefinite unless a certain condition is satisfied, in which case it is always zero. Next, we will use the additivity of the combined clutter cost function to compute its Hessian and prove a similar definiteness result. Finally, we will compute the Hessian of the overall objective and show its indefiniteness by construction. 

\subsubsection{Functions of interest}
Recall that in \eqref{eq:fq_quad_quad_form}, we defined the per-patch clutter function is $f_q =  \mathbf{g}^{H}_{q}(\mathbf{b})\mathbf{M}_{q} \mathbf{g}_{q}(\mathbf{b})$, where
\begin{align*}
\mathbf{M}_{q} & =  \overline{R}^{q}_{\gamma} \Identity{2} & \mathbf{g}_{q}(\mathbf{b}) & =  \begin{bmatrix}
\mathbf{b}^{H}\overline{\boldsymbol{\Gamma}}_{q,H}\mathbf{b} \\
\mathbf{b}^{H}\overline{\boldsymbol{\Gamma}}_{q,A}\mathbf{b}
\end{bmatrix}.
\end{align*}
This form of the per-patch clutter function is quite compact and simplifies finding its augmented and actual Hessians, and therefore the augmented and actual Hessians of the total clutter and overall objective functions. Additionally, from this point forward, let $J = N(ML+1)$  denote the joint number of parameters (and thus the length of the combined beamformer $\mathbf{b}$).

\subsection{STAP Step 1: The differentials of $\mathbf{g}_{q}$}
\subsubsection{First derivative}
Using the rules of complex matrix derivatives/differentials outlined above and in Hj{\o}rungnes~\cite{Hjorungnes2011}, we can find the first derivatives of the inner vector function to be
\begin{align*}
	\dif\mathbf{g}_{q} & =  \begin{bmatrix}
		\dif(\mathbf{b}^{H}\overline{\boldsymbol{\Gamma}}_{q,H}\mathbf{b})\\
		\dif(\mathbf{b}^{H}\overline{\boldsymbol{\Gamma}}_{q,A}\mathbf{b})
	\end{bmatrix} = \begin{bmatrix}
	\mathbf{b}^{H}\overline{\boldsymbol{\Gamma}}_{q,H}\dif\mathbf{b} + \mathbf{b}^{T}\overline{\boldsymbol{\Gamma}}_{q,H}^{T}\dif\mathbf{b}^{*} \\
	\mathbf{b}^{H}\overline{\boldsymbol{\Gamma}}_{q,A}\dif\mathbf{b} + \mathbf{b}^{T}\overline{\boldsymbol{\Gamma}}_{q,A}^{T}\dif\mathbf{b}^{*} 
\end{bmatrix}\\
& =  \begin{bmatrix}
	\mathbf{b}^{H}\overline{\boldsymbol{\Gamma}}_{q,H} & \mathbf{b}^{T}\overline{\boldsymbol{\Gamma}}_{q,H}^{T} \\
	\mathbf{b}^{H}\overline{\boldsymbol{\Gamma}}_{q,A} & \mathbf{b}^{T}\overline{\boldsymbol{\Gamma}}_{q,A}^{T}
\end{bmatrix}
\begin{bmatrix}
	\dif\mathbf{b}^{\phantom{*}} \\
	\dif\mathbf{b}^{*}
\end{bmatrix}
=  \mathcal{D}_{\mathcal{Z}} \mathbf{g} \dif\vecmat(\mathcal{Z}).
\end{align*}
\subsubsection{Hessian}
Let us first observe that $\mathbf{g}_q$ is a stacking of scalar functions of $\mathbf{b}$:
\begin{equation*}
\mathbf{g}_{q}(\mathbf{b})  =  \begin{bmatrix}
\mathbf{b}^{H}\overline{\boldsymbol{\Gamma}}_{q,H}\mathbf{b} \\
\mathbf{b}^{H}\overline{\boldsymbol{\Gamma}}_{q,A}\mathbf{b}
\end{bmatrix} = \begin{bmatrix}
g_{q,0}\\
g_{q,1}
\end{bmatrix}
\end{equation*} 
Using \cite[Eq.~5.69, 5.70]{Hjorungnes2011} , we see that, as a result of the above form, the augmented Hessian of $\mathbf{g}_{q}$ is just a stacking of the per-element augmented Hessians:
\begin{equation*}
\mathcal{H}_{\mathcal{Z}\mathcal{Z}}(\mathbf{g}_{q}) = \begin{bmatrix}
\mathcal{H}_{\mathcal{Z}\mathcal{Z}}(g_{q,0}) \\
\mathcal{H}_{\mathcal{Z}\mathcal{Z}}(g_{q,1})
\end{bmatrix} = \begin{bmatrix}
1 \\ 0
\end{bmatrix} \kronecker \mathcal{H}_{\mathcal{Z}\mathcal{Z}}(g_{q,0}) + \begin{bmatrix}
0 \\1
\end{bmatrix} \kronecker \mathcal{H}_{\mathcal{Z}\mathcal{Z}}(g_{q,1})
\end{equation*}
In order to find the per-element augmented Hessians, we use the following fact. From \cite[Example 5.2]{Hjorungnes2011}, for any compatible matrix $\boldsymbol{\Phi}$,
\begin{equation*}
\mathcal{H}_{\mathcal{Z}\mathcal{Z}}(\mathbf{b}^{H}\boldsymbol{\Phi}\mathbf{b}) = \begin{bmatrix}
\ZeroMat{J}{J} & \boldsymbol{\Phi}^{T} \\
\boldsymbol{\Phi} & \ZeroMat{J}{J}
\end{bmatrix}.
\end{equation*}

Continuing with this in mind, we find the per-element augmented Hessians to be
\begin{align*}
	\mathcal{H}_{\mathcal{Z}\mathcal{Z}}(g_{q,0})  =  \begin{bmatrix}
		\ZeroMat{J}{J} & \overline{\boldsymbol{\Gamma}}_{q,H}^{T} \\
		\overline{\boldsymbol{\Gamma}}_{q,H} & \ZeroMat{J}{J}
	\end{bmatrix} \quad
	\mathcal{H}_{\mathcal{Z}\mathcal{Z}}(g_{q,1})  =  \begin{bmatrix}
		\ZeroMat{J}{J} & \overline{\boldsymbol{\Gamma}}_{q,A}^{T} \\
		\overline{\boldsymbol{\Gamma}}_{q,A} & \ZeroMat{J}{J}
	\end{bmatrix}
\end{align*}
Therefore, the augmented Hessian  of the vector function $\mathbf{g}_q$ is 
\begin{equation*}
\mathcal{H}_{\mathcal{Z}\mathcal{Z}}(\mathbf{g}_{q}) = \begin{bmatrix}
1 \\ 0
\end{bmatrix} \kronecker  \begin{bmatrix}
\ZeroMat{J}{J} & \overline{\boldsymbol{\Gamma}}_{q,H}^{T} \\
\overline{\boldsymbol{\Gamma}}_{q,H} & \ZeroMat{J}{J}
\end{bmatrix} + \begin{bmatrix}
0 \\1
\end{bmatrix} \kronecker  \begin{bmatrix}
\ZeroMat{J}{J} & \overline{\boldsymbol{\Gamma}}_{q,A}^{T} \\
\overline{\boldsymbol{\Gamma}}_{q,A} & \ZeroMat{J}{J}
\end{bmatrix}.
\end{equation*}

\subsection{STAP Step 2: The differentials of $h$ w.r.t. $\mathbf{g}_{q}$}
\subsubsection{First derivative}
Recall $h =  \mathbf{g}^{H}_{q}\mathbf{M}_{q} \mathbf{g}_{q}$. Thus,
\begin{equation*}
\dif h = \dif \left(\mathbf{g}^{H}_{q}\mathbf{M}_{q} \mathbf{g}_{q}\right) =  \mathbf{g}^{H}_{q}\mathbf{M}_{q} \dif\mathbf{g}_{q} + \mathbf{g}^{T}_{q}\mathbf{M}_{q}^{T} \dif\mathbf{g}_{q}^{*}
\end{equation*}
This means that
\begin{eqnarray*}
	\mathcal{D}_{\mathbf{g}_{q}}(h) & = &  \mathbf{g}^{H}_{q}\mathbf{M}_{q} \\
	\mathcal{D}_{\mathbf{g}_{q}^{*}}(h) & = &  \mathbf{g}^{T}_{q}\mathbf{M}_{q}^{T}
\end{eqnarray*}
\subsubsection{Hessian}
From \cite[Example~5.2]{Hjorungnes2011} , we know that
\begin{equation*}
\mathcal{H}_{\mathcal{Z}\mathcal{Z}}(\mathbf{b}^{H}\boldsymbol{\Phi}\mathbf{b}) = \begin{bmatrix}
\ZeroMat{J}{J} & \boldsymbol{\Phi}^{T} \\
\boldsymbol{\Phi} & \ZeroMat{J}{J}
\end{bmatrix}
\end{equation*}
for any compatible matrix $\boldsymbol{\Phi}$.
That means that, by inspection, $\mathcal{H}_{\mathbf{g}\mathbf{g}}(h) = \ZeroMat{J}{J}$!
\subsection{STAP Step 3: The per-patch clutter Hessian}
Applying the chain-rule for the augmented Hessian and the derivatives found above, the augmented Hessian of the per-patch clutter cost function is 
\begin{eqnarray*}
	\mathcal{H}_{\mathcal{Z}\mathcal{Z}}(\textsf{f}_q) & = & ( \mathcal{D}_{\mathbf{g}_{q}} (h) \kronecker \Identity{2P}) \mathcal{H}_{\mathcal{Z}\mathcal{Z}}(\mathbf{g}_{q}) + ( \mathcal{D}_{\mathcal{Z}}(\mathbf{g}_{q}))^{T} \mathcal{H}_{\mathbf{g}_{q}\mathbf{g}_{q}}(h)  (\mathcal{D}_{\mathcal{Z}}(\mathbf{g}_{q})) \\
	& = & ( \mathcal{D}_{\mathbf{g}_{q}} (h) \kronecker \Identity{2P}) \mathcal{H}_{\mathcal{Z}\mathcal{Z}}(\mathbf{g}_{q})  \\
	& = & (\mathbf{g}^{H}_{q}\mathbf{M}_{q} \kronecker \Identity{2P}) \left(\begin{bmatrix}
		1 \\ 0
	\end{bmatrix} \kronecker  \begin{bmatrix}
	\ZeroMat{J}{J} & \overline{\boldsymbol{\Gamma}}_{q,H}^{T} \\
	\overline{\boldsymbol{\Gamma}}_{q,H} & \ZeroMat{J}{J}
\end{bmatrix} + \begin{bmatrix}
0 \\1
\end{bmatrix} \kronecker  \begin{bmatrix}
\ZeroMat{J}{J} & \overline{\boldsymbol{\Gamma}}_{q,A}^{T} \\
\overline{\boldsymbol{\Gamma}}_{q,A} & \ZeroMat{J}{J}
\end{bmatrix}\right) \\
& = & \mathbf{g}^{H}_{q}\mathbf{M}_{q}\begin{bmatrix}
	1 \\ 0
\end{bmatrix}  \begin{bmatrix}
\ZeroMat{J}{J} & \overline{\boldsymbol{\Gamma}}_{q,H}^{T} \\
\overline{\boldsymbol{\Gamma}}_{q,H} & \ZeroMat{J}{J}
\end{bmatrix} + \mathbf{g}^{H}_{q}\mathbf{M}_{q}\begin{bmatrix}
0 \\ 1
\end{bmatrix}  \begin{bmatrix}
\ZeroMat{J}{J} & \overline{\boldsymbol{\Gamma}}_{q,A}^{T} \\
\overline{\boldsymbol{\Gamma}}_{q,A} & \ZeroMat{J}{J}
\end{bmatrix}
\end{eqnarray*}
In order to reduce this to a simpler form, we need to know what the scalars in each term are:
\begin{eqnarray*}
	\mathbf{g}^{H}_{q}\mathbf{M}_{q}\begin{bmatrix}
		1 \\ 0
	\end{bmatrix} & = & \overline{R}^{q}_{\gamma} \begin{bmatrix}
	\mathbf{b}^{H}\overline{\boldsymbol{\Gamma}}_{q,H}\mathbf{b} \\
	\mathbf{b}^{H}\overline{\boldsymbol{\Gamma}}_{q,A}\mathbf{b}
\end{bmatrix}^{H} 
\begin{bmatrix}
	1 \\ 0
\end{bmatrix} = \mathbf{b}^{H}\overline{\boldsymbol{\Gamma}}_{q,H}\mathbf{b} \\
\mathbf{g}^{H}_{q}\mathbf{M}_{q}\begin{bmatrix}
	0 \\ 1
\end{bmatrix} & = & \overline{R}^{q}_{\gamma} \begin{bmatrix}
\mathbf{b}^{H}\overline{\boldsymbol{\Gamma}}_{q,H}\mathbf{b} \\
\mathbf{b}^{H}\overline{\boldsymbol{\Gamma}}_{q,A}\mathbf{b}
\end{bmatrix}^{H} 
\begin{bmatrix}
	0 \\ 1
\end{bmatrix} = -\mathbf{b}^{H}\overline{\boldsymbol{\Gamma}}_{q,A}\mathbf{b}
\end{eqnarray*}
Thus, the per-patch augmented Hessian reduces to 
\begin{equation*}
\mathcal{H}_{\mathcal{Z}\mathcal{Z}}(\textsf{f}_q)  = \overline{R}^{q}_{\gamma} \left(  \mathbf{b}^{H}\overline{\boldsymbol{\Gamma}}_{q,H}\mathbf{b}  \begin{bmatrix}
\ZeroMat{J}{J} & \overline{\boldsymbol{\Gamma}}_{q,H}^{T} \\
\overline{\boldsymbol{\Gamma}}_{q,H} & \ZeroMat{J}{J}
\end{bmatrix} -   \mathbf{b}^{H}\overline{\boldsymbol{\Gamma}}_{q,A}\mathbf{b} \begin{bmatrix}
\ZeroMat{J}{J} & \overline{\boldsymbol{\Gamma}}_{q,A}^{T} \\
\overline{\boldsymbol{\Gamma}}_{q,A} & \ZeroMat{J}{J}
\end{bmatrix} \right)
\end{equation*}
Based on the form, we observe that the only non-zero unique block of this matrix is the (2,1) block (since the (1,2) block is its transpose).  What is the (2,1) block? First, recall that 
\begin{eqnarray*}
	\mathbf{b}^{H}\overline{\boldsymbol{\Gamma}}_{q,H}\mathbf{b} & = & \real(\mathbf{w}^{H}\boldsymbol{\Gamma}_{q}\mathbf{s}) \quad
	\mathbf{b}^{H}\overline{\boldsymbol{\Gamma}}_{q,H}\mathbf{b} = \jmath\imag(\mathbf{w}^{H}\boldsymbol{\Gamma}\mathbf{s}) \\
	\overline{\boldsymbol{\Gamma}}_{q,H} & = & \frac{1}{2}(\overline{\boldsymbol{\Gamma}}_{q}+\overline{\boldsymbol{\Gamma}}_{q}^{H}) \quad \overline{\boldsymbol{\Gamma}}_{q,A}  =  \frac{1}{2}(\overline{\boldsymbol{\Gamma}}_{q}-\overline{\boldsymbol{\Gamma}}_{q}^{H}).
\end{eqnarray*}
Temporarily omitting the $\overline{R}^{q}_{\gamma}$ term, the (2,1) block is
\begin{eqnarray*}
	(\mathbf{b}^{H}\overline{\boldsymbol{\Gamma}}_{q,H}\mathbf{b}) \overline{\boldsymbol{\Gamma}}_{q,H} - (\mathbf{b}^{H}\overline{\boldsymbol{\Gamma}}_{q,A}\mathbf{b}) \overline{\boldsymbol{\Gamma}}_{q,A} & = &  \frac{1}{2}\left(\real(\mathbf{w}^{H}\boldsymbol{\Gamma}_{q}\mathbf{s})(\overline{\boldsymbol{\Gamma}}_{q}+\overline{\boldsymbol{\Gamma}}_{q}^{H}) -  \jmath\imag(\mathbf{w}^{H}\boldsymbol{\Gamma}_{q}\mathbf{s})(\overline{\boldsymbol{\Gamma}}_{q}-\overline{\boldsymbol{\Gamma}}_{q}^{H})\right) \\
	& = & \frac{1}{2}\bigg(\left(\real(\mathbf{w}^{H}\boldsymbol{\Gamma}_{q}\mathbf{s})-\jmath\imag(\mathbf{w}^{H}\boldsymbol{\Gamma}_{q}\mathbf{s})\right)\overline{\boldsymbol{\Gamma}}_{q} \\
	&  & {}+ \left(\real(\mathbf{w}^{H}\boldsymbol{\Gamma}_{q}\mathbf{s})+\jmath\imag(\mathbf{w}^{H}\boldsymbol{\Gamma}_{q}\mathbf{s})\right)\overline{\boldsymbol{\Gamma}}_{q}^{H}\bigg) \\
	& = & \frac{1}{2}\left((\mathbf{w}^{H}\boldsymbol{\Gamma}_{q}\mathbf{s})^{*}\overline{\boldsymbol{\Gamma}}_{q} + (\mathbf{w}^{H}\boldsymbol{\Gamma}_{q}\mathbf{s})\overline{\boldsymbol{\Gamma}}_{q}^{H}\right) \\
	& = & \left((\mathbf{w}^{H}\boldsymbol{\Gamma}_{q}\mathbf{s})^{*}\overline{\boldsymbol{\Gamma}}_{q}\right)_{H}
\end{eqnarray*}
Therefore, the final form of the augmented per-patch Hessian is
\begin{equation*}
\mathcal{H}_{\mathcal{Z}\mathcal{Z}}(\textsf{f}_q)  =  \overline{R}^{q}_{\gamma}  \begin{bmatrix}
\ZeroMat{J}{J} & \left((\mathbf{w}^{H}\boldsymbol{\Gamma}_{q}\mathbf{s})^{*}\overline{\boldsymbol{\Gamma}}_{q}\right)_{H}^{T} \\
\left((\mathbf{w}^{H}\boldsymbol{\Gamma}_{q}\mathbf{s})^{*}\overline{\boldsymbol{\Gamma}}_{q}\right)_{H} & \ZeroMat{J}{J}
\end{bmatrix}
\end{equation*}
Using the second fact in Section 1, the final form of the \emph{true} Hessian is
\begin{equation*}
\mathcal{\widetilde{H}}[\textsf{f}_{q}]  = \overline{R}^{q}_{\gamma}  \begin{bmatrix}
\left((\mathbf{w}^{H}\boldsymbol{\Gamma}_{q}\mathbf{s})^{*}\overline{\boldsymbol{\Gamma}}_{q}\right)_{H} & \ZeroMat{J}{J} \\
\ZeroMat{J}{J}& \left((\mathbf{w}^{H}\boldsymbol{\Gamma}_{q}\mathbf{s})^{*}\overline{\boldsymbol{\Gamma}}_{q}\right)_{H}^{T}
\end{bmatrix} = \begin{bmatrix}
\mathcal{H}_{\mathbf{b}\mathbf{b}^{*}}(\textsf{f}_q) & \ZeroMat{J}{J} \\
\ZeroMat{J}{J}& \mathcal{H}_{\mathbf{b}\mathbf{b}^{*}}(\textsf{f}_q)^{T}
\end{bmatrix}
\end{equation*}
where we have defined $\mathcal{H}_{\mathbf{b}\mathbf{b}^{*}}(\textsf{f}_q) = \overline{R}^{q}_{\gamma}   \left((\mathbf{w}^{H}\boldsymbol{\Gamma}_{q}\mathbf{s})^{*}\overline{\boldsymbol{\Gamma}}_{q}\right)_{H}$. 
\subsection{STAP Step 4: Definiteness of $\mathcal{\widetilde{H}}[\textsf{f}_{q}]$}
In order for $\textsf{f}_q$ to be convex in $\mathbf{b}$ , then $\mathcal{\widetilde{H}}[\textsf{f}_{q}]$ must be positive-semidefinite for all valid $\mathbf{b}_{o}$. Since this Hessian is block diagonal, it is PSD if and only if each of the diagonal blocks are PSD. Furthermore, since the (2,2) block is the transpose of the (1,1) block, the (2,2) block's definiteness is the same as the (1,1) block's definiteness. Therefore, $\mathcal{\widetilde{H}}[\textsf{f}_{q}]$ is PSD iff $\mathcal{H}_{\mathbf{b}\mathbf{b}^{*}}(\textsf{f}_{q}) =  \overline{R}^{q}_{\gamma}\left((\mathbf{w}^{H}\boldsymbol{\Gamma}_{q}\mathbf{s})^{*}\overline{\boldsymbol{\Gamma}}_{q}\right)_{H}$  is PSD. We will show definitively that $\mathcal{H}_{\mathbf{b}\mathbf{b}^{*}}(\textsf{f}_{q})$ \emph{cannot be} PSD solely by its form.

First, $\mathcal{H}_{\mathbf{b}\mathbf{b}^{*}}(\textsf{f}_{q})$ is given by 
\begin{eqnarray*}
	\mathcal{H}_{\mathbf{b}\mathbf{b}^{*}}(\textsf{f}_{q}) & = & \overline{R}^{q}_{\gamma}\left((\mathbf{w}^{H}\boldsymbol{\Gamma}_{q}\mathbf{s})^{*}\overline{\boldsymbol{\Gamma}}_{q}\right)_{H} \\
	& = & \frac{\overline{R}^{q}_{\gamma}}{2}\left((\mathbf{w}^{H}\boldsymbol{\Gamma}_{q}\mathbf{s})^{*}\overline{\boldsymbol{\Gamma}}_{q} + (\mathbf{w}^{H}\boldsymbol{\Gamma}_{q}\mathbf{s})\overline{\boldsymbol{\Gamma}}_{q}^{H}\right)\\
	& = &  \begin{bmatrix}
		\ZeroMat{NML}{NML} &\frac{\overline{R}^{q}_{\gamma}}{2} (\mathbf{w}^{H}\boldsymbol{\Gamma}_{q}\mathbf{s})^{*}\boldsymbol{\Gamma}_{q} \\
		\frac{\overline{R}^{q}_{\gamma}}{2} (\mathbf{w}^{H}\boldsymbol{\Gamma}_{q}\mathbf{s})\boldsymbol{\Gamma}_{q}^{H} & \ZeroMat{N}{N}
	\end{bmatrix}
\end{eqnarray*}
It is well known (see \cite[Lemma~6.32, p.~124]{Seber2007}) that a complex matrix with this form is indefinite. Why? 
Due to the structure, its eigenvalues are:
\begin{itemize}
	\item the $N$ singular values of $\frac{\overline{R}^{q}_{\gamma}}{2} (\mathbf{w}^{H}\boldsymbol{\Gamma}_{q}\mathbf{s})^{*}\boldsymbol{\Gamma}_{q}$ (the upper right corner matrix), which are positive;
	\item the \emph{negatives} of the singular values of $\frac{\overline{R}^{q}_{\gamma}}{2} (\mathbf{w}^{H}\boldsymbol{\Gamma}_{q}\mathbf{s})^{*}\boldsymbol{\Gamma}_{q}$, which are negative;
	\item and zeroes for the rest. 
\end{itemize}
What are these singular values of $\frac{\overline{R}^{q}_{\gamma}}{2} (\mathbf{w}^{H}\boldsymbol{\Gamma}_{q}\mathbf{s})^{*}\boldsymbol{\Gamma}_{q}$? First, 
\begin{eqnarray*}
	\left(\frac{\overline{R}^{q}_{\gamma}}{2} (\mathbf{w}^{H}\boldsymbol{\Gamma}_{q}\mathbf{s})^{*}\boldsymbol{\Gamma}_{q}\right)^{H}\left(\frac{\overline{R}^{q}_{\gamma}}{2} (\mathbf{w}^{H}\boldsymbol{\Gamma}_{q}\mathbf{s})^{*}\boldsymbol{\Gamma}_{q}\right) & = & 
	\left(\frac{\overline{R}^{q}_{\gamma}}{2}\right)^{2} (\mathbf{w}^{H}\boldsymbol{\Gamma}_{q}\mathbf{s})(\mathbf{w}^{H}\boldsymbol{\Gamma}_{q}\mathbf{s})^{*}\boldsymbol{\Gamma}_{q}^{H}\boldsymbol{\Gamma}_{q}\\
	& = & \left(\frac{\overline{R}^{q}_{\gamma}}{2}\right)^{2} |\mathbf{w}^{H}\boldsymbol{\Gamma}_{q}\mathbf{s}|^2\boldsymbol{\Gamma}_{q}^{H}\boldsymbol{\Gamma}_{q} \\
	& = & \left(\frac{\overline{R}^{q}_{\gamma}}{2}\right)^{2} |\mathbf{w}^{H}\boldsymbol{\Gamma}_{q}\mathbf{s}|^2 (\mathbf{c}_q^{H} \otimes \Identity{N})\Commut{LN}{M}^{T}\Commut{LN}{M}(\mathbf{c}_q \otimes \Identity{N}) \\
	& = & \left(\frac{\overline{R}^{q}_{\gamma}}{2}\right)^{2} |\mathbf{w}^{H}\boldsymbol{\Gamma}_{q}\mathbf{s}|^2 \normsq{\mathbf{c}_q}\Identity{N}
\end{eqnarray*}
It is obvious that the eigenvalues of this matrix are $\left(\frac{\overline{R}^{q}_{\gamma}}{2}\right)^{2} |\mathbf{w}^{H}\boldsymbol{\Gamma}_{q}\mathbf{s}|^2 \normsq{\mathbf{c}_q}$ with multiplicity $N$. The singular values of $\frac{\overline{R}^{q}_{\gamma}}{2} (\mathbf{w}^{H}\boldsymbol{\Gamma}_{q}\mathbf{s})^{*}\boldsymbol{\Gamma}_{q}$ are then the positive square roots of this eigenvalue, or $\frac{\overline{R}^{q}_{\gamma}}{2}|\mathbf{w}^{H}\boldsymbol{\Gamma}_{q}\mathbf{s}| \norm{\mathbf{c}_q}$.

If a matrix has negative eigenvalues, it cannot be PSD; similarly, if it has positive eigenvalues, it cannot be negative semidefinite.  The only way this matrix has no negative eigenvalues is if they are all \emph{zero}. Since two of the terms in the singular value expression are always positive, this eigenvalue is zero only if the optimal signal/beamformer pair nulls the $q$th clutter patch. Therefore, the per-patch clutter Hessian is indefinite and the per-patch clutter cost is nonconvex. This is true for any choice of signal or beamformer on any choice of set.

\subsection{STAP Step 5: Definiteness of $\mathcal{\widetilde{H}}[\textsf{f}_{C}]$}
Since the overall clutter cost is a sum of the per-patch costs, its Hessian is the sum of the per-patch clutter Hessians. In other words,
\begin{eqnarray*}
	\mathcal{\widetilde{H}}[\textsf{f}_{C}] & = & \sum_{q=1}^{Q} \mathcal{\widetilde{H}}[\textsf{f}_{q}] \\
	& = & \sum_{q=1}^{Q} \overline{R}^{q}_{\gamma}  \begin{bmatrix}
		\left((\mathbf{w}^{H}\boldsymbol{\Gamma}_{q}\mathbf{s})^{*}\overline{\boldsymbol{\Gamma}}_{q}\right)_{H} & \ZeroMat{J}{J} \\
		\ZeroMat{J}{J}& \left((\mathbf{w}^{H}\boldsymbol{\Gamma}_{q}\mathbf{s})^{*}\overline{\boldsymbol{\Gamma}}_{q}\right)_{H}^{T}
	\end{bmatrix} \\
	& = &   \begin{bmatrix}\sum_{q=1}^{Q}\overline{R}^{q}_{\gamma} \left((\mathbf{w}^{H}\boldsymbol{\Gamma}_{q}\mathbf{s})^{*}\overline{\boldsymbol{\Gamma}}_{q}\right)_{H} & \ZeroMat{J}{J} \\
		\ZeroMat{J}{J}& \sum_{q=1}^{Q}\overline{R}^{q}_{\gamma}\left((\mathbf{w}^{H}\boldsymbol{\Gamma}_{q}\mathbf{s})^{*}\overline{\boldsymbol{\Gamma}}_{q}\right)_{H}^{T}
	\end{bmatrix} \\
	& = & \begin{bmatrix}
		\sum_{q=1}^{Q}\mathcal{H}_{\mathbf{b}\mathbf{b}^{*}}(\textsf{f}_{q}) & \ZeroMat{J}{J} \\
		\ZeroMat{J}{J}& \sum_{q=1}^{Q}\mathcal{H}_{\mathbf{b}\mathbf{b}^{*}}(\textsf{f}_{q})^{T} 
	\end{bmatrix} = \begin{bmatrix}
	\mathcal{H}_{\mathbf{b}\mathbf{b}^{*}}(\textsf{f}_{C}) & \ZeroMat{J}{J} \\
	\ZeroMat{J}{J}& \mathcal{H}_{\mathbf{b}\mathbf{b}^{*}}(\textsf{f}_{C})^{T} 
\end{bmatrix}
\end{eqnarray*}
Similar to the per-patch clutter Hessian, the definiteness of the overall clutter Hessian is determined by the definiteness of $\mathcal{H}_{\mathbf{b}\mathbf{b}^{*}}(\textsf{f}_{C})$. Therefore, let us take a closer look at $\mathcal{H}_{\mathbf{b}\mathbf{b}^{*}}(\textsf{f}_{C})$:
\begin{eqnarray*}
	\mathcal{H}_{\mathbf{b}\mathbf{b}^{*}}(\textsf{f}_{C}) & = & \sum_{q=1}^{Q} \begin{bmatrix}
		\ZeroMat{NML}{NML} &\frac{\overline{R}^{q}_{\gamma}}{2} (\mathbf{w}^{H}\boldsymbol{\Gamma}_{q}\mathbf{s})^{*}\boldsymbol{\Gamma}_{q} \\
		\frac{\overline{R}^{q}_{\gamma}}{2} (\mathbf{w}^{H}\boldsymbol{\Gamma}_{q}\mathbf{s})\boldsymbol{\Gamma}_{q}^{H} & \ZeroMat{N}{N}
	\end{bmatrix} \\
	& = & \begin{bmatrix}
		\ZeroMat{NML}{NML} &\sum_{q=1}^{Q}\frac{\overline{R}^{q}_{\gamma}}{2} (\mathbf{w}^{H}\boldsymbol{\Gamma}_{q}\mathbf{s})^{*}\boldsymbol{\Gamma}_{q} \\
		\sum_{q=1}^{Q}\frac{\overline{R}^{q}_{\gamma}}{2} (\mathbf{w}^{H}\boldsymbol{\Gamma}_{q}\mathbf{s})\boldsymbol{\Gamma}_{q}^{H} & \ZeroMat{N}{N}
	\end{bmatrix} .
\end{eqnarray*}
Define the $NML\times N$ complex matrix $\boldsymbol{\Pi}_{Q} = \sum_{q=1}^{Q} \frac{\overline{R}^{q}_{\gamma}}{2} (\mathbf{w}^{H}\boldsymbol{\Gamma}_{q}\mathbf{s})^{*}\boldsymbol{\Gamma}_{q}$. Hence, 
\begin{equation*}
\mathcal{H}_{\mathbf{b}\mathbf{b}^{*}}(\textsf{f}_{C}) = \begin{bmatrix}
\ZeroMat{NML}{NML} &\boldsymbol{\Pi}_{Q}\\
\boldsymbol{\Pi}_{Q}^{H} & \ZeroMat{N}{N}
\end{bmatrix}
\end{equation*}
Once again, the previous structure occurs. By the lemma, the eigenvalues of $\mathcal{H}_{\mathbf{b}\mathbf{b}^{*}}(\textsf{f}_{C})$ are the $N$ singular values of $\boldsymbol{\Pi}_{Q}$ (i.e., the square roots of the eigenvalues of $\boldsymbol{\Pi}_{Q}^{H}\boldsymbol{\Pi}_{Q}$), the negatives of those singular values, and $N(ML-1)$ zeros. 

What are these singular values? First, let us examine $\boldsymbol{\Pi}_{Q}^{H}\boldsymbol{\Pi}_{Q}$:
\begin{eqnarray*}
	\boldsymbol{\Pi}_{Q}^{H}\boldsymbol{\Pi}_{Q} & = & \left	(\sum_{q=1}^{Q} \frac{\overline{R}^{q}_{\gamma}}{2} (\mathbf{w}^{H}\boldsymbol{\Gamma}_{q}\mathbf{s})\boldsymbol{\Gamma}_{q}^{H}\right)\left(\sum_{l=1}^{Q} \frac{\overline{R}^{l}_{\gamma}}{2} (\mathbf{w}^{H}\boldsymbol{\Gamma}_{l}\mathbf{s})^{*}\boldsymbol{\Gamma}_{l}\right) \\
	& = & \frac{1}{4} \sum_{q=1}^{Q} \sum_{l=1}^{Q}\overline{R}^{q}_{\gamma}\overline{R}^{l}_{\gamma}(\mathbf{w}^{H}\boldsymbol{\Gamma}_{q}\mathbf{s})(\mathbf{w}^{H}\boldsymbol{\Gamma}_{l}\mathbf{s})^{*}\boldsymbol{\Gamma}_{q}^{H}\boldsymbol{\Gamma}_{l}
\end{eqnarray*}
To simplify things slightly, we can find a different form of  $\boldsymbol{\Gamma}_{q}^{H}\boldsymbol{\Gamma}_{l}$:
\begin{eqnarray*}
	\boldsymbol{\Gamma}_{q}^{H}\boldsymbol{\Gamma}_{l} & = & (\mathbf{c}_q^{H} \otimes \Identity{N})\Commut{LN}{M}^{T}\Commut{LN}{M}(\mathbf{c}_l \otimes \Identity{N}) \\
	& = & \left(\mathbf{c}_q^{H}\mathbf{c}_l \right)\Identity{N}
\end{eqnarray*}
Returning to the original matrix $\boldsymbol{\Pi}_{Q}^{H}\boldsymbol{\Pi}_{Q}$, we find
\begin{eqnarray*}
	\boldsymbol{\Pi}_{Q}^{H}\boldsymbol{\Pi}_{Q} & = & \frac{1}{4} \sum_{q=1}^{Q} \sum_{l=1}^{Q}\overline{R}^{q}_{\gamma}\overline{R}^{l}_{\gamma}(\mathbf{w}^{H}\boldsymbol{\Gamma}_{q}\mathbf{s})(\mathbf{w}^{H}\boldsymbol{\Gamma}_{l}\mathbf{s})^{*} \left(\mathbf{c}_q^{H}\mathbf{c}_l \right)\Identity{N} \\
	& = & \left(\frac{1}{4} \sum_{q=1}^{Q} \sum_{l=1}^{Q}\overline{R}^{q}_{\gamma}\overline{R}^{l}_{\gamma}(\mathbf{w}^{H}\boldsymbol{\Gamma}_{q}\mathbf{s})(\mathbf{w}^{H}\boldsymbol{\Gamma}_{l}\mathbf{s})^{*} \left(\mathbf{c}_q^{H}\mathbf{c}_l \right)\right)\Identity{N}
\end{eqnarray*}
If we define the matrix $\mathbf{D} = \begin{bmatrix}(\mathbf{w}^{H}\boldsymbol{\Gamma}_{1}\mathbf{s})^{*}\mathbf{c}_1 \cdots (\mathbf{w}^{H}\boldsymbol{\Gamma}_{Q}\mathbf{s})^{*}\mathbf{c}_Q \end{bmatrix}$, then the scalar inside the double sum above is the $(q,l)$th element of the matrix $\mathbf{D}^{H}\mathbf{D}$, which is a Gramian matrix and is thus positive semidefinite. The double sum, then, is the scalar $\mathbf{1}_{Q}^{T} \mathbf{D}^{H}\mathbf{D}\mathbf{1}_{Q}$, which is clearly positive. Therefore, a final form of $\boldsymbol{\Pi}_{Q}^{H}\boldsymbol{\Pi}_{Q}$ is 
\begin{equation*}
\boldsymbol{\Pi}_{Q}^{H}\boldsymbol{\Pi}_{Q} = \frac{\mathbf{1}_{Q}^{T} \mathbf{D}^{H}\mathbf{D}\mathbf{1}_{Q}}{4}\Identity{N}
\end{equation*}
It is immediately obvious that the matrix has only one eigenvalue -- $ \tfrac{\mathbf{1}_{Q}^{T} \mathbf{D}^{H}\mathbf{D}\mathbf{1}_{Q}}{4}$ -- with multiplicity $N$. This eigenvalue is either always positive or zero. From its form, we can surmise that it is zero only in the case when $\mathbf{w}^{H}\boldsymbol{\Gamma}_{q}\mathbf{s} = 0$ for all clutter patches $q$. Why is this so? 

First, if $\mathbf{w}^{H}\boldsymbol{\Gamma}_{q}\mathbf{s} \neq 0$ for all $q$, then $\mathbf{1}_{Q}^{T} \mathbf{D}^{H}\mathbf{D}\mathbf{1}_{Q} = 0$ only if the inner products $\mathbf{c}_q^{H}\mathbf{c}_l = 0$ for all $q, l$. However, if this is true, then $\mathbf{c}_q^{H}\mathbf{c}_q = \normsq{\mathbf{c}_q} = 0$ for all $q$; in other words, \emph{there is no clutter}, which is a ridiculous requirement.  
What if we relax the inner product condition to $\mathbf{c}_q^{H}\mathbf{c}_l = 0$ for $q \neq  l$? This doesn't change things much, because then $\mathbf{1}_{Q}^{T} \mathbf{D}^{H}\mathbf{D}\mathbf{1}_{Q} = \sum_{q=1}^{Q}\frac{(\overline{R}^{q}_{\gamma})^2}{2}|\mathbf{w}^{H}\boldsymbol{\Gamma}_{q}\mathbf{s}|^{2} \normsq{\mathbf{c}_q}$ . This is absolutely positive unless $\mathbf{w}^{H}\boldsymbol{\Gamma}_{q}\mathbf{s} = 0$ for all $q$, which is the condition we were trying to avoid.

The singular values of $\boldsymbol{\Pi}_{Q}$ are therefore clearly only positive, implying $\mathcal{H}_{\mathbf{b}\mathbf{b}^{*}}(\textsf{f}_{C})$ has positive, negative, and zero eigenvalues. Hence, the clutter Hessian is indefinite, and the clutter cost function is non-convex in $\mathbf{b}$.

\subsection{STAP Step 6: Definiteness of $\mathcal{\widetilde{H}}[\textsf{f}_{o}]$}\label{sec:}
At this point, one might think adding a full rank PSD matrix (in this case, the combined noise-and-interference correlation matrix) somewhere will break up the useful structure from the previous sections.  However, this is not the case, and a similarly useful structure appears in the Hessian of the complete cost function.
We begin by noting that, since $\textsf{f}_o$ is a sum of $\textsf{f}_{NI}$ and $f_C$, the Hessian is also a sum: $\mathcal{\widetilde{H}}[\textsf{f}_{o}] = \mathcal{\widetilde{H}}[f_{NI}]+ \mathcal{\widetilde{H}}[\textsf{f}_{C}]$. 
Since $f_{NI} = \mathbf{b}^{H}\begin{bmatrix}
\mathbf{R_{ni}} & \ZeroMat{NML}{N} \\
\ZeroMat{N}{NML} & \ZeroMat{N}{N}
\end{bmatrix}\mathbf{b} = \mathbf{b}^{H}\mathbf{\widetilde{R}_{ni}}\mathbf{b}$ is a real quadratic form, we once again turn to Hj{\o}rungnes' Example 5.2 as cited above, which means the Hessian $ \mathcal{\widetilde{H}}[f_{NI}]$ is 
\begin{equation*}
\mathcal{\widetilde{H}}[f_{NI}] = \begin{bmatrix}
\mathbf{\widetilde{R}_{ni}} & \ZeroMat{J}{J} \\
\ZeroMat{J}{J} & \mathbf{\widetilde{R}_{ni}}^{T}
\end{bmatrix}
\end{equation*}
or, more compactly, $\mathcal{H}_{\mathbf{b}\mathbf{b}^{*}}(f_{NI}) = \mathbf{\widetilde{R}_{ni}}$. 

Hence, the total Hessian of the cost function $\mathcal{\widetilde{H}}[\textsf{f}_{o}]$ is 
\begin{eqnarray*}
	\mathcal{\widetilde{H}}[\textsf{f}_{o}]  & = & \begin{bmatrix}
		\mathbf{\widetilde{R}_{ni}} & \ZeroMat{J}{J} \\
		\ZeroMat{J}{J} & \mathbf{\widetilde{R}_{ni}}^{T}
	\end{bmatrix} +  \begin{bmatrix}
	\mathcal{H}_{\mathbf{b}\mathbf{b}^{*}}(\textsf{f}_{C}) & \ZeroMat{J}{J} \\
	\ZeroMat{J}{J}& \mathcal{H}_{\mathbf{b}\mathbf{b}^{*}}(\textsf{f}_{C})^{T} 
\end{bmatrix} \\
& = & \begin{bmatrix}
	\mathcal{H}_{\mathbf{b}\mathbf{b}^{*}}(\textsf{f}_{o}) & \ZeroMat{J}{J} \\
	\ZeroMat{J}{J}& \mathcal{H}_{\mathbf{b}\mathbf{b}^{*}}(\textsf{f}_{o})^{T} 
\end{bmatrix}
\end{eqnarray*}
where $\mathcal{H}_{\mathbf{b}\mathbf{b}^{*}}(\textsf{f}_{o})  = \mathbf{\widetilde{R}_{ni}} + \mathcal{H}_{\mathbf{b}\mathbf{b}^{*}}(\textsf{f}_{C})$ . Again, due to the structure, the definiteness of $\mathcal{H}_{\mathbf{b}\mathbf{b}^{*}}(\textsf{f}_{o})$ determines the definiteness of the complete Hessian. 
In order to ascertain the definiteness of $\mathcal{H}_{\mathbf{b}\mathbf{b}^{*}}(\textsf{f}_{o})$, we need to observe its structure:
\begin{eqnarray*}
	\mathcal{H}_{\mathbf{b}\mathbf{b}^{*}}(\textsf{f}_{o}) & = & \mathbf{\widetilde{R}_{ni}} + \mathcal{H}_{\mathbf{b}\mathbf{b}^{*}}(\textsf{f}_{C}) \\
	& = & \begin{bmatrix}
		\mathbf{R_{ni}} &\boldsymbol{\Pi}_{Q}\\
		\boldsymbol{\Pi}_{Q}^{H} & \ZeroMat{N}{N}
	\end{bmatrix}.
\end{eqnarray*}

Before continuing, we introduce two essential, yet equivalent, theorems from Kreindler \& Jameson \cite{Kreindler1972} for the definiteness of a partitioned matrix. 
\begin{theorem}[via~\cite{Kreindler1972},Theorems $I_{a^{\prime\prime}}$ \&  $I_{b^{\prime\prime}}$ ]\label{thm:PSDness}
	A matrix $\mathbf{M}$ partitioned as $\mathbf{M} = \begin{bmatrix}
	\mathbf{A} & \mathbf{B} \\
	\mathbf{B}^{H} & \mathbf{C}
	\end{bmatrix}$
	is non-negative definite if and only if the following conditions are satisfied:\\
	\begin{enumerate}[label={\bf A\arabic*}]
		\item $\mathbf{C} \succeq 0$ 
		\item $\mathbf{B} = \mathbf{B}\mathbf{C}^{\dagger}\mathbf{C}$
		\item  $\mathbf{A} - \mathbf{B}\mathbf{C}^{\dagger} \mathbf{B}^{H} \succeq 0$
	\end{enumerate}
	or
	\begin{enumerate}[label={\bf B\arabic*}]
		\item $\mathbf{A} \succeq 0$ 
		\item $\mathbf{B} = \mathbf{A} \mathbf{A}^{\dagger} \mathbf{B}$
		\item  $\mathbf{C} - \mathbf{B}^{H}\mathbf{A}^{\dagger} \mathbf{B} \succeq 0$
	\end{enumerate}
	where $\mathbf{A}^{\dagger}$ indicates the Moore-Penrose pseudoinverse of the matrix $\mathbf{A}$ and $\mathbf{A}\succeq 0$ means that $\mathbf{A}$ is non-negative definite. 
\end{theorem}
In the case of the matrix $\mathcal{H}_{\mathbf{b}\mathbf{b}^{*}}(f_{o})$, $\mathbf{A} = \mathbf{R_{ni}}, \mathbf{B} = \boldsymbol{\Pi}_{Q}$, and $\mathbf{C} = \ZeroMat{N}{N}$.  The proof of definiteness follows by checking necessary conditions \& observing how and if they are violated. We propose that this matrix is not PSD unless we know, a priori, that all stationary points null every clutter patch. We present two equivalent proofs of our proposition from the sets of A and B conditions in the above theorem. 

\begin{proof}[Proof A]
	Any quadratic form involving the all-zero matrix $\ZeroMat{N}{N}$ is zero, implying the zero matrix is simultaneously positive semidefinite, negative semidefinite, and indefinite. Thus, Condition A1 is immediately satisfied. 
	
	Conditions A2 \& A3 require the Moore-Penrose pseudoinverse of the all-zero matrix, which is itself: $\ZeroMat{N}{N}^{\dagger} = \ZeroMat{N}{N}^{\phantom{*}}$. Then, we can immediately restate condition A2 as $\boldsymbol{\Pi}_Q = \ZeroMat{NML}{N}$. As has been previously established, this only occurs if $\mathbf{w}_{o}^{H}\boldsymbol{\Gamma}_{q}\mathbf{s}_{o} = 0$ for all clutter patches $q$ and all stationary points $\mathbf{w}_{o},\mathbf{s}_{o}$ in the feasible set.
	
	After simplification, Condition A3 is equivalent to $\mathbf{R_{ni}} \succeq 0$. Since $\mathbf{R_{ni}}$ is a covariance matrix and all covariance matrices are positive semidefinite by construction, Condition A3 is immediately satisfied.
\end{proof}

\begin{proof}[Proof B] 
	Since $\mathbf{R_{ni}}$ is a covariance matrix and all covariance matrices are positive semidefinite by construction, Condition B1 is immediately satisfied.
	
	As a consequence of this result, we can use \cite[Lemma~7.63(b)]{Seber2007} to immediately state that because $\mathbf{R_{ni}}$ is positive semidefinite and Hermitian, its Moore-Penrose pseudoinverse $\mathbf{R_{ni}}^{\dagger}$ is also positive semidefinite. 
	
	Furthermore, using \cite[Lemma~10.46(b)(i)]{Seber2007}, we can say that for \emph{any} matrix $\mathbf{X} \in \Complex^{N\times NML}$, $\mathbf{X}^{H}\mathbf{R_{ni}}^{\dagger}\mathbf{X} \succeq 0$. Let us set $\mathbf{X} = \boldsymbol{\Pi}_Q$. Then, this lemma allows us to state $\boldsymbol{\Pi}_{Q}^{H}\mathbf{R_{ni}}^{\dagger}\boldsymbol{\Pi}_Q \succeq 0$, regardless of the choice of $\mathbf{b}$ that composes $\boldsymbol{\Pi}_{Q}$. 
	
	However, this statement is, on its face, an immediate contradiction to Condition B3, unless $\boldsymbol{\Pi}_{Q}^{H}\mathbf{R_{ni}}^{\dagger}\boldsymbol{\Pi}_Q = \ZeroMat{N}{N}$. By  \cite[Lemma~10.12(b)]{Seber2007}, this is only true if $\boldsymbol{\Pi}_{Q}^{H}\mathbf{R_{ni}}^{\dagger} = \ZeroMat{N}{NML}$ or, similarly, $\mathbf{R_{ni}}^{\dagger}\boldsymbol{\Pi}_Q = \ZeroMat{NML}{N}$
	
	If, indeed, $\mathbf{R_{ni}}^{\dagger}\boldsymbol{\Pi}_Q = \ZeroMat{N}{NML}$, then that would mean Condition B2 would read
	\begin{align*}
		\boldsymbol{\Pi}_Q & =  \mathbf{R_{ni}} \mathbf{R_{ni}}^{\dagger} \boldsymbol{\Pi}_{Q} \\
		& =  \mathbf{R_{ni}}\ZeroMat{NML}{N} = \ZeroMat{NML}{N}.
	\end{align*} 
	But, as we established before, this is only possible if, for each stationary point $\mathbf{w}_{o},\mathbf{s}_{o}$ in the feasible set, $\mathbf{w}_{o}^{H}\boldsymbol{\Gamma}_{q}\mathbf{s}_{o} = 0$ for all clutter patches $q$. 
\end{proof}

A third proof relies on a similar construction of Theorem~\ref{thm:PSDness} from \cite{HornJohnsonMA2e}, which we paraphrase in Theorem~\ref{thm:PSDnessHJ} below:
\begin{theorem}[via \cite{HornJohnsonMA2e}, Theorem 7.7.9(a, b)]\label{thm:PSDnessHJ}
	Let the matrix $\mathbf{M}$ be partitioned as in Theorem~\ref{thm:PSDness}. The following statements are equivalent:
	\begin{enumerate}
		\item $\mathbf{M}$ is positive semidefinite
		\item $\Abf$ and $\Cbf$ are positive semidefinite and there is a contraction $\Xbf$ shaped identically to $\Bbf$ such that $\Bbf = \Abf^{1/2}\Xbf\Cbf^{1/2}$.
	\end{enumerate}
\end{theorem}
In the above theorem, a contraction is defined any matrix whose largest singular value is less than or equal to one, and the superscript $1/2$ denotes the unique matrix square root defined for all PSD matrices. Observe also that this theorem implies that $\Range{\Bbf} \subseteq \Range{\Abf}, \Range{\Bbf^{H}} \subseteq \Range{\Cbf}$. 
Our proof continues as follows:
\begin{proof}[Proof C]
As established previously, $\Rni \succeq 0$ and $\ZeroMat{N}{N} \succeq 0$. Clearly, the matrix square root of the all zeros matrix is itself. Thus, for $\mathcal{H}_{\mathbf{b}\mathbf{b}^{*}}(f_{o})$ to be PSD, $\boldsymbol{\Pi}_Q = \Rni^{1/2}\Xbf\ZeroMat{N}{N}^{1/2} = \ZeroMat{NML}{N}$, which will be true for any contraction $\Xbf$. As established previously, $\boldsymbol{\Pi}_Q = \ZeroMat{NML}{N}$ only  if, for each stationary point $\mathbf{w}_{o},\mathbf{s}_{o}$ in the feasible set, $\mathbf{w}_{o}^{H}\boldsymbol{\Gamma}_{q}\mathbf{s}_{o} = 0$ for all clutter patches $q$. Therefore, this is the necessary and sufficient condition for $\mathcal{H}_{\mathbf{b}\mathbf{b}^{*}}(\textsf{f}_{o})$ to be PSD. 
\end{proof}

In any case, $\mathbf{w}_{o}^{H}\boldsymbol{\Gamma}_{q}\mathbf{s}_{o} = 0$ for all clutter patches $q$ and stationary points  $\mathbf{w}_{o},\mathbf{s}_{o}$ in order for the Hessian to be positive-semidefinite and the objective to be convex. This is a clearly illogical condition -- if such a beamformer-signal pair existed \emph{a priori}, there is effectively no clutter in the region of the desired target response, the signal design is arbitrary and we would only design a beamformer! Additionally, this condition also requires $\Rni\wbf_{o} = \ZeroMat{NML}{1}$, via the stationary point definitions. If $\Rni$ is full rank, then $\wbf_{o} = \ZeroMat{NML}{1}$, and we do no processing whatsoever.

Since the only possible ways to not obtain a contradiction are themselves contradictions, we must conclude that, regardless of the beamformer-signal pair, the overall cost function Hessian is indefinite (since the same contradiction would be reached if we desired a negative definite $\mathcal{H}_{\mathbf{b}\mathbf{b}^{*}}(\textsf{f}_{o})$ instead) and the problem is not jointly convex.

\section{The Biquadratic Program \& Relaxations}\label{sec:BQP_initial}
In this section, we discuss the relatively unexplored area of biquadratic programming -- that is, joint optimization of two multidimensional variables over a cost function that is quadratic in each variable -- and its application to fully adaptive radar, which arises from joint signal-beamformer design schemes. 

First, we will review the existing literature on biquadratic programming (henceforth, BQP) and related works on tensor approximations, in order to provide a basic mathematical framework. We will then recall the work of Setlur, et al. in order to construct a relevant BQP and its potentially solvable relaxations, with some additional insights into the other unique challenges this problem presents that are unaddressed by the literature.

Interestingly, coverage of the BQP in the literature, from the optimization community or otherwise, can be charitably described as sparse; in fact, there are only seven papers directly addressing the BQP produced by a total of ten authors, all of which have been released within the last six years. This is somewhat puzzling given the depth and breadth of related problems both cited in the works mentioned below and conceivable by an ordinary engineer -- applications range from economics to quantum physics to the fully-adaptive radar concept we wish to pursue here. 
Nevertheless, this limitation allows a reasonable overview of the state-of-the-art without too much difficulty. 
Other research exists on similar problems, mostly in the areas of nearest tensor approximation, but this is extensive and highly unspecialized. Therefore, we refer the reader to the multiple overlapping sources within the papers mentioned below on these matters.

\subsection{The General BQP}
We begin by describing the general biquadratic program analyzed by Ling, et al. in \cite{Ling2009}. Consider the vectors $\mathbf{x} \in \mathbb{R}^{n}$ and $\mathbf{y} \in \mathbb{R}^{m}$. The homogenous biquadratic optimization problem is
\begin{equation}
	\begin{aligned}
	& \underset{\mathbf{x},\mathbf{y}}{\min}
	& & b(\mathbf{x},\mathbf{y}) = \sum_{1\leq i,k \leq n, 1 \leq j,l \leq m} b_{ijkl} x_{i} y_{j} x_{k} y_{l} \\
	& \text{subject to}
	& & \norm{\mathbf{x}} = 1, \norm{\mathbf{y}} = 1,
	\end{aligned}\label{eq:RawBQP}
\end{equation}
where $b_{ijkl}$ is the $(i,j,k,l)$th element of the 4th-order tensor $\mathcal{B} \in \mathbb{R}^{n \times m \times n \times m}$, the subscripted $x$ and $y$ are the appropriate element of the corresponding vector, and $\norm{\cdot}$ is the standard Euclidean 2-norm. As the authors note, for fixed $\mathbf{x}$ (resp. $\mathbf{y}$), this problem is quadratic in $\mathbf{y}$ (resp. $\mathbf{x}$), can be solved quite easily and, in fact, is convex if the appropriate inner matrix is positive semidefinite. This property also leads to the name of this class of problems (cf. \emph{bilinear} optimization). Before we continue, it is worth noting that the objective function $b(\mathbf{x}, \mathbf{y})$ can be rewritten using a tensor operator and the matrix inner product, \emph{viz.}
\begin{equation*}
b(\mathbf{x},\mathbf{y}) = (\mathcal{B}\mathbf{x}\mathbf{x}^{T}) \bullet (\mathbf{y}\mathbf{y}^{T}) = (\mathbf{y}\mathbf{y}^{T}\mathcal{B}) \bullet (\mathbf{x}\mathbf{x}^{T}) 
\end{equation*}
where $\mathbf{X} \bullet \mathbf{Y} = \Tr(\mathbf{X}^T\mathbf{Y})$, and the tensor-operator matrices are $\mathcal{B}\mathbf{x}\mathbf{x}^{T} = \sum_{i,k = 1}^{n}b_{ijkl}x_{i}x_{k}$ and $\mathbf{y}\mathbf{y}^{T}\mathcal{B} = \sum_{j,l = 1}^{m}b_{ijkl}y_{j}y_{l}$. 

The above problem has been modified elsewhere in the literature, mostly in terms of the constraints. Bomze, et al. \cite{Bomze2012} constrain the objective in Equation~\ref{eq:RawBQP} to lie on the simplex for each variable. Ling, Zhang, and Qi have described strategies when the problem is quadratically constrained in \cite{Ling2011, Zhang2011}.  

Regardless of the constraints, \cite{Ling2009} proves that the BQP is nonconvex, that both the BQP and its naive semidefinite relaxation (which we will more explicitly discuss in the next section) are NP-hard problems, and that the initial general problem does not even admit a polynomial time \emph{approximation} algorithm. This would seem to bode ill for any hope of reasonable solutions, but the authors demonstrate that polynomial-time approximation algorithms exist (with bounds approximately inversely proportional to the dimension of the variables) when the objective is square-free or has squared terms in only one of the variables. Finally, they demonstrate the creation of such algorithms, showing a tradeoff between speed and accuracy for sum-of-squares based solvers (better accuracy) versus convex semidefinite relaxation solvers (faster). Additionally, if the structure is sufficiently sparse, then existing solvers can attack the semidefinite relaxation even more efficiently. 

\subsection{The Problem at Hand}
Before continuing, we recall the basics of our problem. We operate on a radar datacube collected over $L$ pulses with $N$ fast time samples per pulse in $M$ spatial bins. Our goal is to find a STAP beamformer vector $\mathbf{w} \in \Complex^{NML}$ and a transmit signal $\mathbf{s} \in \Complex^{N}$ that minimizes the combined effect of the noise and interference represented by the covariance matrix $\mathbf{R_{ni}} \in \Complex^{NML \times NML}$) and the signal-dependent clutter. The signal-dependent clutter is modeled as $Q$ independent clutter patches. Each patch has an individual spatiodoppler response matrix $\boldsymbol{\Gamma}_{q} \in \Complex^{NML\times N}$, which ties into the per-patch clutter covariance $\mathbf{R}_{\boldsymbol{\gamma}}^q(\mathbf{s}) = \boldsymbol{\Gamma}_{q}\mathbf{s}\mathbf{s}^{H}\boldsymbol{\Gamma}_{q}^{H}$. The overall clutter covariance is then $\sum\limits_{q=1}^Q \boldsymbol{\Gamma}_{q}\mathbf{s}\mathbf{s}^{H}\boldsymbol{\Gamma}_{q}^{H}$. 

We constrain this minimization in two ways. First, for a given target space-time-doppler bin, we want a particular filter output, say, $\kappa \in \Complex$. This filter output can be represented by the Capon beamformer equation $\mathbf{w}^H\mathbf{T}\mathbf{s}$, where $\mathbf{T} \in \Complex^{NML \times N}$ is the spatiodoppler response of the target bin. Second, we place an upper bound on the total signal power, say, $P_{o}$. With these constraints in place, the overall STAP problem is given by

\begin{equation*}
\begin{aligned}
& \underset{\mathbf{w}, \mathbf{s}}{\text{min}}
& & \mathbf{w}^H\mathbf{R_{ni}}\mathbf{w} + \mathbf{w}^{H} \left(\sum\limits_{q=1}^Q \boldsymbol{\Gamma}_{q}\mathbf{s}\mathbf{s}^{H}\boldsymbol{\Gamma}_{q}^{H}\right)\mathbf{w}\\
& \text{s.t.}
& & \mathbf{w}^H\mathbf{T}\mathbf{s}=\kappa \\
& & & \mathbf{s}^H \mathbf{s}\leq P_o
\end{aligned} \label{eq:BQPws} \tag{BQP 1}.
\end{equation*}	

If we treat the signal and beamformer as a single stacked variable, say $\mathbf{b} = \begin{bmatrix}
\mathbf{w}^{T} & \mathbf{s}^{T}
\end{bmatrix}^{T} \in \Complex^{N(ML+1)}$, we can find an equivalent form of the above optimization problem in the new variable. Define $\PsiW,\PsiS$ as the matrices that recover $\mathbf{w},\mathbf{s}$ from $\mathbf{b}$, i.e. $\mathbf{w} = \PsiW\mathbf{b}$ and $\mathbf{s}=\PsiS\mathbf{b}$. Then, we can define the optimization problem as
\begin{equation*}
\begin{aligned}
& \underset{\mathbf{b}}{\text{min}}
& & \mathbf{b}^H\mathbf{\widetilde{R}_{ni}}\mathbf{b} + \mathbf{b}^{H}\left(\sum\limits_{q=1}^Q \overline{\boldsymbol{\Gamma}}_{q}\mathbf{b}\mathbf{b}^{H}\overline{\boldsymbol{\Gamma}}_{q}^{H}\right)\mathbf{b}\\
& \text{s.t.}
& & \mathbf{b}^H\mathbf{\widetilde{T}}\mathbf{b}=\kappa \\
& & & \mathbf{b}^H\PsiS^{T}\PsiS\mathbf{b}\leq P_o
\end{aligned} \label{eq:BQPbb} \tag{BQP 2}
\end{equation*}	
where $\mathbf{\widetilde{R}_{ni}} = \PsiW^{T}\mathbf{R_{ni}}\PsiW$, $\overline{\boldsymbol{\Gamma}}_{q} = \PsiW^{T}\boldsymbol{\Gamma}_{q}\PsiS$, and $\mathbf{\widetilde{T}} = \PsiW^{T}\mathbf{T}\PsiS$ are just ``expanded" versions of the matrices seen in the previous problem.

In either case, these are clearly biquadratic programs, which we've shown are non-convex. The question now is: can we find a solvable representation, relaxed or otherwise?

\subsection{Getting to the SDR}
Our first goal is using the notation/mechanisms of the literature to find the semidefinite relaxation of the equivalent programs \ref{eq:BQPws} and \ref{eq:BQPbb}. 
First, assume we have defined a tensor $\mathcal{C}$ such that the following operator relation holds:  $\mathcal{C}\mathbf{s}\mathbf{s}^{H} = \sum_{q=1}^{Q} \mathbf{R}_{\boldsymbol{\gamma}}^q(\mathbf{s})$. Similiarly, assume an expanded form of this tensor exists, say $\widetilde{\mathcal{C}}$, such that $\widetilde{\mathcal{C}}\mathbf{b}\mathbf{b}^{H} = \sum\limits_{q=1}^Q \overline{\boldsymbol{\Gamma}}_{q}\mathbf{b}\mathbf{b}^{H}\overline{\boldsymbol{\Gamma}}_{q}^{H}$. What this tensor looks like, exactly, will be seen in a later section. 

On complex matrices, the inner product becomes $\mathbf{X} \bullet \mathbf{Y} = \real\{\Tr(\mathbf{X}^H\mathbf{Y})\}$. 
We'd like to get everything into a real form so we can use this inner product somehow, as in the literature. 

Note the first constraint (since $\kappa$ is complex) implies its conjugate must also be constrained, i.e.
\begin{align*}
\left(\mathbf{w}^H\mathbf{T}\mathbf{s}\right)^{*}& = \kappa^{*} \\
\mathbf{w}^T\mathbf{T}^{*}\mathbf{s}^{*}& =\kappa^{*} \\
\mathbf{s}^H\mathbf{T}^{H}\mathbf{w}& =\kappa^{*}
\end{align*}
Hence, the target-based constraints are
\begin{align*}
\mathbf{w}^H\mathbf{T}\mathbf{s}& = \kappa^{\phantom*} \\
\mathbf{s}^H\mathbf{T}^{H}\mathbf{w}& =\kappa^{*}
\end{align*}	
We can recast this constraint as two real constraints. First, recall the obvious things about complex numbers:
\begin{align*}
\kappa + \kappa^{*}  & =  2 \real\{\kappa\} = 2\kappa_R  \Rightarrow \frac{1}{2}\left(\kappa + \kappa^{*}\right) = \kappa_R \\
\kappa - \kappa^{*}  & =  \jmath 2 \imag\{\kappa\} = \jmath 2\kappa_I \Rightarrow  \frac{1}{2\jmath}\left(\kappa - \kappa^{*}\right) = \kappa_I		.
\end{align*}
If we substitute in the Capon constraints, we then obtain
\begin{align*}
\frac{1}{2}\left(\kappa + \kappa^{*}\right) & = \frac{1}{2}\left(\mathbf{w}^H\mathbf{T}\mathbf{s} + \mathbf{s}^H\mathbf{T}^{H}\mathbf{w}\right) = \real\{\mathbf{w}^H\mathbf{T}\mathbf{s}\} \\
& = \frac{1}{2}\begin{bmatrix}
\mathbf{w} \\ \mathbf{s}
\end{bmatrix}^{H} \begin{bmatrix}
\ZeroMat{NML}{NML}& \mathbf{T}\\
\mathbf{T}^{H} & \ZeroMat{N}{N}
\end{bmatrix}\begin{bmatrix}
\mathbf{w} \\ \mathbf{s}
\end{bmatrix} = \mathbf{b}^{H}\mathbf{\widetilde{T}}_{H}\mathbf{b} = \kappa_R \\
\intertext{and}\\
\frac{1}{2\jmath}\left(\kappa - \kappa^{*}\right) & = \frac{1}{2\jmath}\left(\mathbf{w}^H\mathbf{T}\mathbf{s} - \mathbf{s}^H\mathbf{T}^{H}\mathbf{w}\right) = \imag\{\mathbf{w}^H\mathbf{T}\mathbf{s}\} \\
& = \frac{1}{2\jmath}\begin{bmatrix}
\mathbf{w} \\ \mathbf{s}
\end{bmatrix}^{H} \begin{bmatrix}
\ZeroMat{NML}{NML}& \mathbf{T}\\
-\mathbf{T}^{H} & \ZeroMat{N}{N}
\end{bmatrix}\begin{bmatrix}
\mathbf{w} \\ \mathbf{s}
\end{bmatrix} = \mathbf{b}^{H}\mathbf{\widetilde{T}}_{A}\mathbf{b} = \kappa_I \\	
\end{align*}
where the subscripts $H$ and $A$ indicate the hermitian and anti-hermitian part of the expanded matrix $\mathbf{\widetilde{T}}$, respectively. (As an aside, recall that $\mathbf{\widetilde{T}}_{H}^{H} = \mathbf{\widetilde{T}}_{H}$ and $\mathbf{\widetilde{T}}_{A}^{H}=-\mathbf{\widetilde{T}}_{A}$ by definition)
This leads us to a new version of the second form of the optimization problem, as a function of the combined beamformer-signal vector:
\begin{equation*}
\begin{aligned}
& \underset{\mathbf{b}}{\text{min}}
& & \mathbf{b}^H\mathbf{\widetilde{R}_{ni}}\mathbf{b} + \mathbf{b}^{H}\left(\sum\limits_{q=1}^Q \overline{\boldsymbol{\Gamma}}_{q}\mathbf{b}\mathbf{b}^{H}\overline{\boldsymbol{\Gamma}}_{q}^{H}\right)\mathbf{b}\\
& \text{s.t.}
& & \mathbf{b}^{H}\mathbf{\widetilde{T}}_{H}\mathbf{b} = \kappa_R \\
& & & \mathbf{b}^{H}\mathbf{\widetilde{T}}_{A}\mathbf{b} = \kappa_I \\
& & & \mathbf{b}^H\PsiS^{T}\PsiS\mathbf{b}\leq P_o
\end{aligned}
\end{equation*}
or, using the tensor form,
\begin{equation*}
\begin{aligned}
& \underset{\mathbf{b}}{\text{min}}
& & \mathbf{b}^H\mathbf{\widetilde{R}_{ni}}\mathbf{b} + \mathbf{b}^{H}\left(\widetilde{\mathcal{C}}\mathbf{b}\mathbf{b}^{H}\right)\mathbf{b}\\
& \text{s.t.}
& & \mathbf{b}^{H}\mathbf{\widetilde{T}}_{H}\mathbf{b} = \kappa_R \\
& & & \mathbf{b}^{H}\mathbf{\widetilde{T}}_{A}\mathbf{b} = \kappa_I \\
& & & \mathbf{b}^H\PsiS^{T}\PsiS\mathbf{b}\leq P_o
\end{aligned}.
\end{equation*}
Using the properties of trace and the fact that all quantities in this optimization problem are now real, we can recast it using the inner product from above:
\begin{equation*}
\begin{aligned}
& \underset{\mathbf{b}}{\text{min}}
& & \mathbf{\widetilde{R}_{ni}}\bullet \mathbf{b}\mathbf{b}^{H} + \left(\widetilde{\mathcal{C}}\mathbf{b}\mathbf{b}^{H}\right)\bullet \mathbf{b}\mathbf{b}^{H}\\
& \text{s.t.}
& & \mathbf{\widetilde{T}}_{H}\bullet\mathbf{b}\mathbf{b}^{H} = \kappa_R \\
& & & \mathbf{\widetilde{T}}_{A} \bullet \mathbf{b}\mathbf{b}^{H} = -\kappa_I \\
& & & \PsiS^{T}\PsiS \bullet \mathbf{b}\mathbf{b}^H\leq P_o
\end{aligned}.
\end{equation*}	
Note that we've used the Hermitian (resp. anti-Hermitian) nature of $\mathbf{\widetilde{R}_{ni}}$ and $\mathbf{\widetilde{T}}_{H}$ (resp. $\mathbf{\widetilde{T}}_{A}$) to avoid the Hermitian superscript everywhere.

A common path to semidefinite relaxations defines a matrix variable, say, $\mathbf{B} =  \mathbf{b}\mathbf{b}^{H}$. It is clear that $\mathbf{Z}$ is Hermitian and positive semidefinite, and $\rank(\mathbf{B}) = 1$. This means our optimization problem above is equivalent to the matrix quadratic program
\begin{equation*}
\begin{aligned}
& \underset{\mathbf{B} \in \{\mathbf{B} = \mathbf{B}^{H}, \mathbf{B} \succeq 0\}}{\text{min}}
& & \mathbf{\widetilde{R}_{ni}}\bullet \mathbf{B} + \left(\widetilde{\mathcal{C}}\mathbf{B}\right)\bullet \mathbf{B}\\
& \text{s.t.}
& & \mathbf{\widetilde{T}}_{H}\bullet\mathbf{B} = \kappa_R &  \mathbf{\widetilde{T}}_{A} \bullet \mathbf{B} = -\kappa_I\\
& & & \PsiS^{T}\PsiS \bullet \mathbf{Z}\leq P_o & \rank(\mathbf{Z}) = 1
\end{aligned}.
\end{equation*}	

Depending on the properties of $\widetilde{\mathcal{C}}$, this could be a convex problem in $\mathbf{Z}$ if not for the rank condition, which is the primary hurdle. This is where the "relaxation" in semidefinite relaxation comes in. If we permit $\mathbf{Z}$ to take any rank, we have relaxed the problem to the (possibly) convex one:
\begin{equation*}
\begin{aligned}
& \underset{\mathbf{Z} \in \{\mathbf{Z}= \mathbf{Z}^{H}, \mathbf{Z} \succeq 0\}}{\text{min}}
& & \mathbf{\widetilde{R}_{ni}}\bullet \mathbf{Z} + \left(\widetilde{\mathcal{C}}\mathbf{Z}\right)\bullet \mathbf{Z}\\
& \text{s.t.}
& & \mathbf{\widetilde{T}}_{H}\bullet\mathbf{Z} = \kappa_R \\  
& & & \mathbf{\widetilde{T}}_{A} \bullet \mathbf{Z} = -\kappa_I\\
& & & \PsiS^{T}\PsiS \bullet \mathbf{Z}\leq P_o .
\end{aligned} \label{eq:SDRproblem}\tag{SDR}
\end{equation*}		

Notice we did not, at any point, progress through a matrix bilinear form in finding this semidefinite relaxation, as seen in \cite{Ling2009}, etc. This is because the constraint set in this problem prevents such a representation -- namely, the Capon constraint, which is bilinear in the vector arguments already.

If we can obtain a solution to this problem, a natural question is how to apply this solution to the original problem. If we are lucky, the rank of the relaxed solution will be one, and there will be nothing left to do. If the rank of the solution is greater than one, then we can use the nearest rank-one approximation that still satisfies the constraints. Some authors have  We will see in the simulations below that for this particular scenario, we can obtain solutions that are nearly rank-one in a numerical sense. 

\section{Solutions of the Relaxed BQP}\label{SDP_relax}
Recall that, when standardized, the semidefinite relaxation of the BQP is:
\begin{equation*}
\begin{aligned}
& \underset{\mathbf{B} \in \{\mathbf{X} \in \Complex^{J\times J} | \mathbf{X} = \mathbf{X}^{H}, \mathbf{X} \succeq 0\}}{\text{min}}
& & \mathbf{\widetilde{R}_{ni}}\bullet \mathbf{B} + \left(\widetilde{\mathcal{C}}\mathbf{B}\right)\bullet \mathbf{B}\\
& \text{s.t.}
& & \mathbf{\widetilde{T}}_{R}\bullet\mathbf{B} - \kappa_R = 0 \\  
& & & \mathbf{\widetilde{T}}_{I} \bullet \mathbf{B} - \kappa_I = 0\\
& & & \PsiS^{T}\PsiS \bullet \mathbf{B} - P_o \leq 0 .
\end{aligned} \label{eq:SSDRproblem}\tag{SSDR}
\end{equation*}	
Before we begin, we reformulate the problem to use the variable $\vecmat(\mathbf{B})$. First, note that $\Tr(\mathbf{A}\mathbf{B}) = \vecmat^{T}(\mathbf{A}^{T})\vecmat(\mathbf{B})$. For Hermitian $\mathbf{A}$, this becomes $\Tr(\mathbf{A}\mathbf{B}) = \vecmat^{H}(\mathbf{A})\vecmat(\mathbf{B})$ (since $\mathbf{A}^{T} = \mathbf{A}^{*}$ and the conjugation operator is linear). This allows us to say that, for Hermitian $\mathbf{A}$, the relevant inner product on $\Complex$ is $\mathbf{A} \bullet \mathbf{B} = \vecmat^{H}(\mathbf{A})\vecmat(\mathbf{B})$. 

Next, recall that the tensor $\widetilde{\mathcal{C}}$ can be unfolded into the vector matrix $\widetilde{\mathbf{C}}_{V} = \sum_{q=1}^{Q}  \vecmat(\overline{\boldsymbol{\Gamma}}_{q})\vecmat(\overline{\boldsymbol{\Gamma}}_{q})^{H}$. Using these, we recast \ref{eq:SSDRproblem} into a vectorized form:

\begin{equation*}
\begin{aligned}
& \underset{\mathbf{B} \in \{\mathbf{X} \in \Complex^{J\times J} | \mathbf{X} = \mathbf{X}^{H}, \mathbf{X} \succeq 0\}}{\text{min}}
& & \vecmat^{H}(\mathbf{\widetilde{R}_{ni}})\vecmat(\mathbf{B}) + \vecmat^{H}(\mathbf{B})\widetilde{\mathbf{C}}_{V}\vecmat(\mathbf{B}) \\
& \text{s.t.}
& & \vecmat^{H}(\mathbf{\widetilde{T}}_{R})\vecmat(\mathbf{B}) - \kappa_R = 0 \\  
& & & \vecmat^{H}(\mathbf{\widetilde{T}}_{I})\vecmat(\mathbf{B}) - \kappa_I = 0\\
& & & \vecmat^{H}(\PsiS^{T}\PsiS)\vecmat(\mathbf{B}) - P_o \leq 0.
\end{aligned} \label{eq:VSSDRproblem}\tag{VSSDR}.
\end{equation*}	

In a further ``development", let us explicitly put the equality constraint back as a complex number and break out the positive semi-definiteness as a separate constraint. This makes \ref{eq:VSSDRproblem} turn into
\begin{equation*}
\begin{aligned}
& \underset{\mathbf{B} \in \{\mathbf{X} \in \Complex^{J\times J} | \mathbf{X} = \mathbf{X}^{H}\}}{\text{min}}
& & \vecmat^{H}(\mathbf{\widetilde{R}_{ni}})\vecmat(\mathbf{B}) + \vecmat^{H}(\mathbf{B})\widetilde{\mathbf{C}}_{V}\vecmat(\mathbf{B}) \\
& \text{s.t.}
& & \vecmat^{H}(\mathbf{B})\vecmat(\mathbf{\widetilde{T}}) - \kappa = 0 \\  
& & & \vecmat^{H}(\PsiS^{T}\PsiS)\vecmat(\mathbf{B}) - P_o \leq 0\\
& & & \mathbf{B} \succeq 0
\end{aligned}
\end{equation*}

According to Hj{\o}rungnes, there are certain methods we need to account for the Hermitian constraint in the following analysis. For notational ease, let us define the following vectors:
\begin{align*}
\boldsymbol{\psi}_{S} & =  \vecmat(\PsiS^{T}\PsiS) & \boldsymbol{\rho} & = \vecmat(\mathbf{\widetilde{R}_{ni}}) .
\end{align*}
Furthermore, if we define the following matrices,
\begin{align*}
\mathbf{P}_{SS} & = \PsiS \otimes \PsiS & \mathbf{P}_{WW} = \PsiW \otimes \PsiW \\ 
\mathbf{P}_{SW} & = \PsiS \otimes \PsiW & \mathbf{P}_{WS} = \PsiW \otimes \PsiS, 
\end{align*}
then these can be rewritten as
\begin{align*}
\boldsymbol{\psi}_{S} & =  \mathbf{P}_{SS}^{T}\vecmat(\Identity{N}) & \boldsymbol{\rho} & = \mathbf{P}_{WW}^{T}\vecmat(\mathbf{R_{ni}}) .
\end{align*}
\subsection{Proof of Convexity}
Before continuing, we demonstrate directly that this relaxed optimization problem is indeed convex. Recall that the relaxed problem is
\begin{equation*}
\begin{aligned}
& \underset{\mathbf{B} \in \{\mathbf{X} \in \Complex^{J\times J} | \mathbf{X} = \mathbf{X}^{H}\}}{\text{min}}
& & \vecmat^{H}(\mathbf{\widetilde{R}_{ni}})\vecmat(\mathbf{B}) + \vecmat^{H}(\mathbf{B})\widetilde{\mathbf{C}}_{V}\vecmat(\mathbf{B}) \\
& \text{s.t.}
& & \vecmat^{H}(\mathbf{B})\vecmat(\mathbf{\widetilde{T}}) - \kappa = 0 \\  
& & & \vecmat^{H}(\PsiS^{T}\PsiS)\vecmat(\mathbf{B}) - P_o \leq 0\\
& & & \mathbf{B} \succeq 0.
\end{aligned}
\end{equation*}	
A convex optimization problem minimizes a convex objective over a convex set. A well-known restriction of this concept is given in \cite[pp. 136-137]{BoydVandenberghe} as follows: for a given optimization problem, if
\begin{enumerate}
	\item the objective function is convex,
	\item the inequality constraint functions are convex, and,
	\item the equality constraint functions are affine (i.e. it looks like $\abf^{H}\xbf - b$.) 
\end{enumerate} 
then the problem is convex. This mostly stems from the fact that affine equality constraints define a polytope, which is convex, and so its intersection with the existing convex set is also convex. 

We start with the objective. A quadratic function is convex if the associated matrix is positive semidefinite. Thus, since $\tilde{\Cbf}_{V} \succeq 0$ by definition (as a sum of rank one Hermitian matrices), then $\vecmat^{H}(\Bbf)\widetilde{\Cbf}_{V}\vecmat(\Bbf)$ is a convex function. Adding an affine function to it does not change the convexity, so the overall objective is convex. Continuing to the constraints, both the power constraint and the semidefiniteness constraints are convex -- the former because it is affine, the latter because it describes a convex cone. Finally, the equality constraint is convex, because it can be broken up into two real affine constraints on the real and imaginary parts. Therefore, the relaxed problem is convex. However, since $\tilde{\Cbf}_{V}$ is typically \emph{not} full rank (indeed, the rank is bounded above by $ML$ in this case), it is not usually strictly convex, and thus there will exist a multitude of solutions.

\subsection{Slater's Condition}
Recall that \emph{strong duality} is said to hold for a given optimization problem  is the primal problem is convex and it satisfies Slater's condition (aka the problem is \emph{strictly feasible}). A consequence of this property is that there is no duality gap between solutions of the primal and dual problems. We will show that our problem satisfies Slater's condition given a relationship between the constraint values. 

Given an optimization problem
\begin{equation*}
\begin{aligned}
& \underset{\mathbf{x} \in \Omega}{\text{min}}
& & f_o(\mathbf{x}) \\
& \text{s.t.}
& & g_i(\mathbf{x}) \leq 0 \quad i = 1,\dots, m\\  
& & & h_j(\mathbf{x}) = 0 \quad j = 1, \dots, n 
\end{aligned} 
\end{equation*}	
(where $\Omega$ is a convex set), let us assume $f_o(\mathbf{x})$ is convex and define the set $\mathcal{D} = \Omega \cap \,\dom(f_o) \,\cap \, (\cap_{i=1}^{m} \dom(g_i))$ to be the total feasible domain of the problem. Slater's condition is satisfied if there exists at least one point in the relative interior (i.e. not on the boundary) of the problem's feasibility set that satisfy all of the equality constraints and strictly satsifies the inequality constraints.  Mathematically, we can represent the condition as follows:
\begin{lemma}[Slater's Condition]
	For the standard optimization problem above, if each $g_i$ is convex and there exists a point $\mathbf{x}_{o} \in \relint(\mathcal{D})$ such that $g_i(\mathbf{x}_o) < 0 \; \forall i$ and $h_j(\mathbf{x}_o) = 0  \; \forall j$, then strong duality holds and there is zero duality gap. 
\end{lemma}
For our problem, the feasibility set is positive semidefinite matrices, which means its relative interior is positive-definite matrices. Our constraint functions are
\begin{align*}
g(\mathbf{B}) & = \boldsymbol{\psi}_{S}^{T}\boldsymbol{\beta} - P_o \\
h_{1}(\mathbf{B}) & = \boldsymbol{\beta}^{H}\tilde{\boldsymbol{\tau}} - \kappa^{\phantom*} = \tilde{\boldsymbol{\tau}}^{T}\Commut{J}{J}\boldsymbol{\beta} - \kappa \\
h_{2}(\mathbf{B}) & = \tilde{\boldsymbol{\tau}}^{H}\boldsymbol{\beta} - \kappa^{*}.
\end{align*}
where $\boldsymbol{\beta} = \vecmat(\Bbf)$ for a given Hermitian matrix $\Bbf$. Thus, we need to find a matrix $\Bbf \succ 0$ that satisfies the above equations. 

We start with solving for $\boldsymbol{\beta}$ in our equality constraints, then ensure that the resulting matrix is both positive definite and strictly satisfies the power inequality. As a combined matrix-vector equation, we have:
\begin{align*}
\begin{bmatrix}
\tilde{\boldsymbol{\tau}}^{T}\Commut{J}{J} \\
\tilde{\boldsymbol{\tau}}^{H}
\end{bmatrix} \boldsymbol{\beta} = \begin{bmatrix}
\kappa \\ \kappa^{*}
\end{bmatrix}
\end{align*}
If solutions of this equation exist (and they should exist), then they are given by
\begin{align*}
\boldsymbol{\beta} = \begin{bmatrix}
\tilde{\boldsymbol{\tau}}^{T}\Commut{J}{J} \\
\tilde{\boldsymbol{\tau}}^{H}
\end{bmatrix}^{\dagger} \begin{bmatrix}
\kappa \\ \kappa^{*}
\end{bmatrix} +
\left(\Identity{N^2(ML+1)^2} - \begin{bmatrix}
\tilde{\boldsymbol{\tau}}^{T}\Commut{J}{J} \\
\tilde{\boldsymbol{\tau}}^{H}
\end{bmatrix}^{\dagger}\begin{bmatrix}
\tilde{\boldsymbol{\tau}}^{T}\Commut{J}{J} \\
\tilde{\boldsymbol{\tau}}^{H}\end{bmatrix}\right) \zbf
\end{align*}
where $\dagger$ indicates the pseudoinverse and $\zbf$ is the vectorization of a matrix $\Zbf$ similarly shaped to $\Bbf$. 

A logical question is, then, what is that pseudoinverse (and consequently, what is the existence condition)? Since the matrix $\begin{bmatrix}
\tilde{\boldsymbol{\tau}}^{T}\Commut{J}{J} \\
\tilde{\boldsymbol{\tau}}^{H}
\end{bmatrix}$ is fat, we might have some hope in the form $\Cbf^{\dagger} = \Cbf^{H}(\Cbf\Cbf^{H})^{-1}$. First, we can find the matrix we hope to invert:
\begin{align*}
\begin{bmatrix}
\tilde{\boldsymbol{\tau}}^{T}\Commut{J}{J} \\
\tilde{\boldsymbol{\tau}}^{H}
\end{bmatrix}\begin{bmatrix}
\tilde{\boldsymbol{\tau}}^{T}\Commut{J}{J} \\
\tilde{\boldsymbol{\tau}}^{H}
\end{bmatrix}^{H} & = \begin{bmatrix}
\tilde{\boldsymbol{\tau}}^{T}\Commut{J}{J}\Commut{J}{J}\tilde{\boldsymbol{\tau}}^{*} &  \tilde{\boldsymbol{\tau}}^{T}\Commut{J}{J}\tilde{\boldsymbol{\tau}}\\
\tilde{\boldsymbol{\tau}}^{H}\Commut{J}{J}\tilde{\boldsymbol{\tau}}^{*} & \tilde{\boldsymbol{\tau}}^{H}\tilde{\boldsymbol{\tau}}
\end{bmatrix} \\
& = \begin{bmatrix}
\frobnorm{\Tbf}{2} & 0 \\ 0 & \frobnorm{\Tbf}{2}
\end{bmatrix} = \frobnorm{\Tbf}{2}\Identity{2}.
\end{align*}
So long as we have an actual target, this is always invertible (and hence, a solution always exists). This means that the pseudoinverse of interest is 
\begin{align*}
\begin{bmatrix}
\tilde{\boldsymbol{\tau}}^{T}\Commut{J}{J} \\
\tilde{\boldsymbol{\tau}}^{H}
\end{bmatrix}^{\dagger} = \frac{1}{\frobnorm{\Tbf}{2}}\begin{bmatrix}
\Commut{J}{J}\tilde{\boldsymbol{\tau}}^{*} & \tilde{\boldsymbol{\tau}}
\end{bmatrix}.
\end{align*}
Additionally, the projection matrix in the solution is 
\begin{align*}
\begin{bmatrix}
\tilde{\boldsymbol{\tau}}^{T}\Commut{J}{J} \\
\tilde{\boldsymbol{\tau}}^{H}
\end{bmatrix}^{\dagger}\begin{bmatrix}
\tilde{\boldsymbol{\tau}}^{T}\Commut{J}{J} \\
\tilde{\boldsymbol{\tau}}^{H}\end{bmatrix} & = \frac{1}{\frobnorm{\Tbf}{2}} \left(\Commut{J}{J}\tilde{\boldsymbol{\tau}}^{*}\tilde{\boldsymbol{\tau}}^{T}\Commut{J}{J} + \tilde{\boldsymbol{\tau}}\tilde{\boldsymbol{\tau}}^{H}\right) \\
& = \frac{1}{\frobnorm{\Tbf}{2}} \left(\Pws^{T}\boldsymbol{\tau}_{H}\boldsymbol{\tau}_{H}^{H}\Pws + \Psw^{T}\boldsymbol{\tau}\boldsymbol{\tau}^{H}\Psw\right)
\end{align*}
where $\boldsymbol{\tau}_{H} = \vecmat(\Tbf^{H})$. 

We can break our solution into the ``min-norm" and ``nullspace" parts. Clearly, the min-norm part is
\begin{align*}
\Bbf_{min} = \frac{1}{\frobnorm{\Tbf}{2}} \begin{bmatrix}
\ZeroMat{NML}{NML} & \kappa^{*}\Tbf \\ \kappa\Tbf^{H} & \ZeroMat{N}{N}
\end{bmatrix}.
\end{align*}
It is clear from the structure that this is not a positive definite matrix (or even a positive semidefinite one), so we require the nullspace component to place us into the relative interior.
If we break the vector $\zbf$ into components as $\zbf = \Pww^{T}\zbf_1 + \Psw^{T}\zbf_2 + \Pws^{T}\zbf_{2H} + \Pss^{T}\zbf_3$, then the nullspace solution is:
\begin{align*}
\Bbf_{null} = \Zbf - \frac{1}{\frobnorm{\Tbf}{2}}\begin{bmatrix}\ZeroMat{NML}{NML} & \trace(
\Tbf^{H}\Zbf_2) \Tbf \\ \trace(\Zbf_2^{H}\Tbf) \Tbf^{H} & \ZeroMat{N}{N}
\end{bmatrix}.
\end{align*} 
Hence, the overall matrix that satisfies the equality constraints is 
\begin{align*}
\Bbf = \Zbf +  \frac{1}{\frobnorm{\Tbf}{2}} \begin{bmatrix}\ZeroMat{NML}{NML} & (\kappa^{*} - \trace(
\Tbf^{H}\Zbf_2)) \Tbf \\ (\kappa - \trace(\Zbf_2^{H}\Tbf)) \Tbf^{H} & \ZeroMat{N}{N}
\end{bmatrix}.
\end{align*}
Our problem now becomes finding a matrix $\Zbf$ such that $\Bbf \succ 0$ and $\trace(\Zbf_3) < P_o$. One method of attack is setting $\Zbf_2 = \ZeroMat{NML}{N}$ and attempting to reach a valid result this way. Under this assumption, our solution matrix is
\begin{align*}
\Bbf =  \begin{bmatrix}
\Zbf_1 & \frac{\kappa^{*}}{\frobnorm{\Tbf}{2}}\Tbf \\ \frac{\kappa}{\frobnorm{\Tbf}{2}}\Tbf^{H} & \Zbf_3 \end{bmatrix}.
\end{align*}
The positive definiteness requirement can then be expressed in one of two ways:
\begin{align*}
\Zbf_1 & \succ 0, \quad \Zbf_3 - \frac{|\kappa|^2}{\frobnorm{\Tbf}{4}}\Tbf^{H}\Zbf_1^{-1}\Tbf \succ 0 \\
\Zbf_3 & \succ 0, \quad \Zbf_1 - \frac{|\kappa|^2}{\frobnorm{\Tbf}{4}}\Tbf\Zbf_3^{-1}\Tbf^{H} \succ 0.
\end{align*}
Here, we rely on another judicious guess, setting $\Zbf_1 = \Identity{NML}$, which is clearly positive definite.\footnotemark  Then, we only need to find a $\Zbf_3$ such that $\Zbf_3 \succ \frac{|\kappa|^2}{\frobnorm{\Tbf}{4}}\Tbf^{H}\Tbf$ \emph{and} $\trace(\Zbf_3) < P_o$. 
\footnotetext{We can make this guess because $NML > N$. If, for whatever reason, the number of transmit resources were greater than the number of receive resources, then we could start with $\Zbf_3$ and continue from there.}
Let us assume that such a matrix exists. Following \cite[Corollary 7.7.4(d)]{HornJohnsonMA2e}, since $\Zbf_3 \succ \frac{|\kappa|^2}{\frobnorm{\Tbf}{2}}\Tbf^{H}\Tbf$, then $\trace(\Zbf_3) > \frac{|\kappa|^2}{\frobnorm{\Tbf}{4}}\trace(\Tbf^{H}\Tbf) = \frac{|\kappa|^2}{\frobnorm{\Tbf}{2}}$. But we already know that $\trace(\Zbf_3) < P_o$. Hence, we have the chained inequality
\begin{align*}
P_o > \trace(\Zbf_3) > \frac{|\kappa|^2}{\frobnorm{\Tbf}{2}},
\end{align*}
which obviously requires that $P_o > \frac{|\kappa|^2}{\frobnorm{\Tbf}{2}}$ or perhaps $\frobnorm{\Tbf}{2} > \frac{|\kappa|^2}{P_o}$.  

With this in hand, we can now construct our matrix. Since $NML > N$ and (generally) $\rank(\Tbf) = N$, then $\Tbf^{H}\Tbf \succ 0$.\footnote{Again, if our available resources are swapped, we can use the outer product instead and get a similar answer.} For any matrix $\Abf \succ 0$ and real numbers $\alpha, \beta$, $\alpha\Abf \succ \beta\Abf$ if and only if $\alpha > \beta$. Let $\Zbf_3 = 
\frac{P_o - \varepsilon}{\frobnorm{\Tbf}{2}} \,\Tbf^{H}\Tbf$, where $\varepsilon > 0$ is some small real number. This guarantees that the trace feasibility is strictly satisfied, since $\trace(\Zbf_3) = P_o - \varepsilon < P_o$. It also satisfies the chained inequality above, so long as $\epsilon < P_o -  \frac{|\kappa|^2}{\frobnorm{\Tbf}{2}}$. Thus, the chain inequality is the \emph{only} controlling element, and thus there is no duality gap so long as $\frobnorm{\Tbf}{2} > \frac{|\kappa|^2}{P_o}$.

In the side-looking STAP case, the target matrix has a Kronecker structure -- that is, $\Tbf = \vbf_t \kronecker \Identity{N} \kronecker \abf_t$. This means that $\Tbf^{H}\Tbf = \normsq{\vbf_t}\normsq{\abf_t} \Identity{N}$. If we assume that the spatial response comes from a uniform linear array and that the Doppler response is similarly uniform, then $\normsq{\abf_t} = M$ and $\normsq{\vbf_t} = L$. This further implies that $\Tbf^{H}\Tbf = ML\Identity{N}$ and $\frobnorm{\Tbf}{2} = NML$. If we apply this knowledge to the positive definiteness condition, we need to find a $\Zbf_3$ such that $\Zbf_3 \succ \frac{|\kappa|^{2}}{N^2ML}\Identity{N}$. This clearly exists if we set $\Zbf_3 = \alpha\Identity{N}$ where $\alpha > \frac{|\kappa|^{2}}{NML}$. Following the path from the general case, this means that the Slater condition is satisfied so long as $NML > \frac{|\kappa|^2}{P_o}$.

\subsection{Obtaining the KKTs}
First, let us adopt the convention following conventions: each inequality constraint (multiplier) is given by $g_{j}(\cdot)$ ($\lambda_{j}$), each equality constraint (multiplier) is given by  $h_{i}(\cdot)$ ($\mu_i$), and an optimal value of a variable is denoted by a superscript o (\emph{i.e.}, $\mathbf{B}^{o}$). The KKT conditions can then be written as 
\begin{enumerate}
	\item $\nabla_{\mathbf{B}} \mathcal{L}(\mathbf{B}^{o},\boldsymbol{\mu}^{o},\boldsymbol{\lambda}^{o}) = \ZeroMat{J}{J}$
	\item $\lambda_{j}^{o} \geq 0 \; \forall j$
	\item $\lambda_{j}^{o} g_{j}(\mathbf{B}^{o}) = 0 \; \forall j$
	\item $g_{j}(\mathbf{B}^{o}) \leq 0 \; \forall j$
	\item $h_{i}(\mathbf{B}^{o}) = 0 \; \forall i$
	\item $\mathcal{H}[\mathcal{L}(\mathbf{B}^{o},\boldsymbol{\mu}^{o},\boldsymbol{\lambda}^{o})] \succeq 0$
\end{enumerate}
That is, for a matrix $\mathbf{B}^{o}$ to be a regular minimizer of the related optimization problem, it is necessary that it satisfies these conditions. If the problem is convex, these are necessary and sufficient conditions for the optimal minimizer and associated Lagrange multipliers. For future notational simplicity, we will drop the superscript $o$ until absolutely necessary (i.e., a final statement of the optimal solution).
According to  \cite{Hjorungnes2011}, the gradient in the first KKT condition is $\nabla_{\mathbf{B}} = \frac{\partial}{\partial B^{*}}$ or, in other words, $\mathcal{D}_{\mathbf{B}^{*}} \mathcal{L} = \vecmat^{T}(\frac{\partial \mathcal{L}}{\partial B^{*}})$. Thus, this vectorized form can take the place of the gradient. 

\subsubsection{KKT Condition 1}
Let $\boldsymbol{\Sigma}$ be the slackness variable associated with the PSD condition on $\mathbf{B}$, and $\boldsymbol{\sigma}$ be its vectorization. 
The Lagrangian is
\begin{equation}
\mathcal{L}(\mathbf{B},\boldsymbol{\Sigma},\mu,\lambda) = \boldsymbol{\beta}^{H}(\widetilde{\mathbf{C}}_{V}\boldsymbol{\beta} + \boldsymbol{\rho} - \tilde{\mu}^{*}\tilde{\boldsymbol{\tau}} + \lambda\boldsymbol{\psi}_{S} -  \boldsymbol{\sigma}) + \tilde{\mu}^{*}\kappa - \lambda P_o.
\end{equation}
Condition 1 is, after taking the derivatives,
\begin{align}
(\widetilde{\mathbf{C}}_{V}+\Commut{J}{J}\widetilde{\mathbf{C}}_{V}^{*}\Commut{J}{J})\boldsymbol{\beta} + \boldsymbol{\rho} - \tilde{\mu}^{*}\tilde{\boldsymbol{\tau}} - \tilde{\mu}\tilde{\boldsymbol{\tau}}_{H} + \lambda\boldsymbol{\psi}_{S} -  \boldsymbol{\sigma} = \ZeroMat{J}{1} \nonumber	\\
(\widetilde{\mathbf{C}}_{V}+\Commut{J}{J}\widetilde{\mathbf{C}}_{V}^{*}\Commut{J}{J})	\boldsymbol{\beta}  =  \boldsymbol{\sigma}-(\boldsymbol{\rho} + \tilde{\mu}^{*}\tilde{\boldsymbol{\tau}} + \tilde{\mu}\tilde{\boldsymbol{\tau}}_{H} + \lambda\boldsymbol{\psi}_{S}).
\end{align}
In general, the full set of solutions to the second form of Condition 1 is given by 
\begin{align}
\boldsymbol{\beta} & = (\widetilde{\mathbf{C}}_{V}+\Commut{J}{J}\widetilde{\mathbf{C}}_{V}^{*}\Commut{J}{J})^{\dagger} (\boldsymbol{\sigma}-(\boldsymbol{\rho} + \tilde{\mu}^{*}\tilde{\boldsymbol{\tau}} + \tilde{\mu}\tilde{\boldsymbol{\tau}}_{H} + \lambda\boldsymbol{\psi}_{S}) \nonumber \\ 
& \qquad + (\Identity{J} - (\widetilde{\mathbf{C}}_{V}+\Commut{J}{J}\widetilde{\mathbf{C}}_{V}^{*}\Commut{J}{J}) (\widetilde{\mathbf{C}}_{V}+\Commut{J}{J}\widetilde{\mathbf{C}}_{V}^{*}\Commut{J}{J})^{\dagger})\mathbf{z}
\end{align}
where $\mathbf{z}$ is an arbitrary vector isomorphic to a Hermitian matrix $\mathbf{Z}$, partitioned identically to $\mathbf{B}$ and $\boldsymbol{\Sigma}$. This solution exists provided the following existence condition holds:
\begin{align}
(\widetilde{\mathbf{C}}_{V}+\Commut{J}{J}\widetilde{\mathbf{C}}_{V}^{*}\Commut{J}{J}) (\widetilde{\mathbf{C}}_{V}+\Commut{J}{J}\widetilde{\mathbf{C}}_{V}^{*}\Commut{J}{J})^{\dagger} (\boldsymbol{\sigma}-(\boldsymbol{\rho} + \tilde{\mu}^{*}\tilde{\boldsymbol{\tau}} + \tilde{\mu}\tilde{\boldsymbol{\tau}}_{H} + \lambda\boldsymbol{\psi}_{S}))  \nonumber \\
\quad = \quad \boldsymbol{\sigma}-(\boldsymbol{\rho} + \tilde{\mu}^{*}\tilde{\boldsymbol{\tau}} + \tilde{\mu}\tilde{\boldsymbol{\tau}}_{H} + \lambda\boldsymbol{\psi}_{S})
\end{align}

This may seem daunting, but we can unpack parts of this quite easily. Recall that the vectors $\boldsymbol{\beta}, \boldsymbol{\sigma}$ can be partitioned into the sums:  
\begin{align*}
\boldsymbol{\beta} & = \mathbf{P}_{WW}^{T}\boldsymbol{\beta}_1 + \mathbf{P}_{SW}^{T}\boldsymbol{\beta}_2 + \mathbf{P}_{WS}^{T}\boldsymbol{\beta}_{2,H} + \mathbf{P}_{SS}^{T}\boldsymbol{\beta}_3 \\
\boldsymbol{\sigma} & = \mathbf{P}_{WW}^{T}\boldsymbol{\sigma}_1 + \mathbf{P}_{SW}^{T}\boldsymbol{\sigma}_2 + \mathbf{P}_{WS}^{T}\boldsymbol{\sigma}_{2,H} + \mathbf{P}_{SS}^{T}\boldsymbol{\sigma}_3		
\end{align*}
and that these partitions are disjoint. Due to the disjointness, a quick examination of the second form of Condition 1 gives us a few results:
\begin{subequations}
	\begin{eqnarray}
	\boldsymbol{\sigma}_{1} - \mathbf{r}_{ni} & = & \ZeroMat{(NML)^2}{1} \label{eq:Sigma1Sol}\\
	\boldsymbol{\sigma}_{3} - \lambda \vecmat(\Identity{N}) & = & \ZeroMat{N^2}{1} \label{eq:Sigma3Sol}\\
	\mathbf{C}_{V}\boldsymbol{\beta}_{2} & = &\boldsymbol{\sigma}_2 + \tilde{\mu}^{*}\boldsymbol{\tau}\label{eq:Sig2VectorForm}.
	\end{eqnarray}	
\end{subequations}	
The first two equations result from the clutter matrix on the left hand side annihilating the $\mathbf{P}_{WW}^{T}$ and $\mathbf{P}_{SS}^{T}$ components of $\boldsymbol{\beta}$, and imply $\boldsymbol{\Sigma}_1 = \mathbf{R_{ni}}$ \& $\boldsymbol{\Sigma}_3 = \lambda\Identity{N}$. The third equation covers both the $\mathbf{P}_{SW}^{T}$ and $\mathbf{P}_{WS}^{T}$ components, because the $\mathbf{P}_{WS}^{T}$ component is a conjugate permutation of the $\mathbf{P}_{SW}^{T}$ component. The solution (if it exists) of that third component, generally, is
\begin{equation}
\boldsymbol{\beta}_{2} = \mathbf{C}_{V}^{\dagger}(\boldsymbol{\sigma}_2 + \tilde{\mu}^{*}\boldsymbol{\tau}) + (\Identity{N^{2}ML} - \mathbf{C}_{V}^{\dagger}\mathbf{C}_{V})\mathbf{z}_{2} \label{eq:Beta2GenSol}
\end{equation}	
where $\mathbf{z}_{2}$ is the $\mathbf{P}_{SW}^{T}$ component of the arbitrary vector $\mathbf{z}$ above. The existence condition for this solution is, unsurprisingly, 
\begin{equation*}
\mathbf{C}_{V}\mathbf{C}_{V}^{\dagger}(\boldsymbol{\sigma}_2 + \tilde{\mu}^{*}\boldsymbol{\tau}) = \boldsymbol{\sigma}_2 + \tilde{\mu}^{*}\boldsymbol{\tau},
\end{equation*}
or, after rearrangement,
\begin{equation}
(\Identity{N^{2}ML} - \mathbf{C}_{V}\mathbf{C}_{V}^{\dagger})\boldsymbol{\sigma}_2 = - \tilde{\mu}^{*}(\Identity{N^{2}ML} - \mathbf{C}_{V}\mathbf{C}_{V}^{\dagger})\boldsymbol{\tau}, \label{eq:Beta2ExistCond}
\end{equation}	
With some algebraic fiddling, one can verify that Equations~\ref{eq:Sigma1Sol}, \ref{eq:Sigma3Sol}, \ref{eq:Beta2GenSol}, and \ref{eq:Beta2ExistCond} are equivalent to most of the much uglier solution above. The remaining necessary mathematical spackle is setting $\mathbf{B}_{1} = \mathbf{Z}_{1}$ and $\mathbf{B}_{3} = \mathbf{Z}_{3}$, again due to the annihilating nature of the clutter matrix on the block-diagonal terms. This more or less says the diagonal blocks of the matrix are arbitrary, subject to satisfying the other KKT conditions. 

We conclude this section by repeating the necessary and sufficient conditions for the first KKT condition to be satisfied, as found above:
\begin{align*}
\boldsymbol{\Sigma}_1 &= \mathbf{R_{ni}} & \boldsymbol{\Sigma}_3 &= \lambda\Identity{N} \\
\mathbf{B}_{1} &= \mathbf{Z}_{1}  & \mathbf{B}_{3} &= \mathbf{Z}_{3} \\
\boldsymbol{\beta}_{2} &= \mathbf{C}_{V}^{\dagger}(\boldsymbol{\sigma}_2 + \tilde{\mu}^{*}\boldsymbol{\tau}) + \mathbf{P}_{\mathbf{C}_V}^{\perp}\mathbf{z}_{2} & \mathbf{P}_{\mathbf{C}_V}^{\perp}\boldsymbol{\sigma}_2 &= -\tilde{\mu}^{*}\mathbf{P}_{\mathbf{C}_V}^{\perp}\boldsymbol{\tau}
\end{align*}
where $\mathbf{P}_{\mathbf{C}_V}^{\perp} = \Identity{N^{2}ML} - \mathbf{C}_{V}\mathbf{C}_{V}^{\dagger} = \Identity{N^{2}ML} - \mathbf{C}_{V}^{\dagger}\mathbf{C}_{V}$ is the orthogonal projector onto the nullspace of $\mathbf{C}_{V}$.

\subsubsection{KKT Conditions 2-4: The Inequality Constraints}
With the gradient condition exhausted, we turn to the inequality constraints (power bound, positive-semidefiniteness of the solution) \& their related conditions. For convenience, we shall attack these somewhat independently in separate subsections, though they will interact.

\paragraph{The power constraint}
The KKT conditions related to the power constraint are as follows:
\begin{align*}
\lambda & \geq 0 \tag{Scalar Positivity}\\
\boldsymbol{\psi}_{S}^{T}\boldsymbol{\beta} - P_{o} & \leq 0 \tag{Scalar Feasibility} \\
\lambda(\boldsymbol{\psi}_{S}^{T}\boldsymbol{\beta} - P_{o}) &= 0 \tag{Scalar comp. slackness} 
\end{align*}
Rearranging with our partitioned variables, we get
\begin{align*}
\lambda & \geq 0 \tag{Scalar Positivity}\\
\Tr(\mathbf{B}_3) - P_{o} & \leq 0 \tag{Scalar Feasibility} \\
\lambda(\Tr(\mathbf{B}_3) - P_{o}) &= 0 \tag{Scalar comp. slackness}.
\end{align*}
A nice result of the slackness condition is $\lambda\Tr(\mathbf{B}_3) = \lambda P_o$, which we will use later. The other ``result" is that $\lambda = 0$ when the solution does not reach the power bound, and $\lambda > 0$ when it does. This will inform interpretations of the matrix situation below.

\paragraph{Positive semidefiniteness of $\mathbf{B}$} 
The conditions for semidefiniteness of the relaxed beamformer-signal basis are slightly more complex, but reveal a significant amount of structure to the solution. First, the direct form of these conditions are
\begin{align*}
\boldsymbol{\Sigma} \succeq 0, & \quad \mathbf{B} \succeq 0, \quad \boldsymbol{\Sigma}\mathbf{B} = \ZeroMat{J}{J}.
\end{align*} Of course, in this form, they are not especially useful. However, recall that from the first KKT condition, we know $\boldsymbol{\Sigma}$ to some extent:
\begin{equation}
\boldsymbol{\Sigma} = \begin{bmatrix}
\mathbf{R_{ni}} & \boldsymbol{\Sigma}_2 \\
\boldsymbol{\Sigma}_2^{H} & \lambda \Identity{N}
\end{bmatrix}.
\end{equation}
Via Theorem~\ref{thm:PSDness}, we have conditions for the positive-semidefiniteness of the basis matrix and its slackness variable. To wit, the basis matrix $\mathbf{B} = \begin{bmatrix}
\mathbf{B}_1 & \mathbf{B}_2 \\ \mathbf{B}_2^{H} &\mathbf{B}_3
\end{bmatrix}$ is PSD if and only if
\begin{align}
\mathbf{B}_{1} &\succeq 0 \label{eq:BP1}\tag{BP1}\\
\mathbf{B}_{2} &= \mathbf{B}_{1}\mathbf{B}_{1}^{\dagger}\mathbf{B}_{2} \label{eq:BP2}\tag{BP2} \\
\mathbf{B}_{3} &- \mathbf{B}_{2}^{H}\mathbf{B}_{1}^{\dagger}\mathbf{B}_{2} \succeq 0 \label{eq:BP3}\tag{BP3},
\end{align}
and the slackness matrix $\boldsymbol{\Sigma}$ is PSD if and only if
\begin{align}
\lambda \Identity{N} &\succeq 0 \label{eq:SigP1}\tag{$\Sigma$P1}\\
\boldsymbol{\Sigma}_{2} &= \lambda\lambda^{\dagger}\boldsymbol{\Sigma}_{2} \label{eq:SigP2}\tag{$\Sigma$P2} \\
\mathbf{R_{ni}} &- \lambda^{\dagger}\mathbf{\Sigma}_{2}\boldsymbol{\Sigma}_{2} \succeq 0 \label{eq:SigP3}\tag{$\Sigma$P3},
\end{align}
where $\lambda^{\dagger}$ is a scalar pseudoinverse, equaling $\lambda^{-1}$ when $\lambda > 0$ and zero otherwise. Using Theorem~\ref{thm:PSDnessHJ}, we can also say that Conditions~\ref{eq:SigP1}--\ref{eq:SigP3} are satisfied if a contraction $\Xbf \in \Complex^{NML \times N}$ exists such that $\Sigws = \sqrt{\lambda}\Rnihalf \Xbf$.
	
The complementary slackness condition for the matrix case reduces to 4 equalities, given below
\begin{align}
\mathbf{B}_{1}\mathbf{R_{ni}} & = - \mathbf{B}_{2}\boldsymbol{\Sigma}_{2}^{H} \label{eq:Slack1}\tag{CS1} \\
\mathbf{B}_{1}\boldsymbol{\Sigma}_{2} & = - \lambda\mathbf{B}_{2} \label{eq:Slack2}\tag{CS2} \\
\mathbf{B}_{2}^{H}\mathbf{R_{ni}} & = - \mathbf{B}_{3}\boldsymbol{\Sigma}_{2}^{H} \label{eq:Slack3}\tag{CS3} \\
\mathbf{B}_{2}^{H}\boldsymbol{\Sigma}_{2} & = -\lambda \mathbf{B}_{3} \label{eq:Slack4}\tag{CS4} 
\end{align}
We can use these conditions to find equivalent forms for $\lambda$. First, taking the trace of Condition~\ref{eq:Slack4}, we have
\begin{align*}
\lambda\trace(\Bss) & = -\trace(\Bws^{H}\Sigws).\\
\intertext{But since $\lambda\trace(\Bss) = \lambda P_o$ from scalar complimentary slackness,}
\lambda & = -\frac{\trace(\Bws^{H}\Sigws)}{P_o}.
\end{align*}
Since $\lambda$ is both real and nonnegative, this means that $\trace(\Bws^{H}\Sigws)$ is real and nonpositive! We also have $\trace(\Bws^{H}\Sigws) = \trace(\Sigws^{H}\Bws)$, which we can apply to the trace of Condition~\ref{eq:Slack1} to obtain another form of $\lambda$:
\begin{align*}
\lambda = \frac{\trace(\Bww\Rni)}{P_o}.
\end{align*}

We can also apply this logic to Equation~(\ref{eq:Sig2VectorForm}) to reveal an interesting consequence about the cost function. When vectorized, $\trace(\Bws^{H}\Sigws) = \boldsymbol{\beta}_2^{H}\boldsymbol{\sigma}_2$. Then, we have
\begin{align*}
\boldsymbol{\beta}_2^{H}\boldsymbol{\sigma}_2 & = \boldsymbol{\beta}_2^{H}	\mathbf{C}_{V}\boldsymbol{\beta}_{2} - \tilde{\mu}^{*}\boldsymbol{\beta}_2^{H}\boldsymbol{\tau} \\
\intertext{Applying the equality constraint, this is equivalent to}
\boldsymbol{\beta}_2^{H}\boldsymbol{\sigma}_2 & = \boldsymbol{\beta}_2^{H}	\mathbf{C}_{V}\boldsymbol{\beta}_{2} - \tilde{\mu}^{*}\kappa.
\end{align*}
(Incidentally, this means that $\tilde{\mu}^{*}\kappa$ is real, and hence the optimal phase of $\tilde{\mu}$ is that of $\kappa$.)
From the above, however, we can see that $-\boldsymbol{\beta}_2^{H}\boldsymbol{\sigma}_2 = \lambda P_o$, and so
\begin{align*}
\tilde{\mu}^{*}\kappa - \lambda P_o & = \boldsymbol{\beta}_2^{H}	\mathbf{C}_{V}\boldsymbol{\beta}_{2} \\
\tilde{\mu}^{*}\kappa & = \boldsymbol{\beta}_2^{H}\mathbf{C}_{V}\boldsymbol{\beta}_{2}  + \lambda P_o. \\
\intertext{But, $\lambda P_o = \trace(\Bww\Rni)$ as well, and thus}
\tilde{\mu}^{*}\kappa & = \boldsymbol{\beta}_2^{H}\mathbf{C}_{V}\boldsymbol{\beta}_{2}  + \trace(\Bww\Rni).
\end{align*}
The right hand side of the final equation is immediately recognizable as our objective function, which in some sense implies that this would be the only remaining aspect of the dual. 

\subsection{KKT Condition 5: Equality Constraints}
This is the final major KKT condition left to examine, because KKT Condition 6 is trivially satisfied by $\Cbf_{V}$ being positive semidefinite. Here, our primary concern is the equality constraint $\boldsymbol{\beta}^{H}_2\boldsymbol{\tau} = \kappa$. According to the first KKT condition, we know that 
\begin{align}
\boldsymbol{\beta}_{2} = \mathbf{C}_{V}^{\dagger}(\boldsymbol{\sigma}_2 + \tilde{\mu}^{*}\boldsymbol{\tau}) + \mathbf{P}_{\mathbf{C}_V}^{\perp}\mathbf{z}_{2}.
\end{align}
Substituting this into the equality constraint gives us
\begin{align}
\boldsymbol{\sigma}_2^{H}\mathbf{C}_{V}^{\dagger}\boldsymbol{\tau} + \tilde{\mu}\boldsymbol{\tau}^{H}\mathbf{C}_{V}^{\dagger}\boldsymbol{\tau} + \mathbf{z}_{2}^{H}\mathbf{P}_{\mathbf{C}_V}^{\perp} \boldsymbol{\tau} = \kappa. \label{eq:EquConst}\tag{EquC}
\end{align} 
Additionally, recall that, for this solution to exist, the following condition on $\boldsymbol{\sigma}_2$ must hold:
\begin{align}
\mathbf{P}_{\mathbf{C}_V}^{\perp}\boldsymbol{\sigma}_2 = -\tilde{\mu}^{*}\mathbf{P}_{\mathbf{C}_V}^{\perp}\boldsymbol{\tau}. \label{eq:ExistCond}\tag{XC}
\end{align}

\section{Consequences of the KKTs}\label{sec:KKTConsequences}
The optimality conditions shown above are, at first blush, a complicated set of matrix equations to solve. However, we can derive some insight into the nature of the relaxed solution by attacking small pieces of it. In this section, we will first demonstrate generic properties of every solution to the KKTs, then show a potential power-bounded solution under certain conditions. Finally, we will provide a roadmap for non-power-bounded solutions and their feasibility. 

\subsection{General properties of the relaxed solution}

First, we show that the solution does not reach the power bound $P_o$ iff the slackness matrix $\Sigws = \ZeroMat{NML}{N}$. 
\begin{lemma}
	$\lambda = 0 \iff \Sigws = \ZeroMat{NML}{N} $ \label{lem:LambdaZeroSigma2Zero}
\end{lemma} 	
\begin{IEEEproof}
	First, we proceed in the forward direction. If $\lambda = 0$, then the slackness matrix becomes $$\Sigbf = \begin{bmatrix}
	\Rni & \Sigws \\ \Sigws^{H} & \ZeroMat{N}{N}
	\end{bmatrix}.$$ To be part of a feasible solution, this must be positive semidefinite, which is only possible if $\Sigws = \Sigws(\ZeroMat{N}{N})^{\dagger}(\ZeroMat{N}{N}) = \ZeroMat{NML}{N}$ (see Theorem~\ref{thm:PSDness}). Hence, the forward direction is proved. 
	
	In the reverse direction, we prove through contradiction. Assume that $\Sigws = \ZeroMat{NML}{N}$ and $\lambda > 0$. The matrix slackness condition \ref{eq:Slack2} dictates that $\lambda\Bws = - \Bww \Sigws$. Under our first assumption, this becomes $\lambda\Bws = \ZeroMat{NML}{N}$, which simplifies to $\Bws = \ZeroMat{NML}{N}$ under the second assumption. However, any feasible solution must also satisfy the Capon constraint $\trace(\Bws^{H}\Tbf) = \kappa \neq 0$. Clearly, $\Bws = \ZeroMat{NML}{N}$ violates this constraint, which leads to our contradiction and completes the proof. 
\end{IEEEproof}

Next, we show that every feasible solution reaches the power bound if and only if the noise-and-interference correlation matrix $\Rni$ is full rank. This is a considerably longer and more complex proof. 
\begin{prop}
	$\lambda > 0 \iff \rank(\Rni) = NML$ \label{prop:FullNILambdaPos}
\end{prop}
\begin{IEEEproof}[Proof of $\rank(\Rni) = NML \Rightarrow \lambda > 0$]
	We will prove this by contradiction as well. Assume that $\Rni$ is full rank and $\lambda = 0$. Given the second condition, Lemma~\ref{lem:LambdaZeroSigma2Zero} requires that $\Sigws = \ZeroMat{NML}{N}$. If we apply this to the slackness condition \ref{eq:Slack1}, then $\Bww\Rni = \ZeroMat{NML}{NML}$. However, since $\Rni$ is full rank, this implies $\Bww = \ZeroMat{NML}{NML}$. This violates our non-triviality, but we will continue with the proof to show we reach a further contradiction. Since the overall solution matrix must be PSD, $\Bws = \ZeroMat{NML}{N}$ as well. As in the proof of Lemma~\ref{lem:LambdaZeroSigma2Zero}, we have reached a contradiction because the Capon constraint is violated, which completes the proof. 
\end{IEEEproof}
\begin{IEEEproof}[Proof of $\lambda > 0 \Rightarrow \rank(\Rni) = NML$]
	First, observe that for a non-trivial PSD solution matrix, $\Range{\Bws^{H}} \subseteq \Range{\Bss}$ by Theorem~\ref{thm:PSDnessHJ}. Next, we turn to slackness condition \ref{eq:Slack4}, which states $\lambda\Bss = -\Bws^{H}\Sigws$. If $\lambda > 0$, then clearly $\Range{\Bss} \subseteq \Range{\Bws^{H}}$. By the standard rules of subset inclusion, then, $\Range{\Bws^{H}} = \Range{\Bss}$ and $\rank(\Bws^{H}) = \rank(\Bss)$. This fact will become useful later.
	
	We will now use the results of \cite{KhatriMitra1976} on another slackness condition, \ref{eq:Slack3}, and use a substitution of \ref{eq:Slack4} to get to our destination. Using \cite[Theorem 2.2]{KhatriMitra1976} on \ref{eq:Slack3} to solve for $\Rni$, we have the following requirements for $\Rni$ to be at least PSD (which it is):
	\begin{enumerate}
		\item $-\Bss\Sigws^{H}\Bws \succeq 0$: If we substitute \ref{eq:Slack4} into this, we obtain $\lambda\Bss^{2} \succeq 0$, which is satisfied because $\lambda > 0$ and $\Bss$ is PSD.
		\item $\Range{\Bss\Sigws^{H}} \subseteq \Range{\Bws^{H}}$: This is satisfied because $\Bss\Sigws^{H} = - \frac{1}{\lambda}\Bws^{H}\Sigws\Sigws^{H}$ via \ref{eq:Slack4}, and the range inclusion follows directly.
		\item $\rank(-\Bss\Sigws^{H}\Bws) = \rank(\Bss\Sigws^{H})$: This is not immediately satisfied by the other conditions, but it does imply that $\rank(\Bss) = \rank(\Bss\Sigws^{H})$. 
	\end{enumerate}
	The more interesting requirement comes from \cite{Baksalary1984}, which adds the following: $\Rni$ is positive definite (and thus full rank) if and only if $\rank(-\Bss\Sigws^{H}\Bws) = \rank(\Bws^{H})$. We know from the third PSD requirement above that $\rank(-\Bss\Sigws^{H}\Bws) = \rank(\Bss)$, and thus $\Rni$ is full rank if and only if $\rank(\Bss) = \rank(\Bws^{H})$. However, as seen from above, this condition is already satisfied if $\lambda > 0$, and so our proof is complete.
\end{IEEEproof}

Since the proof of Proposition~\ref{prop:FullNILambdaPos} provides us with the evidence to show that power-bounded solutions occur only in non-singular noise-and-interference environments, we can also demonstrate an additional property of any power-bounded solution: namely, the rank of the overall solution matrix. Recall that for $\Bbf$ to be PSD, $\Range{\Bws} \subseteq \Range{\Bww}$. However, slackness condition \ref{eq:Slack1} provides that $\Bww\Rni = -\Bws\Sigws^{H}$. Since $\Rni$ is full rank, \ref{eq:Slack1} becomes $\Bww = -\Bws\Sigws^{H}\Rniinv$, which implies $\Range{\Bww} \subseteq \Range{\Bws}$. Thus, $\Range{\Bww} = \Range{\Bws}$, and $\rank(\Bww) = \rank(\Bws) = \rank(\Bss) \leq N$, where the last equality is implied by Proposition~\ref{prop:FullNILambdaPos}'s proof and the inequality is obvious. 

In the sidelooking STAP case, the noise \& interference covariance is \emph{always} full-rank, so we conclude that most practical solutions will achieve the power bound. However, for numerical purposes, these are worthwhile observations, because computation might result in a rank-deficient estimate of $\Rni$. We will explore this in more detail in Section~\ref{ssec:NonPowerBoundedSolns} below.

\begin{lemma}
	If $\lambda > 0$, $\tilde{\mu} \neq 0$.
\end{lemma}
\begin{IEEEproof}
	We prove by contradiction. Assume $\tilde{\mu} = 0$. When applied to \eqref{eq:Sig2VectorForm}, we then have $\boldsymbol{\sigma}_2 =	\mathbf{C}_{V}\boldsymbol{\beta}_{2}$. Premultiplying with $\boldsymbol{\beta}_{2}^{H}$, we have $\boldsymbol{\beta}_{2}^{H}\boldsymbol{\sigma}_2 = \boldsymbol{\beta}_{2}^{H}\mathbf{C}_{V}\boldsymbol{\beta}_{2} \geq 0$ as the quadratic form of a PSD matrix. But we've already established that for $\lambda > 0$, $\boldsymbol{\beta}_{2}^{H}\boldsymbol{\sigma}_2 < 0$, which is a contradiction and our proof is complete.
\end{IEEEproof}

\begin{lemma}
	In a power-bounded solution, $\rank(\Bbf) = N - \rank(\lambda\Identity{N} - \Sigws^{H}\Rniinv\Sigws)$. 
\end{lemma}
\begin{IEEEproof}
Our above note on Proposition~\ref{prop:FullNILambdaPos} is our starting point.  Since the matrix product $\Bbf\Sigbf$ is zero, so is its rank. This further implies $\rank(\Bbf) = J - \rank(\Sigbf)$. In general, since $\Sigbf$ is PSD, $\rank(\Sigbf) = \rank(\Rni) + \rank(\lambda\Identity{N} - \Sigws^{H}\Rnipinv\Sigws)$. Therefore, generally, $\rank(\Bbf) = (NML - \rank(\Rni)) + (N - \rank(\lambda\Identity{N} - \Sigws^{H}\Rnipinv\Sigws))$. In a power bounded solution, $\rank(\Rni) = NML$ and the inverse exists, thus $\rank(\Bbf) = N - \rank(\lambda\Identity{N} - \Sigws^{H}\Rniinv\Sigws)$.
\end{IEEEproof}

\subsection{A connection with waterfilling}\label{ssec:Waterfilling}
Using some of these general results from above, we can show that the optimal solution to the KKTs at the power bound satisfies equations that resemble the well-known "waterfilling" concept.

First, we begin with a decomposition of the clutter matrix $\Clutmat$. Recall that the rank of the clutter "subspace" (i.e., the rank of $\Rbf_{\mathtt{c}}(\sbf)$ and thus $\Clutmat$)is limited by both the physical extent and nature of non-target scatterers and our overall ability to observe this state of nature. For side-looking airborne arrays, the well-known Brennan rule~\cite{BrennanRule,Ward1994} is a reasonable approximation if certain conditions hold, but as \cite{GoodmanStiles2007} showed, a more robust result is obtained by applying the Landau-Pollak theorem. In any case, let us assume that the "true" rank of the clutter is $Q_{\mathrm{eff}} \leq Q \leq N^2ML$ (the subscript denoting the \emph{effective} number of clutter patches). Then, we can use the economy eigendecomposition to find two equivalent representations of $\Clutmat$, namely:
\begin{align*}
	\Clutmat & = \breve{\Ubf}_{C}\Dbf_{C}\breve{\Ubf}_{C}^{H} = \sum_{i = 1}^{Q_{\mathrm{eff}}}\nu_i \breve{\ubf}_{i}\breve{\ubf}_{i}^{H}.
\end{align*}
Here, $\breve{\Ubf}_{C} \in \Complex^{N^2 ML \times Q_{\mathrm{eff}}}$ is the matrix that forms the basis for the $Q_{\mathrm{eff}}$-dimensional vectorized clutter subspace, whose $i$th column is the eigenvector $\breve{\ubf}_{i} \in \Complex^{N^2 ML}$, $i \in \{1,\, \dots,\, Q_{\mathrm{eff}}\}$. $\Dbf_{C} \in \Complex^{Q_{\mathrm{eff}} \times Q_{\mathrm{eff}}}$ is a diagonal matrix whose $(i,i)$th element is the nonzero eigenvalue $\nu_i$. We can also consider the full eigendecomposition $\Clutmat = \breve{\Ubf}\Dbf\breve{\Ubf}$. Here, the unitary matrix $\breve{\Ubf} = [\breve{\Ubf}_{C} \; \breve{\Ubf}_{N}]$, where $\breve{\Ubf}_{C}$ is as above and $\breve{\Ubf}_{N}$ collects the $N^2ML - Q_{\textrm{eff}}$ eigenvectors in the nullspace of $\Clutmat$, which we can regard as the vectors $\breve{\ubf}_{i}, i \in \{Q_{\mathrm{eff}}+1,\, \dots,\, N^2ML\}$. $\Dbf$ is just the direct sum of $\Dbf_{C}$ and a $N^2ML-Q_{\mathrm{eff}} \times N^2ML-Q_{\mathrm{eff}}$ all-zeros matrix. This decomposition also provides us with an alternative representation of the signal-dependent clutter covariance matrix $\Rbf_{\mathtt{c}}(\sbf)$. Let us assume that there is a set of $N^2ML$ matrices $\Ubf_i \in \Complex^{NML \times N}$ whose vectorizations are $\vecmat(\Ubf_i) = \breve{\ubf}_i$ -- that is, they correspond to the eigenvectors of $\Clutmat$. Then, the signal-dependent clutter covariance can also be given by 
\begin{align}
\Rbf_{\mathtt{c}}(\sbf) = \sum_{i = 1}^{Q_{\mathrm{eff}}}\nu_i\Ubf_i\sbf\sbf^{H}\Ubf_i^{H}.
\end{align}
We note that $\Ubf_i^{H}\Ubf_i = \frac{1}{N}\Identity{N}$ for all $i$, not just those in the clutter indices. Thus, $\sqrt{N}\Ubf_i$ is a rank-$N$ partial isometry.

Now, given this formulation, we can show a waterfilling-like effect by combining the matrix slackness condition \eqref{eq:Slack3} and one of the Lagrangian conditions. Recall that the Lagrangian requires $\vecSigws = \Clutmat\vecBws - \tilde{\mu}^{*} \vecTarg$. In matrix form, and using the set of partial isometries, this expands to
\begin{align}
\Sigws = \sum_{i = 1}^{Q_{\mathrm{eff}}} \nu_i \trace(\Ubf_i^{H}\Bws) \, \Ubf_i - \tilde{\mu}^{*} \Tbf.
\end{align}
Plugging this into \eqref{eq:Slack3} gives us
\begin{align}
\Rni\Bws & = - \Sigws\Bss \\
		 & = \tilde{\mu}^{*} \Tbf\Bss - \sum_{i = 1}^{Q_{\mathrm{eff}}} \nu_i \trace(\Ubf_i^{H}\Bws) \, \Ubf_i\Bss.
\end{align}
Let us assume that we are power-bounded and everything that implies from the lemmas above. Thus, we can ``directly" find $\Bws$ by applying $\Rniinv$ above to obtain:
\begin{align}
\Bws & = \tilde{\mu}^{*}\Rniinv\Tbf\Bss - \sum_{i = 1}^{Q_{\mathrm{eff}}} \nu_i \trace(\Ubf_i^{H}\Bws) \, \Rniinv\Ubf_i\Bss \label{eq:WF_Bws}
\end{align}
From here, we make a judicious guess. Premultiplying both sides with another normalized channel matrix $\Ubf_j^{H}$ and taking the trace gives us
\begin{align}
\trace(\Ubf_j^{H}\Bws) & = \tilde{\mu}^{*} \trace(\Ubf_j^{H}\Rniinv\Tbf\Bss) - \sum_{i = 1}^{Q_{\mathrm{eff}}} \nu_i \trace(\Ubf_i^{H}\Bws) \,\trace(\Ubf_j^{H}\Rniinv\Ubf_i\Bss).
\end{align}
We can extract the $j$th term from the sum and collect it on the left hand side to obtain
\begin{align}
\trace(\Ubf_j^{H}\Bws)& (1 + \nu_j \,\trace(\Ubf_j^{H}\Rniinv\Ubf_j\Bss)) \nonumber\\
& = \tilde{\mu}^{*} \trace(\Ubf_j^{H}\Rniinv\Tbf\Bss)
- \sum_{\substack{i = 1 \\ i \neq j}}^{Q_{\mathrm{eff}}} \nu_i \trace(\Ubf_i^{H}\Bws) \,\trace(\Ubf_j^{H}\Rniinv\Ubf_i\Bss).
\end{align}
Now, the second term on the left-hand side of the above equation could be divided out as long as we had a guarantee it was always positive. First, if $\nu_j = 0$, this term is 1, which is positive. Otherwise, $\nu_j > 0$ since they are the non-zero eigenvalues of a positive semidefinite matrix. Next, we need examine the trace statement. $\Bss$ is positive semidefinite by construction, and $\Ubf_j^{H}\Rniinv\Ubf_j$ is positive definite because each $\Ubf_j$ is a full rank scaled partial isometry and $\Rni$ is positive definite (see \cite[Theorem 7.7.2]{HornJohnsonMA2e}). Thus, the trace of this matrix product is always positive and real.
With this in hand, we can now say
\begin{align}
\trace(\Ubf_j^{H}\Bws)
 = \tilde{\mu}^{*} \frac{\trace(\Ubf_j^{H}\Rniinv\Tbf\Bss)}{1 + \nu_j \,\trace(\Ubf_j^{H}\Rniinv\Ubf_j\Bss)}
- \sum_{\substack{i = 1 \\ i \neq j}}^{Q_{\mathrm{eff}}} \frac{\nu_i \,\trace(\Ubf_j^{H}\Rniinv\Ubf_i\Bss)}{1 + \nu_j \,\trace(\Ubf_j^{H}\Rniinv\Ubf_j\Bss)} \trace(\Ubf_i^{H}\Bws). \label{eq:Waterfilling1}
\end{align}
Since $\Bss$ is a relaxed version of $\sbf\sbf^{H}$, we can regard the expression $\nu_j \,\trace(\Ubf_j^{H}\Rniinv\Ubf_j\Bss)$ to be a relaxed form of $\nu_j \,\sbf^{H}\Ubf_j^{H}\Rniinv\Ubf_j\sbf$, which is (effectively) a clutter-to-noise-and-interference ratio for the $j$th basis matrix. Similarly, $\trace(\Ubf_j^{H}\Rniinv\Tbf\Bss)$ is a joint target-and-clutter to noise-and-interference ratio, and the cross term $\nu_i \,\trace(\Ubf_j^{H}\Rniinv\Ubf_i\Bss)$ captures the ratio of patch-to-patch interaction and the noise-and-interference level. Thus, the first term on the right hand side is effectively a normalized measure of the clutter-matched target spectrum given a signal basis $\Bss$, and the right hand side is a normalized measure of the intraclutter spectrum given that same basis. 

Traditional waterfilling dictates that power is preferentially injected to bands where the overall signal-to-noise ratio is high, proceeding to ``worse''-off bands until the available power is exhausted. This is somewhat \emph{inverted} in \eqref{eq:Waterfilling1} because the left-hand side is an unscaled version of the subspace alignment between the solution $\Bws$ and the $j$th canonical clutter transfer matrix $\Ubf_j$. Naively, we would like to \emph{minimize} this alignment for $j \in \{1, \dots, Q_{\mathrm{eff}}\}$, since aligning with the clutter would nominally degrade our matching of the target. However, the coupled nature of \eqref{eq:Waterfilling1} requires more nuance than that. First, observe that for the non-clutter directions, i.e., $j \in \{Q_{\mathrm{eff}} + 1, ..., N^2ML\}$, $\nu_j = 0$ and \eqref{eq:Waterfilling1} becomes
\begin{align}
\trace(\Ubf_j^{H}\Bws)
= \tilde{\mu}^{*} \trace(\Ubf_j^{H}\Rniinv\Tbf\Bss)
- \sum_{i = 1}^{Q_{\mathrm{eff}}} \nu_i \,\trace(\Ubf_j^{H}\Rniinv\Ubf_i\Bss) \trace(\Ubf_i^{H}\Bws). 
\end{align}
This means that the solution's alignment in the non-clutter directions follows that of the whitened target's, less the combined crossover between non-clutter and clutter in the whitened spectrum. If the target is strong in these directions, we are all set, because the solution should align towards them. Otherwise, the solution must match the target in-spectrum as near as possible in directions both where the whitened clutter power is the lowest and the alignment with other directions is minimized. This corresponds to the findings in \cite{SetlurDevroyeRangaswamy2014}, who showed a similar behavior in two-step mutual information waveform design over consecutive transmit epochs. In this case, the ``two steps'' can be regarded as the trans-receive pair instead of sequential temporal designs. 
	
But what of the value $\tilde{\mu}$? This clearly relates to the total available resources --  in this case, $\kappa$ and $P_o$ -- and provides us with the ``water'' in waterfilling. We can find a form of $\tilde{\mu}$ by reexamining \eqref{eq:WF_Bws} as follows. If we premultiply by the target $\Tbf^{H}$, we obtain
\begin{align}
\Tbf^{H}\Bws & = \tilde{\mu}^{*}\Tbf^{H}\Rniinv\Tbf\Bss - \sum_{i = 1}^{Q_{\mathrm{eff}}} \nu_i \trace(\Ubf_i^{H}\Bws) \, \Tbf^{H}\Rniinv\Ubf_i\Bss
\end{align}
Taking the trace of both sides, and recognizing that a feasible solution satisfies $\trace(\Tbf^{H}\Bws) = \kappa^{*}$, we now have
\begin{align}
\kappa^{*} = \tilde{\mu}^{*}\trace(\Tbf^{H}\Rniinv\Tbf\Bss) -&\sum_{i = 1}^{Q_{\mathrm{eff}}} \nu_i \trace(\Ubf_i^{H}\Bws) \, \trace(\Tbf^{H}\Rniinv\Ubf_i\Bss) .
\end{align}
\emph{This} is where the traditional waterfilling appears, since we are saying that the gain across the target (which we know in this case to be $\kappa^{*}$) is the upper bound on the available resources (given by the first term) minus the overall impact of the clutter weighted by its alignment (the second term). Continuing on, we can rearrange this form to be
\begin{align}
\tilde{\mu}^{*}\trace(\Tbf^{H}\Rniinv\Tbf\Bss)  =  \kappa^{*} + &\sum_{i = 1}^{Q_{\mathrm{eff}}} \nu_i \trace(\Ubf_i^{H}\Bws) \, \trace(\Tbf^{H}\Rniinv\Ubf_i\Bss).
\end{align}
Since according to the above argument, $\trace(\Tbf^{H}\Rniinv\Tbf\Bss)$ is never zero,  $\tilde{\mu}^{*}$ is therefore:
\begin{align}
\tilde{\mu}^{*} = \frac{\kappa^{*} + \sum_{i = 1}^{Q_{\mathrm{eff}}} \nu_i \trace(\Ubf_i^{H}\Bws) \, \trace(\Tbf^{H}\Rniinv\Ubf_i\Bss)}{\trace(\Tbf^{H}\Rniinv\Tbf\Bss)}
\end{align}
We can insert this into \eqref{eq:Waterfilling1}, collect terms, and solve for a matrix equation that we will leave for later study. Thus, the optimal solution to the relaxed problem describes a generalized whiten-and-match trans-receive filter process that exhibits waterfilling behavior, shaping the transmit process to match the target's response in clutter as much as possible.  

\subsection{A Potential Power-Bounded Solution}
If $\Cbf_{V}$ is less than full rank (which, in all practical scenarios, is always true), there is a potential path to a solution. First, assume that $\Rni \succ 0$, so $\lambda > 0$. Let us propose that the optimal vectorized corner matrix $\boldsymbol{\beta}_2$ is completely orthogonal to the clutter, i.e. $\Cbf_{V}\boldsymbol{\beta}_2 = \ZeroMat{N^2ML}{1}$. A natural requirement for this to be a feasible solution is that $\boldsymbol{\tau}$ is not completely embedded in the clutter spectrum, i.e. $\boldsymbol{\tau} \notin \Range{\Cbf_{V}}$. First, this means that $\boldsymbol{\sigma}_{2} = - \tilde{\mu}^{*} \boldsymbol{\tau}$, or $\Sigws = -\tilde{\mu}^{*}\Tbf$. 
Since we know that $\boldsymbol{\beta}_{2}^{H}\Cbf_{V}\boldsymbol{\beta}_2 = \tilde{\mu}^{*}\kappa - \lambda P_o$, a nullspace solution implies $\lambda = \frac{\tilde{\mu}^{*}\kappa}{P_o}$. Solving directly for $\tilde{\mu}$, we have $\tilde{\mu} = \frac{\trace(\Bww\Rni)}{\kappa^{*}}$.

Applied to Condition~\ref{eq:Slack2}, we then obtain $\Bws = \frac{\tilde{\mu}^{*}}{\lambda}\Bww\Tbf$, which, given the substitution above implies $\Bws = \frac{P_o}{\kappa}\Bww\Tbf$. In order for this to be completely orthogonal to the clutter, $\Cbf_{V}\vecmat(\Bww\Tbf) = \Cbf_{V}(\Tbf^{T}\kronecker\Identity{NML})\vecmat(\Bww) = \ZeroMat{N^2ML}{1}$. 

Following on to Condition~\ref{eq:Slack4}, we also have $\Bss = \frac{P_o^2}{\lvert\kappa\rvert^{2}}\Tbf^{H}\Bww\Tbf$. Since scalar feasibility requires that $\trace(\Bss) = P_o$, this implies that $\trace(\Tbf^{H}\Bww\Tbf) = \frac{\lvert\kappa\rvert^{2}}{P_o}$, which also satisfies the equality constraints. These results can also be shown to satisfy Conditions~\ref{eq:Slack1} and \ref{eq:Slack3}, as well as all of the positive semidefiniteness conditions, so long as $\trace(\Bww\Rni) \leq \frac{\lvert\kappa\rvert^2}{\frobnorm{\Tbf}{2}} \frac{\trace(\Rni)}{P_o}$. If the problem satisfies Slater's condition, we also have, directly, $\trace(\Bww\Rni) < \trace(\Rni)$, so the previous necessary condition is immediate. 
Hence, we have as a possible incomplete solution
\begin{align}
\Bbf = \begin{bmatrix}
\Bww & \frac{P_o}{\kappa}\Bww\Tbf \\
(\frac{P_o}{\kappa}\Bww\Tbf)^{H} & \frac{P_o^2}{\lvert\kappa\rvert^{2}}\Tbf^{H}\Bww\Tbf
\end{bmatrix}\label{eq:B1IncompleteMat}
\end{align}
where $\Bww$ minimizes $\trace(\Bww\Rni)$ subject to the requirements 
\begin{align*}
\Cbf_{V}(\Tbf^{T}\kronecker\Identity{NML})\vecmat(\Bww) & = \ZeroMat{N^2ML}{1} \\
\trace(\Tbf\Tbf^{H}\Bww) & = \frac{\lvert\kappa\rvert^{2}}{P_o} \\
\Bww \succeq 0.
\end{align*}

The eagle-eyed reader will note correctly that this sequence of equations resembles another optimization problem. Indeed, we can reframe the implicit matrix completion in (\ref{eq:B1IncompleteMat}) as the solution to
\begin{equation*}
\begin{aligned}
& \underset{\Bww}{\text{min}}
& & \trace(\Bww\Rni) \\
& \text{s.t.}
& & \Cbf_{V}(\Tbf^{T}\kronecker\Identity{NML})\vecmat(\Bww) = \ZeroMat{N^2ML}{1}\\  
& & & \trace(\Tbf\Tbf^{H}\Bww) = \frac{\lvert\kappa\rvert^{2}}{P_o}\\
& & & \Bww \succeq 0.
\end{aligned} 
\end{equation*}	
or, perhaps, after collapsing the nullspace requirement into its necessary constituent parts, 
\begin{equation*}
\begin{aligned}
& \underset{\Bww}{\text{min}}
& & \trace(\Bww\Rni) \\
& \text{s.t.}
& & \trace(\Tbf\boldsymbol{\Gamma}_{q}^{H}\Bww) = 0 \quad \forall q \in \{1, \dots, Q\}\\  
& & & \trace(\Tbf\Tbf^{H}\Bww) = \frac{\lvert\kappa\rvert^{2}}{P_o}\\
& & & \Bww \succeq 0.
\end{aligned} 
\end{equation*}
This is a standard semidefinite program so long as we use the Hermitian part of  $\Tbf\boldsymbol{\Gamma}_{q}^{H}$, which should be a semidefinite matrix, for all $q$. For now, we will not directly solve this subproblem, but we will note that some preliminary numerical analysis has shown that this is a reasonable path forward for the future.  

\subsection{Non-Power-Bounded Solutions}\label{ssec:NonPowerBoundedSolns}
As mentioned above, there may be situations when $r_{\mathtt{NI}} = \rank(\Rni) < NML$ -- for example, if we actually use an estimate $\Rniest$ of the noise-and-interference covariance, or if we consider a noise-free case for analysis. In these cases, we know from Proposition~\ref{prop:FullNILambdaPos} that $\lambda = 0$. This implies that the solution is not power bounded, and $\trace(\Bss) < P_o$. Furthermore, due to Lemma~\ref{lem:LambdaZeroSigma2Zero}, $\Sigws = \ZeroMat{NML}{N}$. With these in mind, we can produce, at the very least, a flowchart of solution properties in the non-power-bounded case. 

As a simple beginning, we note that the matrix complementary slackness conditions reduce to two:
\begin{align}
\Rni\Bww & = \ZeroMat{NML}{NML} \\
\Rni\Bws & = \ZeroMat{NML}{N}
\end{align}
This implies that if the solution is not power-bounded, it must null the entire noise-and-interference spectrum. Hence, we know that the matrices $\Bww$ and $\Bws$ have the general form:
\begin{align}
 \Bww & = (\Identity{NML} - \Rni\Rni^{\dagger})\Vbf_1(\Identity{NML} - \Rni\Rni^{\dagger}) = \NIProjPerp\Vbf_1\NIProjPerp \\
 \Bws & = \NIProjPerp\Vbf_{2}
\end{align}
where $\Vbf_1$ is an arbitrary $NML \times NML$ PSD matrix, $\Vbf_2$ is an arbitrary $NML \times N$ matrix, and $\NIProjPerp = \Identity{NML} - \Rni\Rni^{\dagger}$ is the orthogonal projection matrix onto the nullspace of $\Rni$. Additionally, because of the first slackness condition, we know that $\rank(\Bww) \leq NML - r_{\mathtt{NI}}$. 
 
 This form of $\Bws$ has a feasibility consequence. Recall that a solution is feasible only if the Capon constraint $\trace(\Bws^{H}\Tbf) = \kappa \neq 0$ is satisfied. If we substitute our new form of $\Bws$ into the constraint, it becomes 
 \begin{equation}
 \trace(\Bws^{H}\Tbf) = \trace(\Vbf_{2}^{H}\NIProjPerp\Tbf) = \kappa.
 \end{equation}
 Since $\Vbf_{2}$ is arbitrary (but not trivial), feasibility is violated when $\NIProjPerp\Tbf = \ZeroMat{NML}{N}$, which only occurs when $\Range{\Tbf} \subseteq \Range{\Rni}$ (or, when vectorized, $\vecTarg \notin \Range{\Rnirep}$). Hence, there is no feasible solution if the target is embedded in the noise \& interference spectrum. We state this directly in the following Lemma.
 \begin{lemma}
 	If $\lambda = 0$ and $\Range{\Tbf} \subseteq \Range{\Rni}$, there is no feasible solution to (\ref{eq:VSSDRproblem}) or its equivalent problems.\label{lem:OrthNIFeasibility}
 \end{lemma}
 
 Continuing on, our Lagrangian minimization becomes
 \begin{align}
 \Clutmat\vecBws & = \tilde{\mu}^{*} \vecTarg \\
 \intertext{or, given the form of $\Bws$ above, }
 \Clutmat\NIBreveProjPerp\vbf_{2} & = \tilde{\mu}^{*} \vecTarg
 \end{align}
 where $\NIBreveProjPerp = \Identity{N} \otimes \NIProjPerp$ and $\vbf_{2} = \vecmat(\Vbf_2)$. 
 
 Let us momentarily consider the generally unrealistic case when $\Clutmat$ is full rank (i.e. $Q_{eff} = N^2ML$). Under the $\lambda = 0$ hypothesis, we can directly solve for $\vecBws$:
 \begin{equation}
 \vecBws = \tilde{\mu}^{*} \Clutmat^{-1}\vecTarg.
 \end{equation}
 However, since $\vecBws \in \Nullspace{\Rnirep}$, $\NIBreveProj\vecBws = \ZeroMat{N^2ML}{1}$, so we can also say that 
 \begin{equation}
 \tilde{\mu}^{*} \NIBreveProj\Clutmat^{-1}\vecTarg = \ZeroMat{N^2ML}{1}
 \end{equation}
 If we want this to be a feasible result (i.e. $\tilde{\mu} \neq 0$), then the whitened target $\Clutmat^{-1}\vecTarg \in \Nullspace{\Rnirep}$ (and by extension, $\vecTarg \in \Nullspace{\Rnirep}$). Hence, if the clutter is full rank, the target must be clear of the noise and interference (and the dimensionality of the available resources must be such that this is possible) for a feasible solution to exist. This is effectively a further restriction of Lemma~\ref{lem:OrthNIFeasibility}.
 
 Assuming $\Clutmat$ is less than full rank, we can find another condition on the target response for the Lagrange multipliers by premultiplying the above equation by the  clutter nullspace projection matrix $\ClutProjPerp$, which becomes:
 \begin{equation*}
 \tilde{\mu}^{*}\ClutProjPerp\vecTarg = \ZeroMat{N^2ML}{1}
 \end{equation*}
 This means that either $\tilde{\mu} = 0$ or $\vecTarg \in \Range{\Clutmat}$. 	Thus, if $\vecTarg \notin \Range{\Clutmat}$ (that is, it has a non-zero component outside of the clutter), then $\tilde{\mu} = 0$. 
 
 The latter scenario, which we treat first, simplifies things considerably, because now:
 \begin{equation}
 \Clutmat\NIBreveProjPerp\vbf_{2}  = \ZeroMat{N^2ML}{1}
 \end{equation}
 Hence, either $\vbf_2$ (and thus $\vecBws$) lies in $\Nullspace{\Clutmat\NIBreveProjPerp}$ or $\vbf_2 = \vecBws = \ZeroMat{N^2ML}{1}$. The second possibility violates the Capon constraint, therefore $\vbf_2,\; \vecBws \in \Nullspace{\Clutmat\NIBreveProjPerp}$. However, if (for whatever reason), the target does not lie in this space, then we have reached another infeasiblity result via the Capon constraint. 
 
 If $\vecTarg$ \emph{does} lie entirely within the clutter, then there are a few complex scenarios. First, consider a situation where $\Nullspace{\Rnirep} \subseteq \Nullspace{\Clutmat}$, which means the clutter is subsumed entirely into interference and noise (and thus $\Range{\Clutmat} \subseteq \Range{\Rnirep}$ \& $Q_{eff} \leq Nr_{\mathtt{NI}}$). In this case, $\Clutmat\NIBreveProjPerp = \ZeroMat{N^2ML}{N^2ML}$. However, this also means that the \emph{target} is now also embedded in the noise and interference. By Lemma~\ref{lem:OrthNIFeasibility}, there is no feasible solution! 
 In any other case, a solution exists if and only if $\tilde{\mu}^{*}\vecTarg \in \Range{\Clutmat\NIBreveProjPerp}$.
 

\section{Simulations and Results}\label{sec:simul}

\subsection{The Solvers}
While we have preliminarily reduced solving (\ref{eq:SSDRproblem}) to a matrix completion problem, it is possible to solve (in some sense) the problem numerically with commercial solvers. Most of the analysis presented here is enabled by the modeling package CVX\cite{CVXSoftware,GrantBoyd08}, which permitted us to construct two equivalent representations of the problem with rather different results. In all cases, we considered only the solver \emph{package} SDPT3 \cite{Toh1999,Tutuncu2003}. 

The major difference occurs in how the quadratic form in the problem is presented to CVX. The first method (hereinafter the ``QuadSolver") uses the \texttt{quad\_form()} function, which preserves the form $\vecmat(\Bbf)^{H}\tilde{\Cbf}_{V}\vecmat(\Bbf)$. The second method (hereinafter the ``NormSolver") uses a composition of the \texttt{pow\_pos()} and \texttt{norm()} functions, which implements the equivalent form $\normsq{\tilde{\Cbf}_{V}^{1/2}\vecmat(\Bbf)}$. Though these forms are symbolically equivalent, the latter is supposedly more amenable to optimization by a conic representation which is more efficiently processed by the default solvers available to CVX. We will demonstrate that while this might be true, the solvers behave very differently under certain conditions. 

\paragraph*{A note on dimensionality}
Regardless of the solver particulars, it is important to note that all presented scenarios and solutions are necessarily constrained by available computing power and memory constraints. While CVX and MATLAB combine to make a powerful tool, they are limited in the size of problems that can be solved on a regular workstation. To wit, we have found that when $NML > 400$, our current workstation (AMD Athlon\textsuperscript{TM} II X2 B24 processor at 2.3GHz with 8GB memory) throws out-of-memory errors during CVX's setup phase. We anticipate performing this analysis on more robust systems in the future, which will allow us to extend it to more realistic radar scenarios. 

\subsection{Computational Analysis of the KKTs}
In the following simulations, we used certain common parameters for ease of comparison. As noted above, computational limitations constrain our overall problem size; therefore, in this case, we assume $N = 5$ fast time samples, $M = 5$ array elements, and $L = 16$ pulses. The radar operates on a carrier frequency of 1 GHz and transmits pulses with a bandwidth of 50 MHz. The receive array has elements spaced at half a wavelength, i.e. $d = \lambda_o/2$. We recognize that these are not particularly realistic parameters (at least for the dimensionality), but as stated above, computational limitations currently conspire against a more representative simulation. 

Unless otherwise stated, the scenarios presented to the solver were as follows: the noise covariance matrix was a scaled correlation matrix with correlation function $\exp(-0.05|n|)$ for $n \in \{1, \dots, NML\}$. Interferers were placed at the azimuth-elevation pairs (0.3941, 0.3) radians and (-0.4951,0.3), with correlation matrices given by the Toeplitz matrix associated with the correlation function $\exp(0.2|l|)$ for $l \in \{1, \dots, NL\}$. The clutter was simulated by placing $Q = 25$ patches of $P = 5$ scatterers each, equally spaced in azimuth over the interval $(-\pi/2, \pi/2)$ at an elevation angle of $\tfrac{\pi}{4}$ radians. Adjacent scatterers within each patch are correlated with coefficient -0.2. 
In all cases, we have made the Capon constraint real. 

To obtain signals and beamformers from the relaxed problem, we generated the best "rank-1" approximation by summing the first $B$ singular vectors of the solution matrix $\Bbf$ weighted by their associated singular value. In the Frobenius sense, only the largest singular vector/value pair is the best approximation, but we have found that this approximation does, in fact, sketch out the nature of the unrelaxed problem. In the results below, we will often sweep the value $B$ from 1 to $J$, capturing, in some sense, the entire space. We recognize, however, that this is merely a heuristic that in this case happens to work well. 

\subsubsection{Variation in $\kappa$}
First, we examine varying the Capon constraint with a fixed power constraint. In these scenarios, $\kappa$ ranges from $10^{-2}$ to $NML \times 10^{4}$ while $P_o = 10^{7}$. Before continuing, we note that Slater's condition dictates that the duality gap should be zero so long as $|\kappa|^2 < NMLP_o = 4 \times 10^{9}$, which means that our parameter sweep should lie within this region. For reasonable comparison, we assumed no interference in this scenario.
Plotted scenarios exclude solver results where either the solver failed or the solution violated the KKTs significantly despite the solver claiming success. In practice, this means we excluded ``solutions" where $\lambda < 0$. 

\begin{figure}[!ht]
	\centering
	\includegraphics[width=\textwidth]{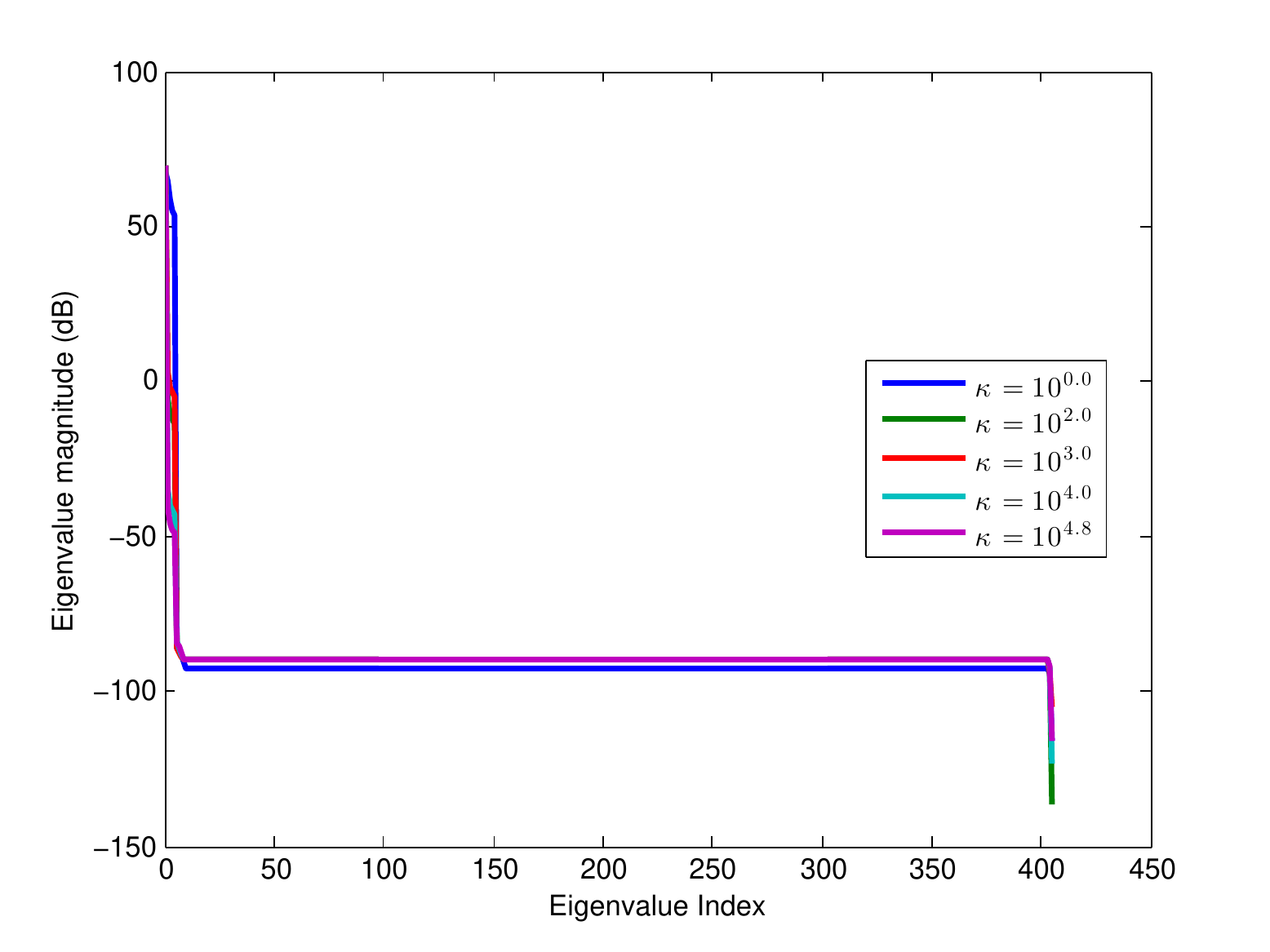}
	\caption{Eigenvalue spread of $\Bbf$ solution matrix using QuadSolver}
	\label{fig:CCS_Quad_ESpread}
\end{figure}

\begin{figure}[!ht]
	\centering
	\includegraphics[width=\textwidth]{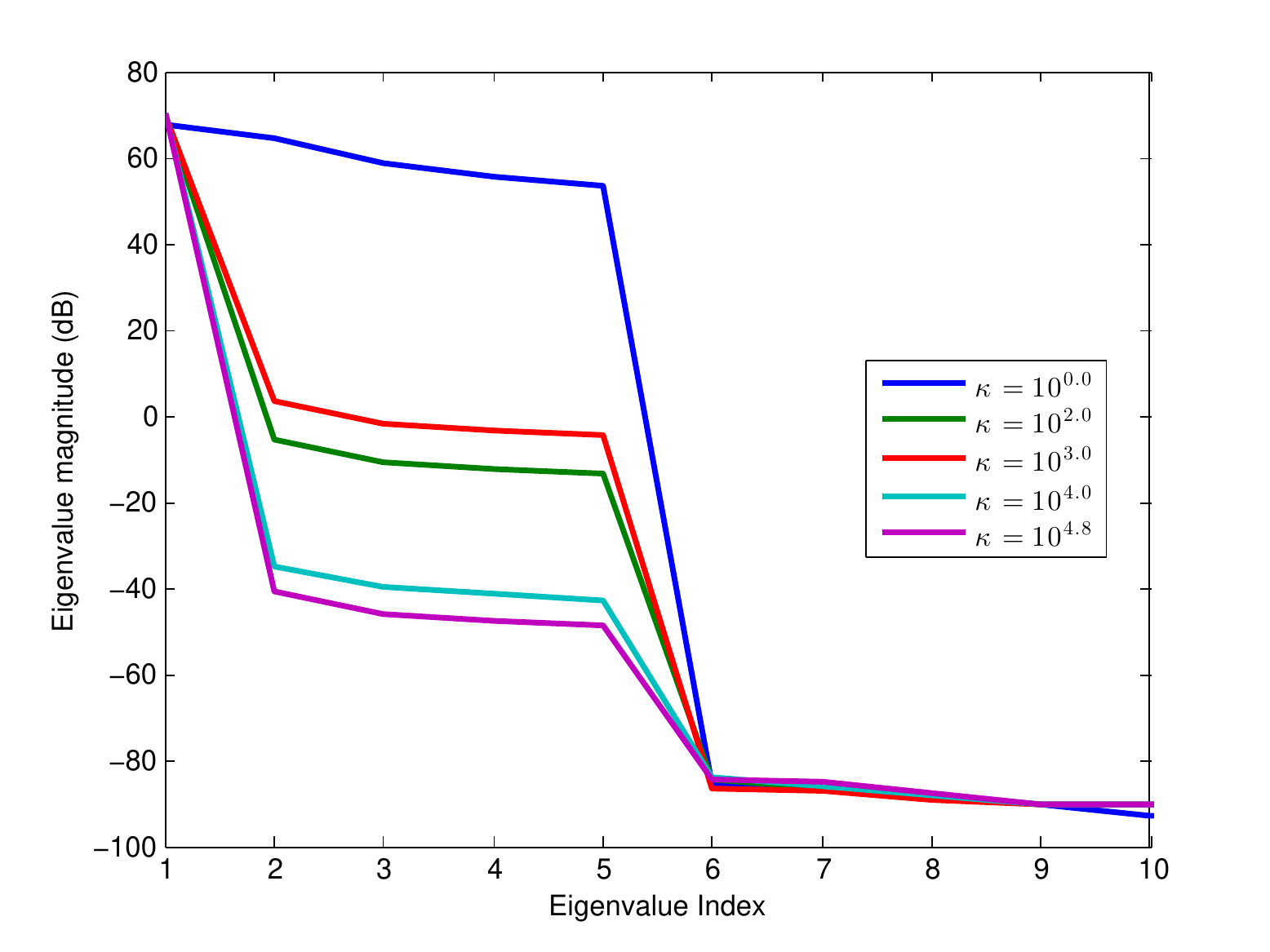}
	\caption{Closeup of the first $2N$ eigenvals. of $\Bbf$ using QuadSolver}
	\label{fig:CCS_Quad_ESpread_Zoom}
\end{figure}

We begin with the eigenspread of the output solution matrix $\Bbf$ for each solver and each value of $\kappa$. Figure~\ref{fig:CCS_Quad_ESpread} shows the eigenspreads produced by Quad Solver, with Figure~\ref{fig:CCS_Quad_ESpread_Zoom} showing finer detail of the first $2N$ eigenvalues. Clearly, all scenarios produce functionally low-rank solutions -- the "true" rank never exceeds $N$. Indeed, the \emph{effective} numeric rank only exceeds one in the $\kappa = 1$ scenario.  Since the power is constant, the peak eigenvalue is nearly same in all scenarios, but increasing $\kappa$ drives the non-peak eigenvalues lower.

\begin{figure}[!ht]
	\centering
	\includegraphics[width=\textwidth]{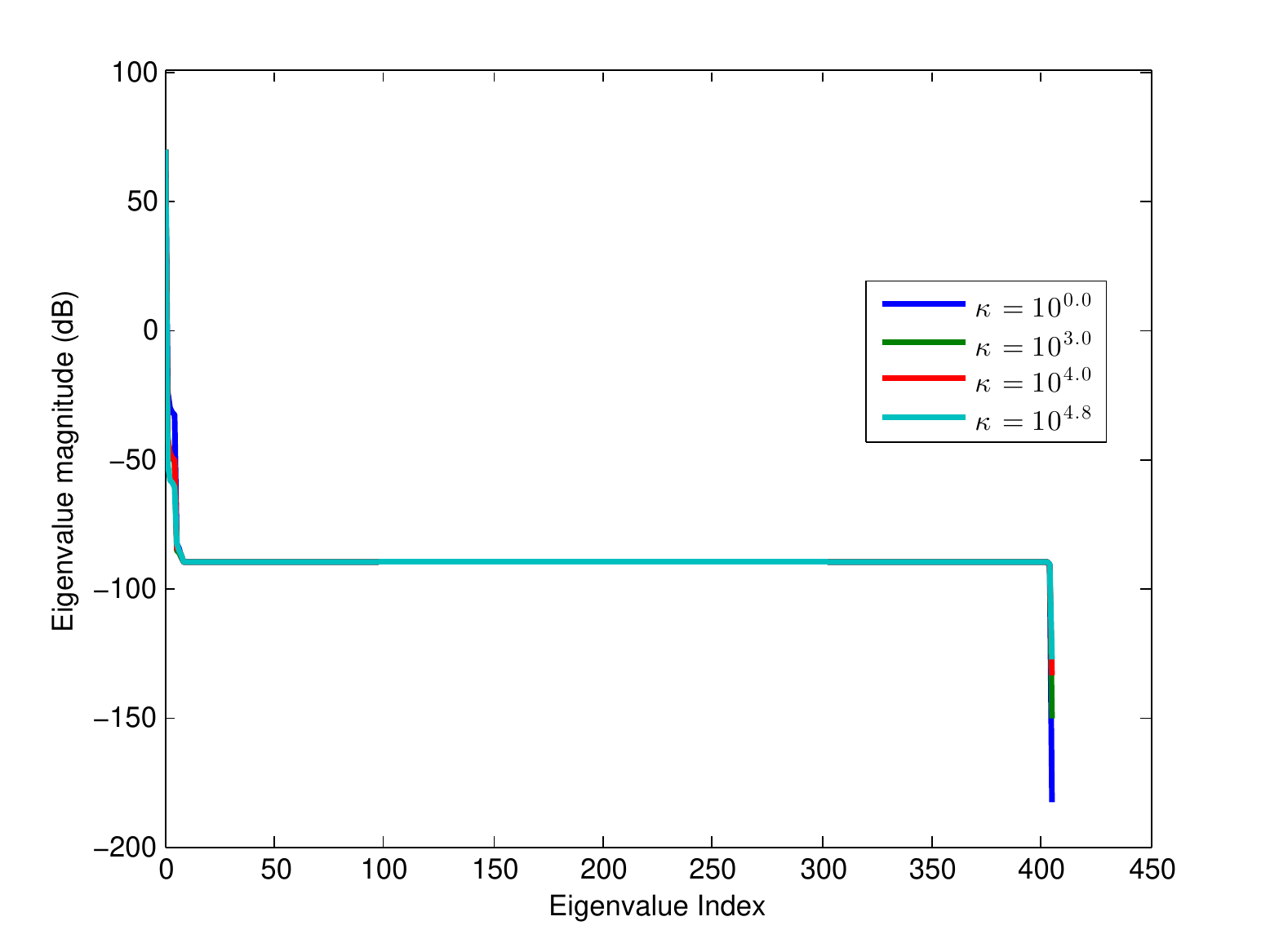}
	\caption{Eigenvalue spread of $\Bbf$ solution matrix using NormSolver}
	\label{fig:CCS_Norm_ESpread}
\end{figure}

\begin{figure}[!ht]
	\centering
	\includegraphics[width=\textwidth]{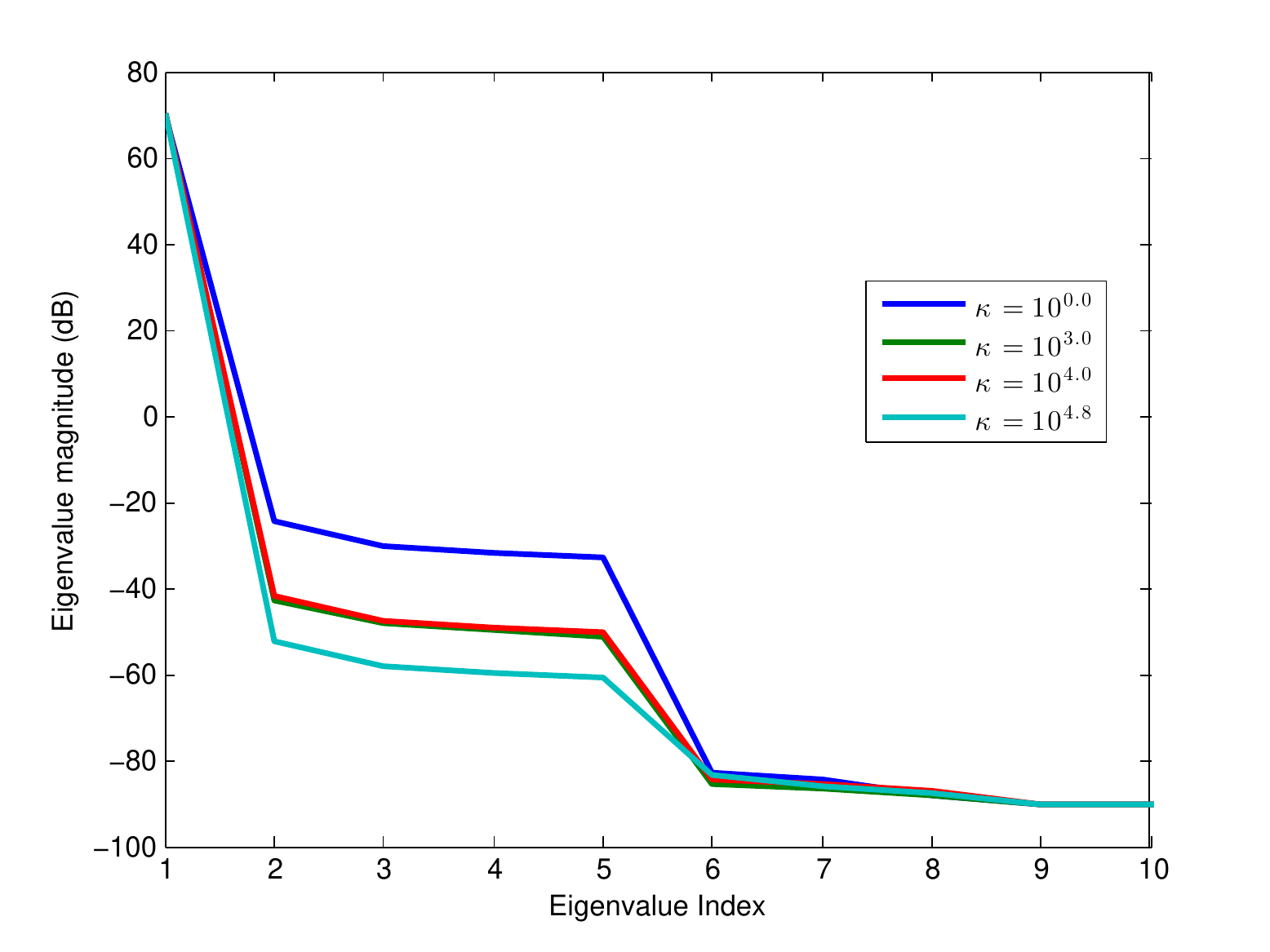}
	\caption{Closeup of the first $2N$ eigenvals. of $\Bbf$ using NormSolver}
	\label{fig:CCS_Norm_ESpread_Zoom}
\end{figure}

This dependence on $\kappa$ essentially repeats for the NormSolver-produced solutions, as can be seen in Figure~\ref{fig:CCS_Norm_ESpread} (the overall eigenspread) and Figure~\ref{fig:CCS_Norm_ESpread_Zoom} (the first $2N$ eigenvalues). The primary difference here is that all valid solutions produced have an effective rank of one, and the overall dropoff from the $N$th eigenvalue to the $N-1$th eigenvalue is not nearly as steep. This can potentially be attributed to less accumulated numerical error in NormSolver's process, leading to a cleaner solution. 

\begin{figure}[!ht]
	\centering
	\includegraphics[width=\textwidth]{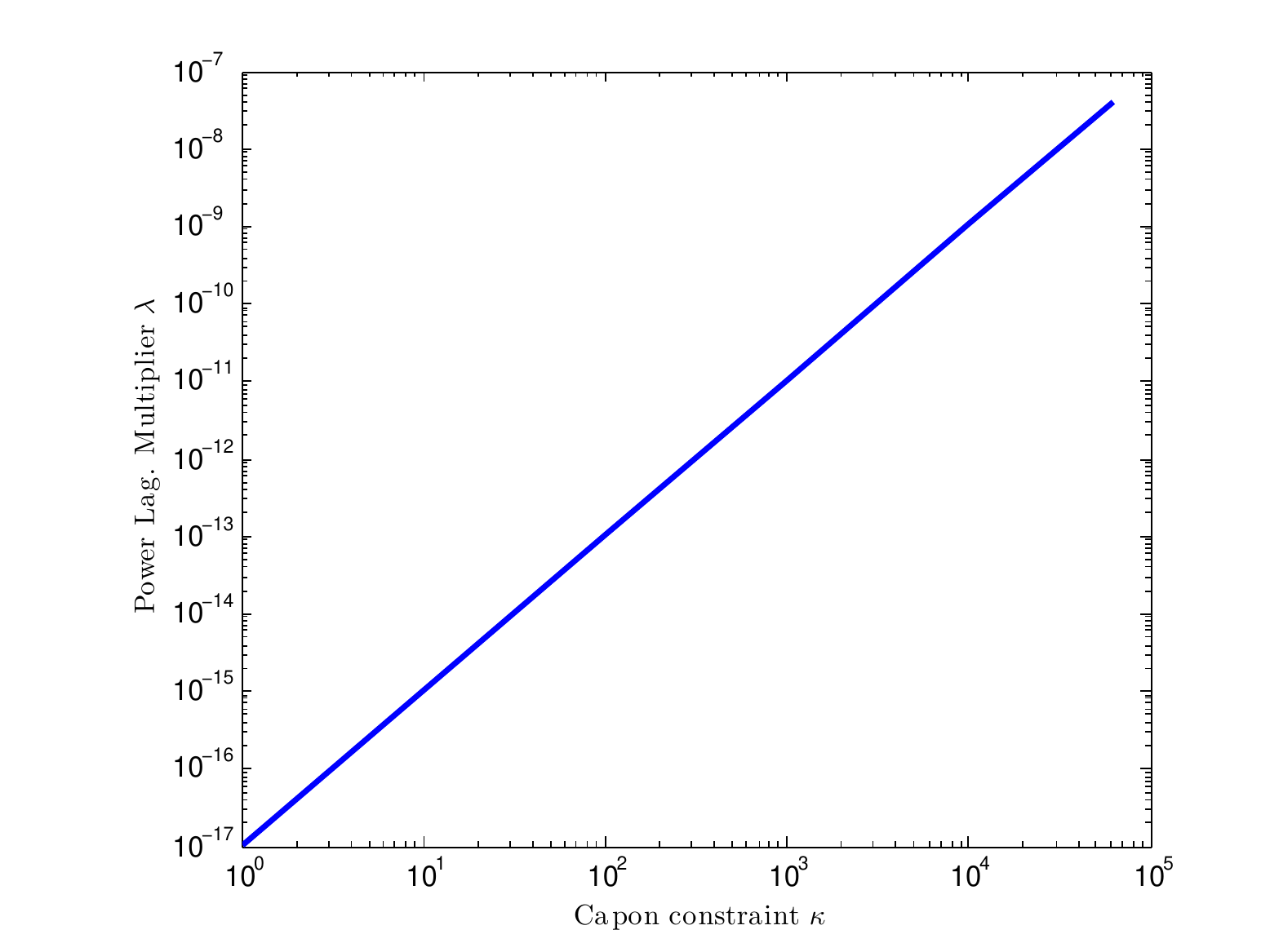}
	\caption{Optimal $\lambda$ as a function of the Capon constraint, QuadSolver}
	\label{fig:CCS_Quad_Lambda}
\end{figure}

\begin{figure}[!ht]
	\centering
	\includegraphics[width=\textwidth]{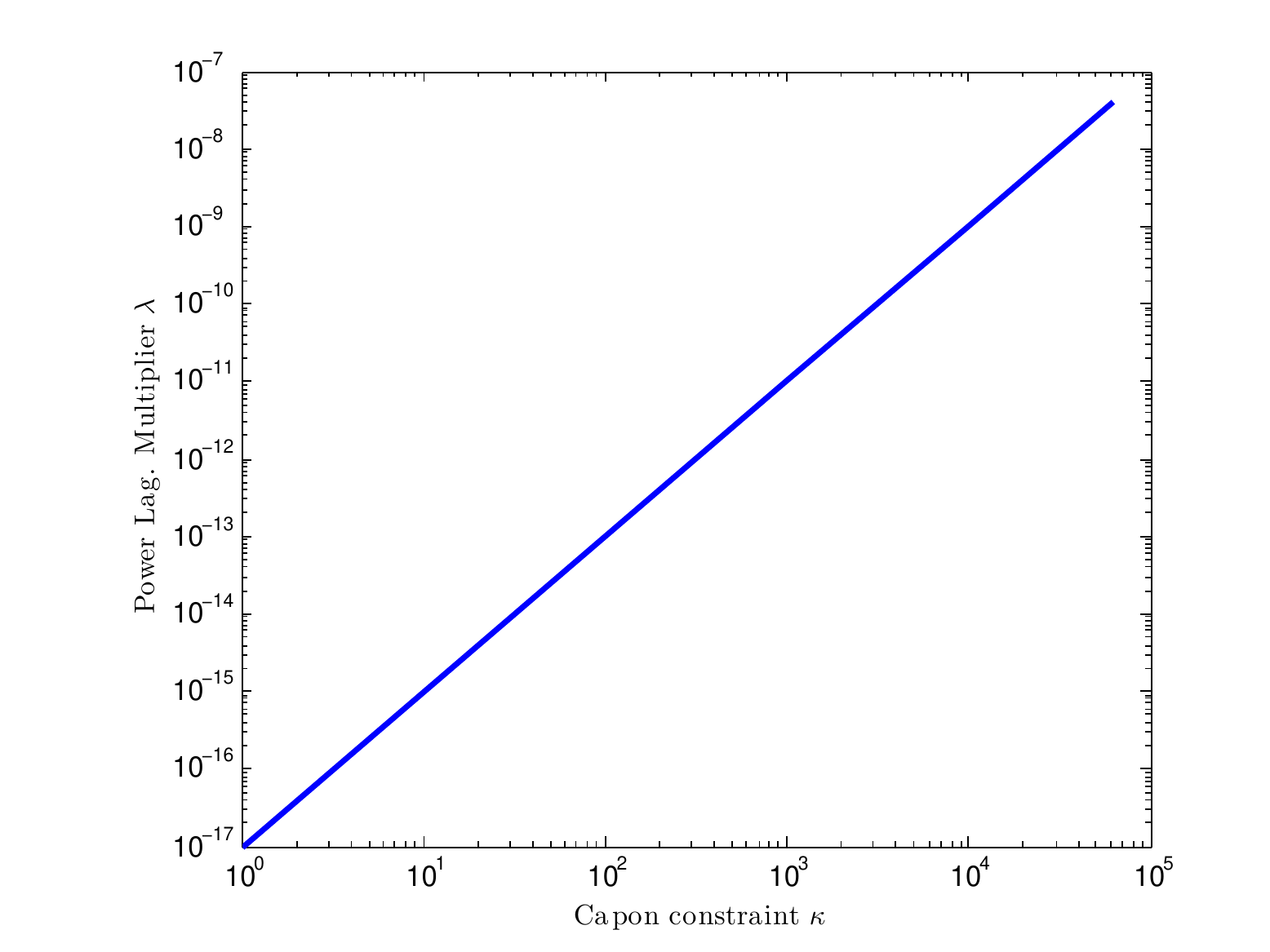}
	\caption{Optimal $\lambda$ as a function of the Capon constraint, NormSolver}
		\label{fig:CCS_Norm_Lambda}
\end{figure}

As predicted by the KKTs, increasing $\kappa$ increases the Lagrange multiplier $\lambda$ in equal measures for a fixed power level. This effect is demonstrated directly in Figures~\ref{fig:CCS_Quad_Lambda} and \ref{fig:CCS_Norm_Lambda}. We see that the returned $\lambda$ values appear to be similar for both solvers across the entire $\kappa$ parameter space which indicates they converge to similar solutions, at least in terms of noise-and-interference suppression. 

Next, we examine the effectiveness of our low-rank approximation by analyzing the cost function of \eqref{eq:OriginalProblem}/\eqref{eq:BQPws}. In the following figures, we use the term "basis length" to indicate how many of the $J$ singular vectors of $\Bbf_{opt}$ are combined to form our estimate of $\mathbf{b}$. That is, if $\sigma_{\Bbf,i}$ is the $i$th largest singular value of the solution matrix $\Bbf$ and $\ubf_{\Bbf,i}$ is the associated singular vector, a basis length of $B$ produces an approximate vector of $\mathbf{b}_{\mathrm{appx},B} = \sum_{i=1}^{B} \sigma_{\Bbf,i}\ubf_{\Bbf,i}$, with the associated beamformer $\wbf_{\mathrm{appx},B}$ and signal $\sbf_{\mathrm{appx}, B}$ recovered from the stacked vector as in \eqref{eq:SplittingMatrices}.

\begin{figure}[!ht]
	\centering
	\includegraphics[width=\textwidth]{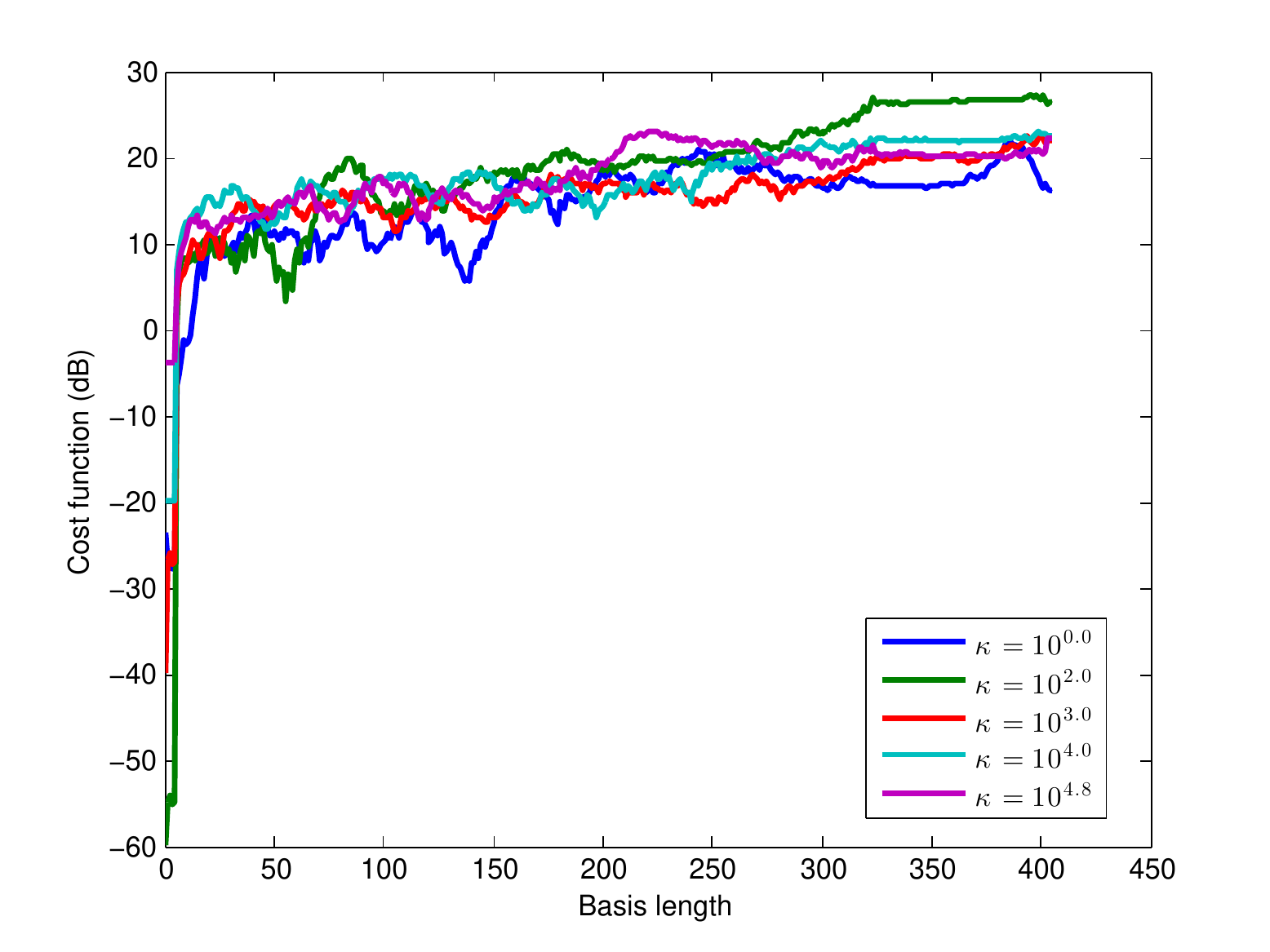}
	\caption{Original cost as function of basis length, QuadSolver}
	\label{fig:CCSQuadCostfunc}
\end{figure}

\begin{figure}[!ht]
	\centering
	\includegraphics[width=\textwidth]{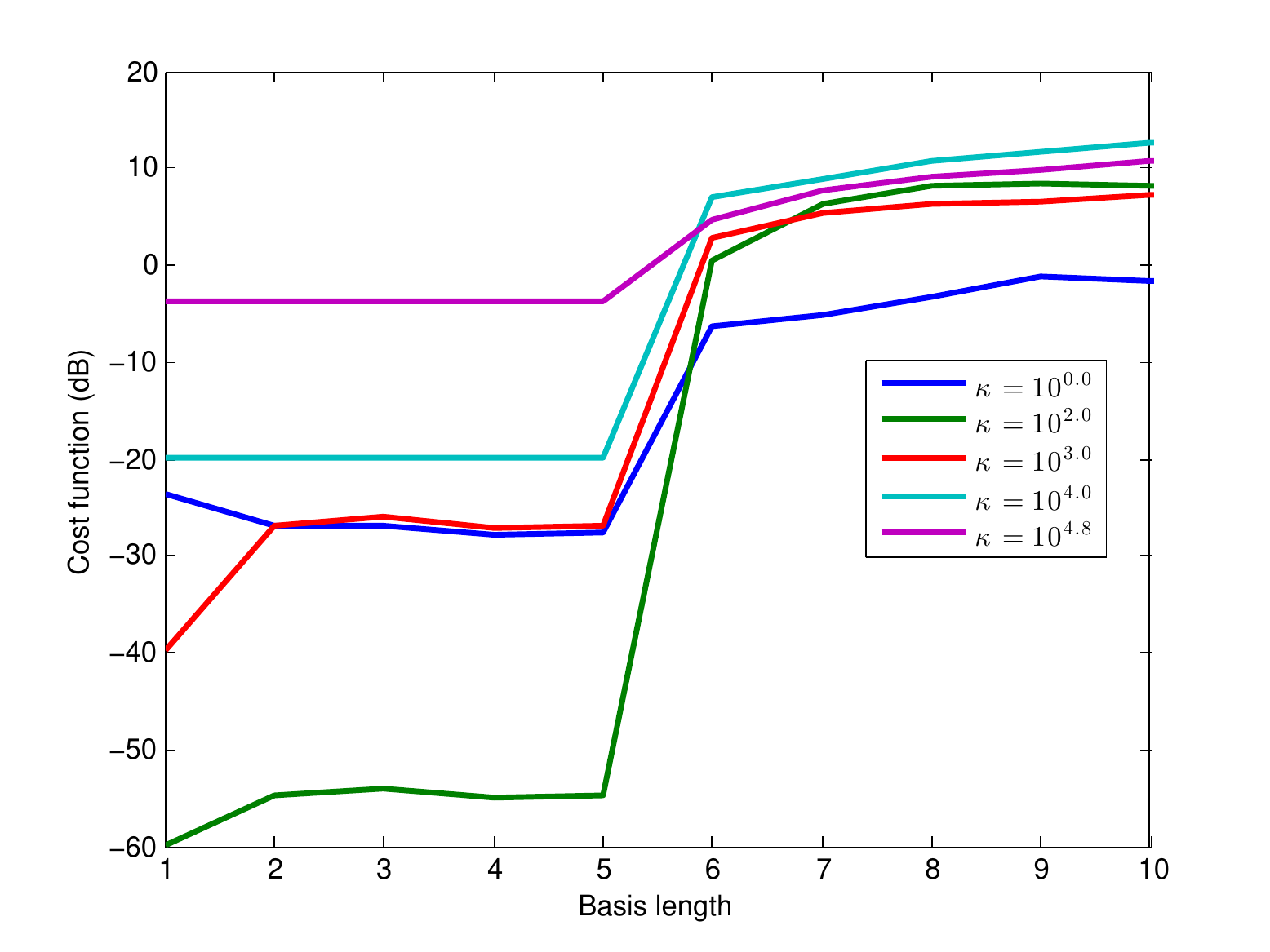}
	\caption{Original cost as function of basis length, first $2N$, QuadSolver}
	\label{fig:CCSQuadCostfuncZoom}
\end{figure}

Figure~\ref{fig:CCSQuadCostfunc} shows the overall sweep of basis length for the approximate solutions recovered from QuadSolver as a function of $\kappa$, with Figure~\ref{fig:CCSQuadCostfuncZoom} focusing only on the first $2N$ values. 
With the notable exception of $\kappa = 1$, as $\kappa$ increases, the minimum original cost increases and the difference over the first $N$ basis vectors becomes less pronounced. We can regard the result at $\kappa = 1$ to be a threshhold scenario where, since the effective numerical rank of the relaxed solution is closer to $N$, more of the basis is needed for better approximation.

\begin{figure}[!ht]
	\centering
	\includegraphics[width=\textwidth]{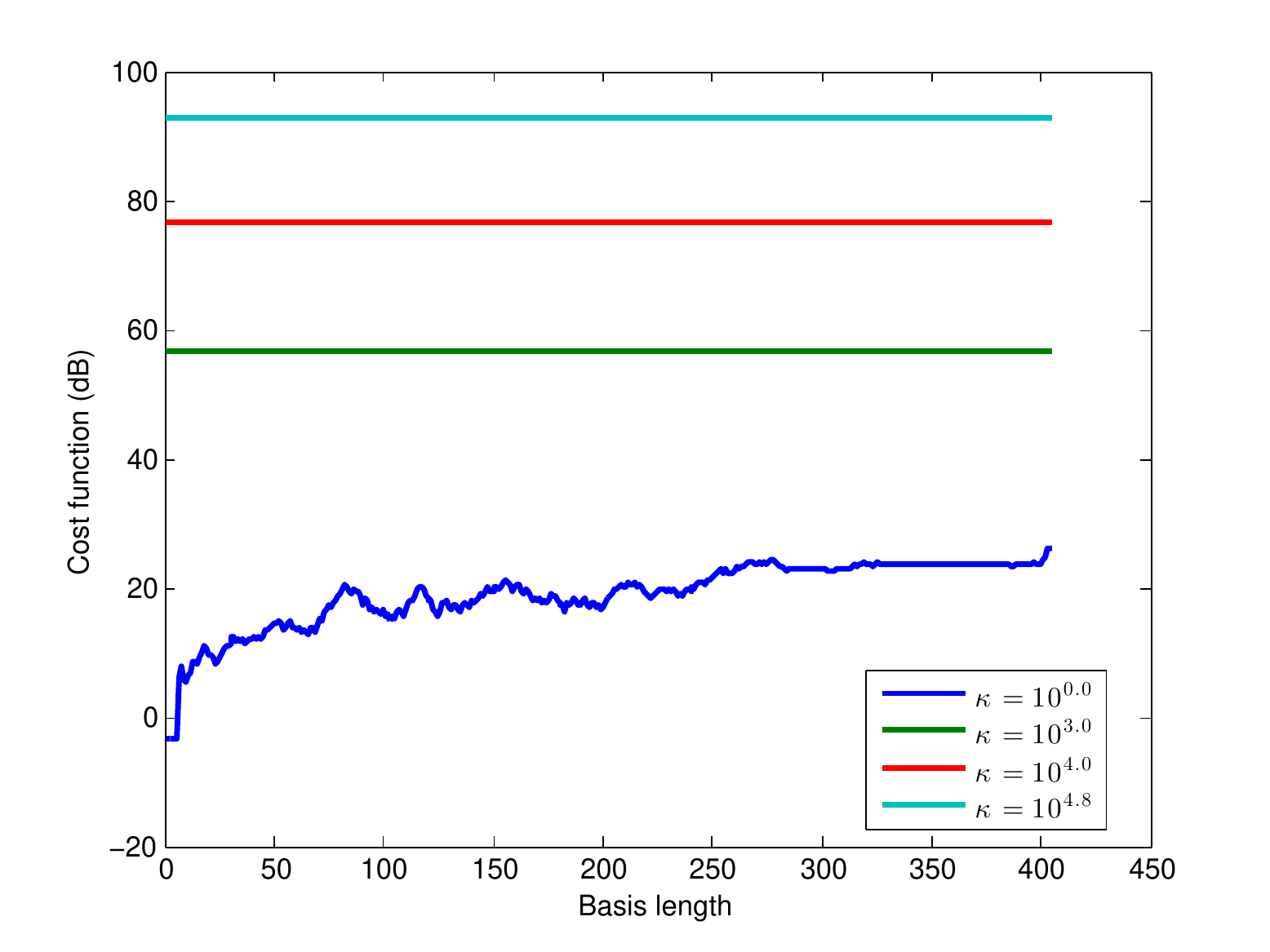}
	\caption{Original cost as function of basis length, NormSolver}
	\label{fig:CCSNormCostfunc}
\end{figure}

\begin{figure}[!ht]
	\centering
	\includegraphics[width=\textwidth]{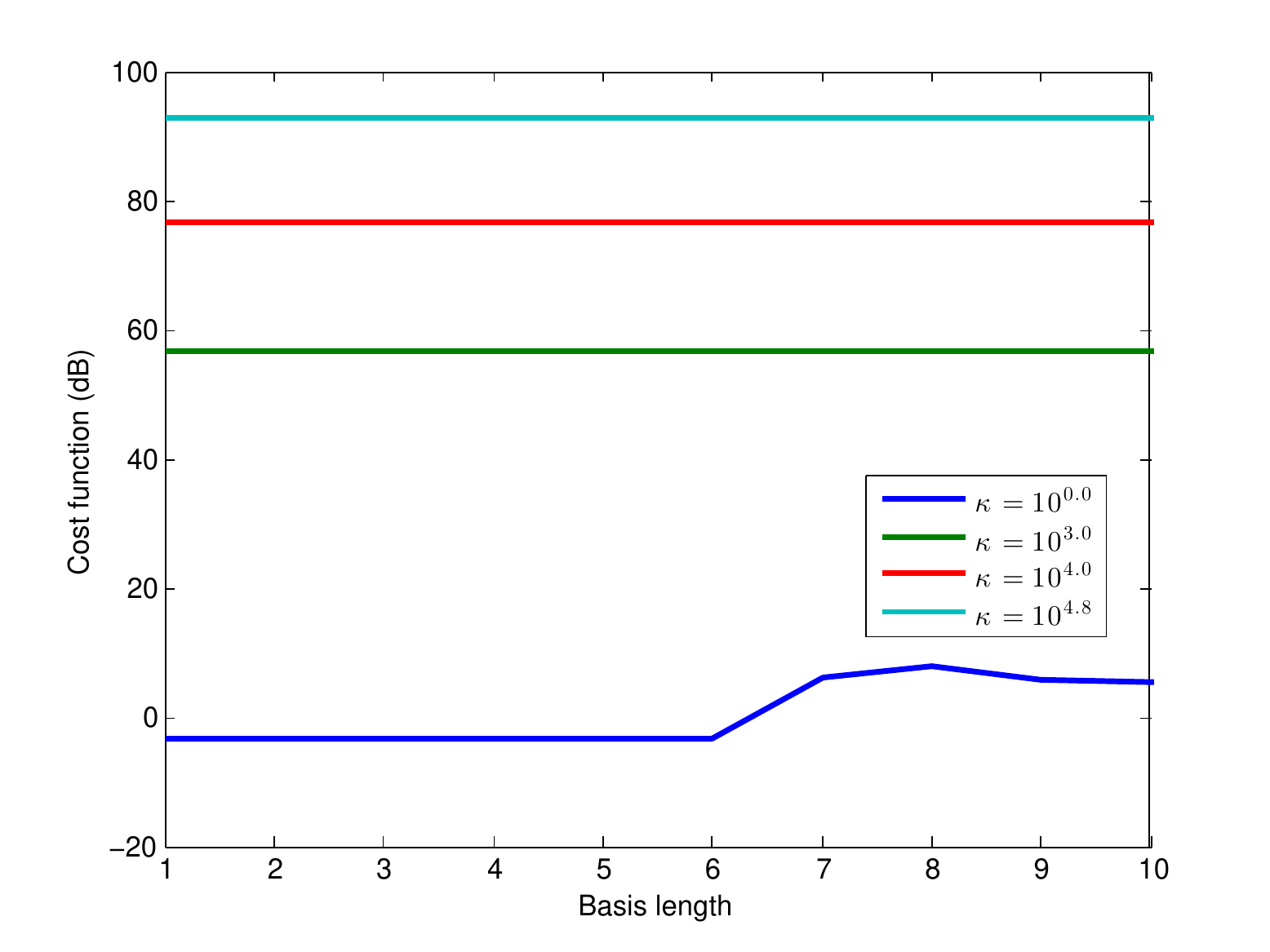}
	\caption{Original cost as function of basis length, first $2N$, NormSolver}
	\label{fig:CCSNormCostfuncZoom}
\end{figure}

In contrast, NormSolver produces solutions that generally perform poorly in the original cost function. As seen in Figures~\ref{fig:CCSNormCostfunc} and \ref{fig:CCSNormCostfuncZoom}, only the $\kappa = 1$ scenario replicates the general behavior seen from QuadSolver, with the NormSolver-produced minimum value exceeding its QuadSolver counterpart by nearly 20 dB. For higher values of $\kappa$, the previously observed flattening effect is in full force, indicating that this solver produces identically structured solutions that merely scale with $\kappa$. 

\begin{figure}[!ht]
	\centering
	\includegraphics[width=\textwidth]{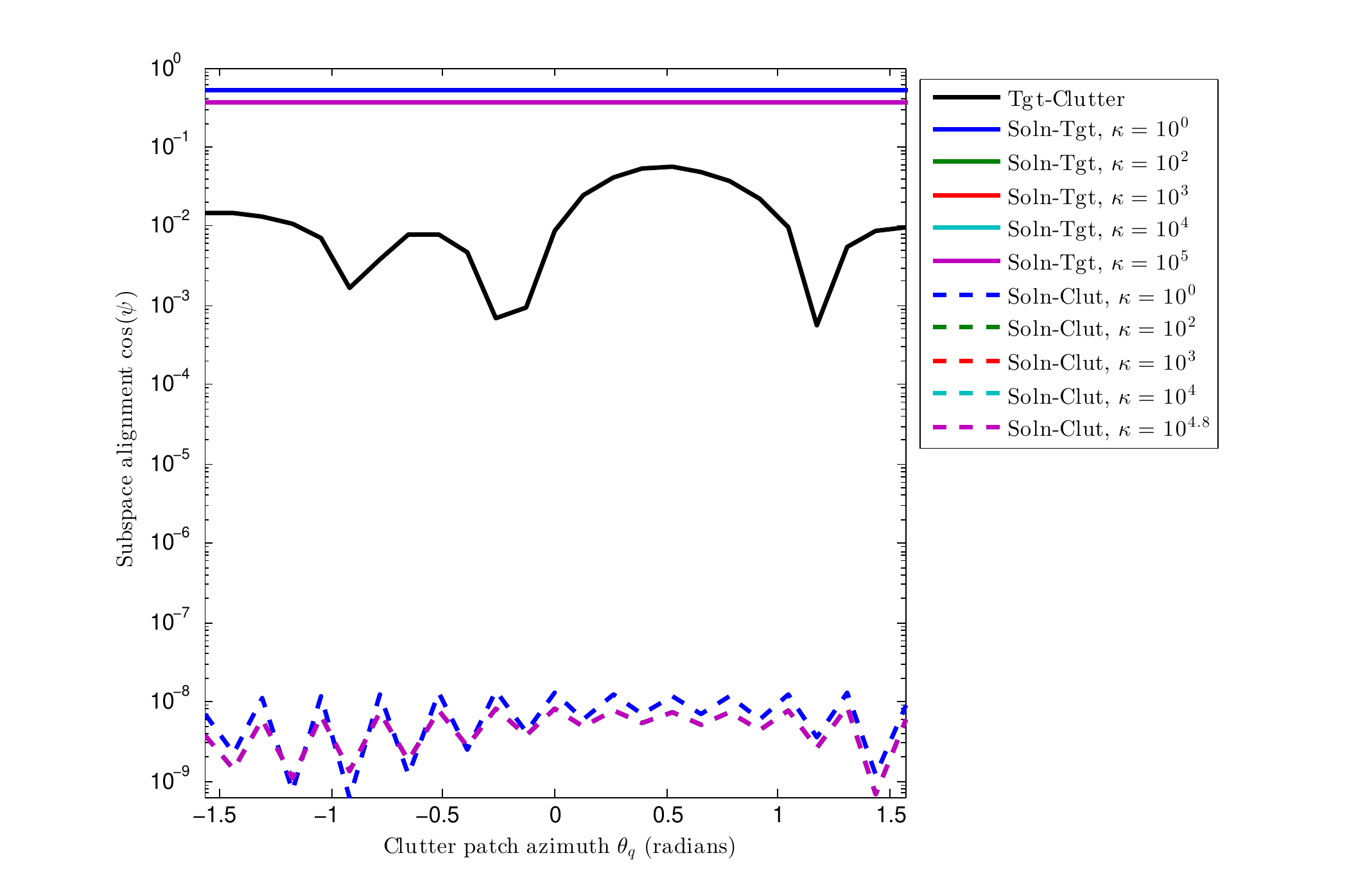}
	\caption{Subspace Cosines, QuadSolver}
	\label{fig:CCSQuadSubspaceC}
\end{figure}

\begin{figure}[!ht]
	\centering
	\includegraphics[width=\textwidth]{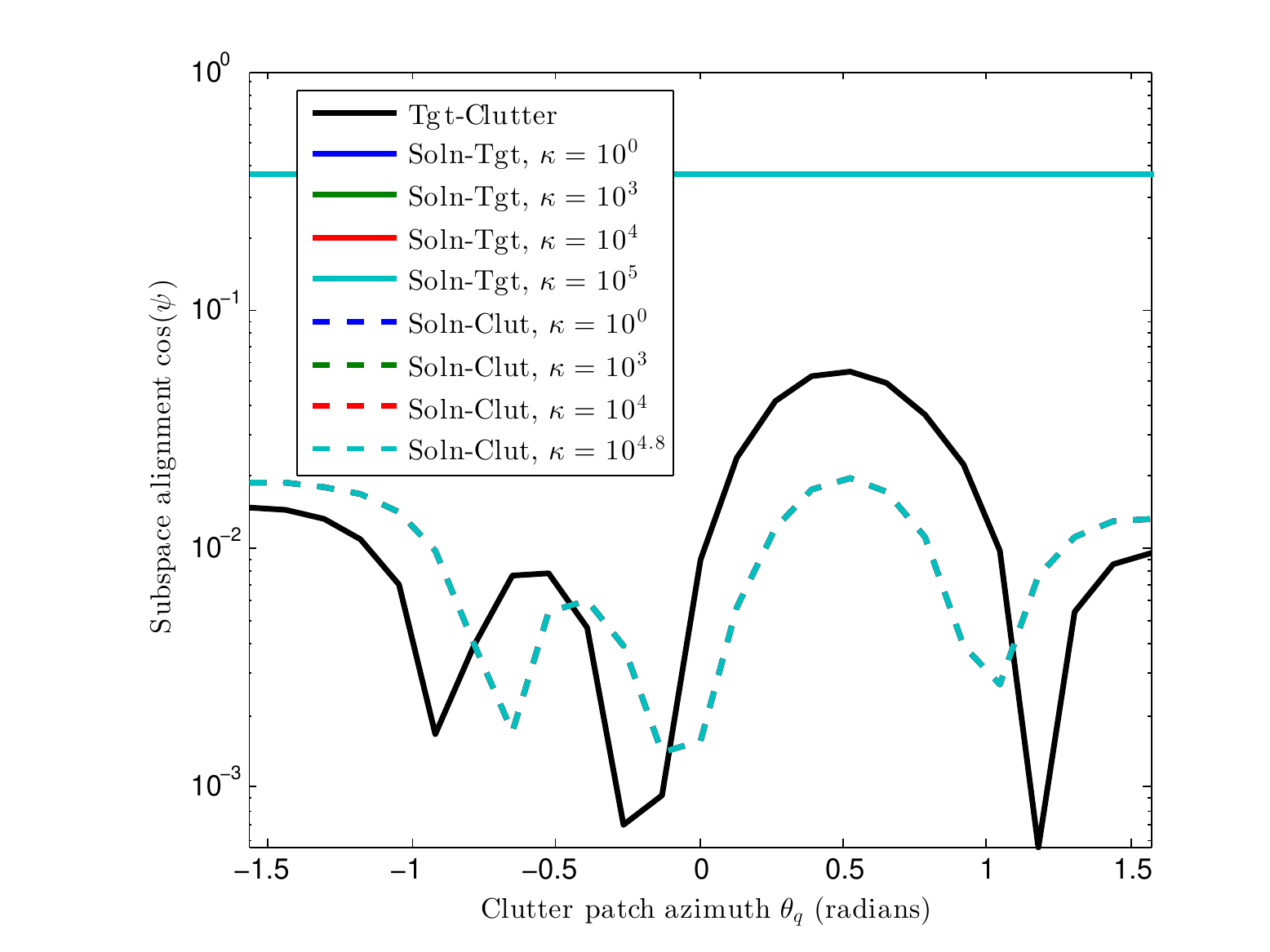}
	\caption{Subspace Cosines, NormSolver}
	\label{fig:CCSNormSubspaceC}
\end{figure}

Finally, our investigation of the effect of $\kappa$ concludes by examining the alignment between part of the relaxed solution and the clutter patches. Since we established in Proposition~\ref{prop:FullNILambdaPos} that the submatrix $\Bws$'s left eigenspace spans the columns of $\Bww$ and its right eigenspace spans those of $\Bss$, this effectively means that the subspace alignment between $\Bws$ and each clutter transfer matrix $\boldsymbol{\Gamma}_{q}$ is a proxy for the combined ability of the trans-recieve pair to suppress clutter. We can measure subspace alignment through the cosine of their \emph{principal angles} as follows. For given matrices $\Xbf, \Ybf \in \Complex^{p \times q}$, the cosine of the principal angle $\psi_{\Xbf,\Ybf}$ between the subspaces $\Span\{\Xbf\}$ and $\Span\{\Ybf\}$ is 
\begin{equation}
\cos(\psi_{\Xbf,\Ybf}) = \frac{|\trace(\Xbf^{H}\Ybf)|}{\frobnorm{\Xbf}{\phantom*}\frobnorm{\Ybf}{\phantom*}}.
\end{equation}
The closer $\cos(\psi_{\Xbf,\Ybf})$ is to 1, the more aligned the subspaces. In Figures~\ref{fig:CCSQuadSubspaceC} and \ref{fig:CCSNormSubspaceC}, the resulting subspace alignments for both target and clutter channel matrices are plotted for the solution from QuadSolver and NormSolver, respectively. For reference, we denote the existing alignment between the target channel matrix $T$ and each clutter channel matrix. For both solvers, variation in $\kappa$ seems to have little effect on the overall alignment. However, QuadSolver's solution for $\kappa = 1$, the solution's subspace is slightly better aligned with the target at the cost of marginally higher alignment to the clutter. In any case, we see a significant difference between the two solvers: QuadSolver attempts to avoid any alignment with the clutter, while NormSolver attempts to match the target's alignment spectrum (in shape, if not magnitude). This behavior in NormSolver appears to come at a cost to the overall solution-target alignment. 

\subsubsection{Variation in $P_o$}
Next, we examine the scenario opposite the above: varying the power constraint while keeping the Capon constraint fixed. In these scenarios, $P_o$ ranges from $10^{3}$ to $10^{9}$ while $\kappa = 100$. The Slater condition in all cases is easily satisfied; however, we do note that the solver failed to satisfy the KKTs for the final two values. This is generally attributable to scaling issues. For reasonable comparison, we assumed no interference in this scenario.

\begin{figure}[!ht]
	\centering
	\includegraphics[width=\textwidth]{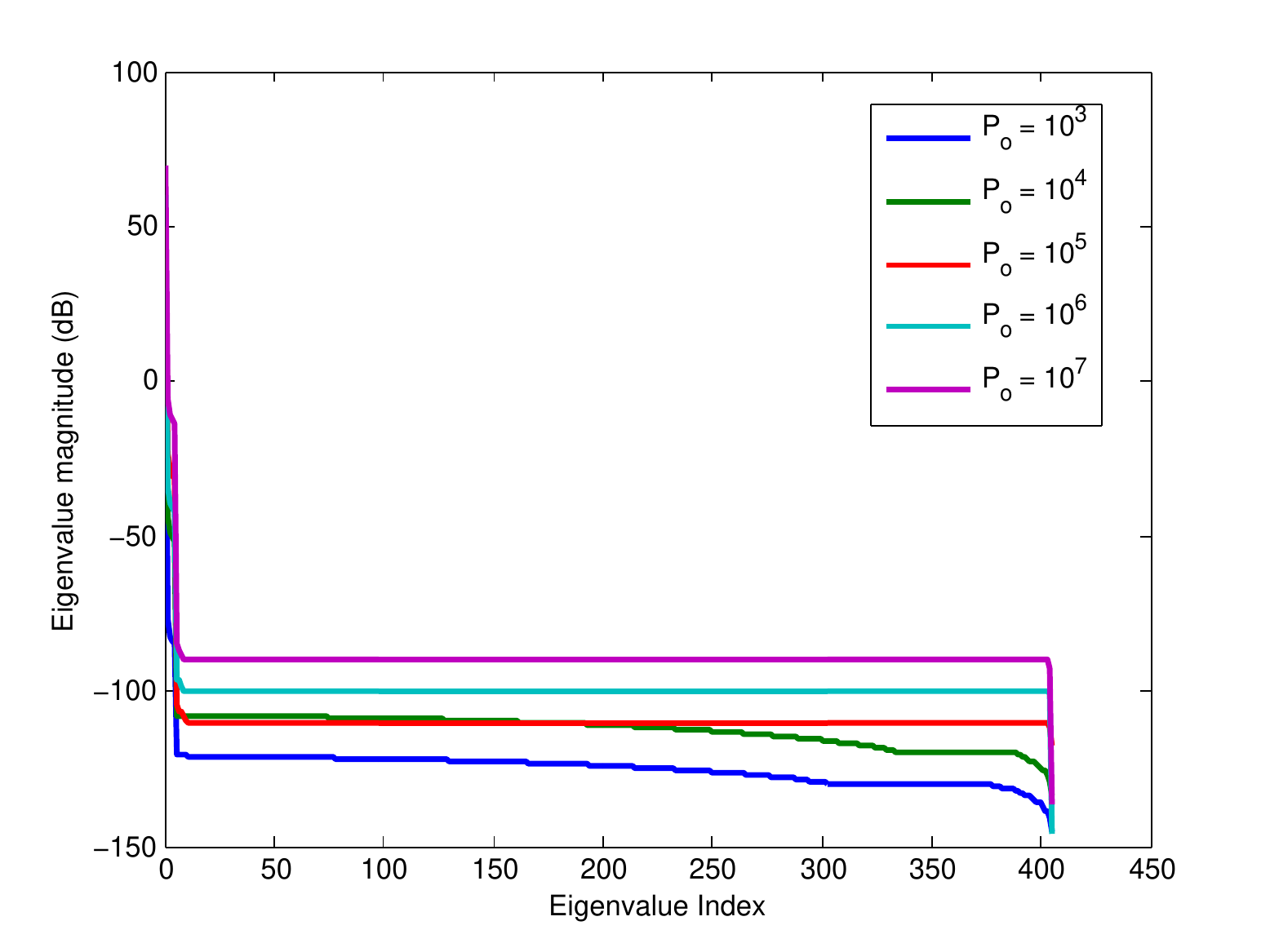}
	\caption{Eigenvalue spread of $\Bbf$ solution matrix using QuadSolver}
	\label{fig:PCS_Quad_ESpread}
\end{figure}

\begin{figure}[!ht]
	\centering
	\includegraphics[width=\textwidth]{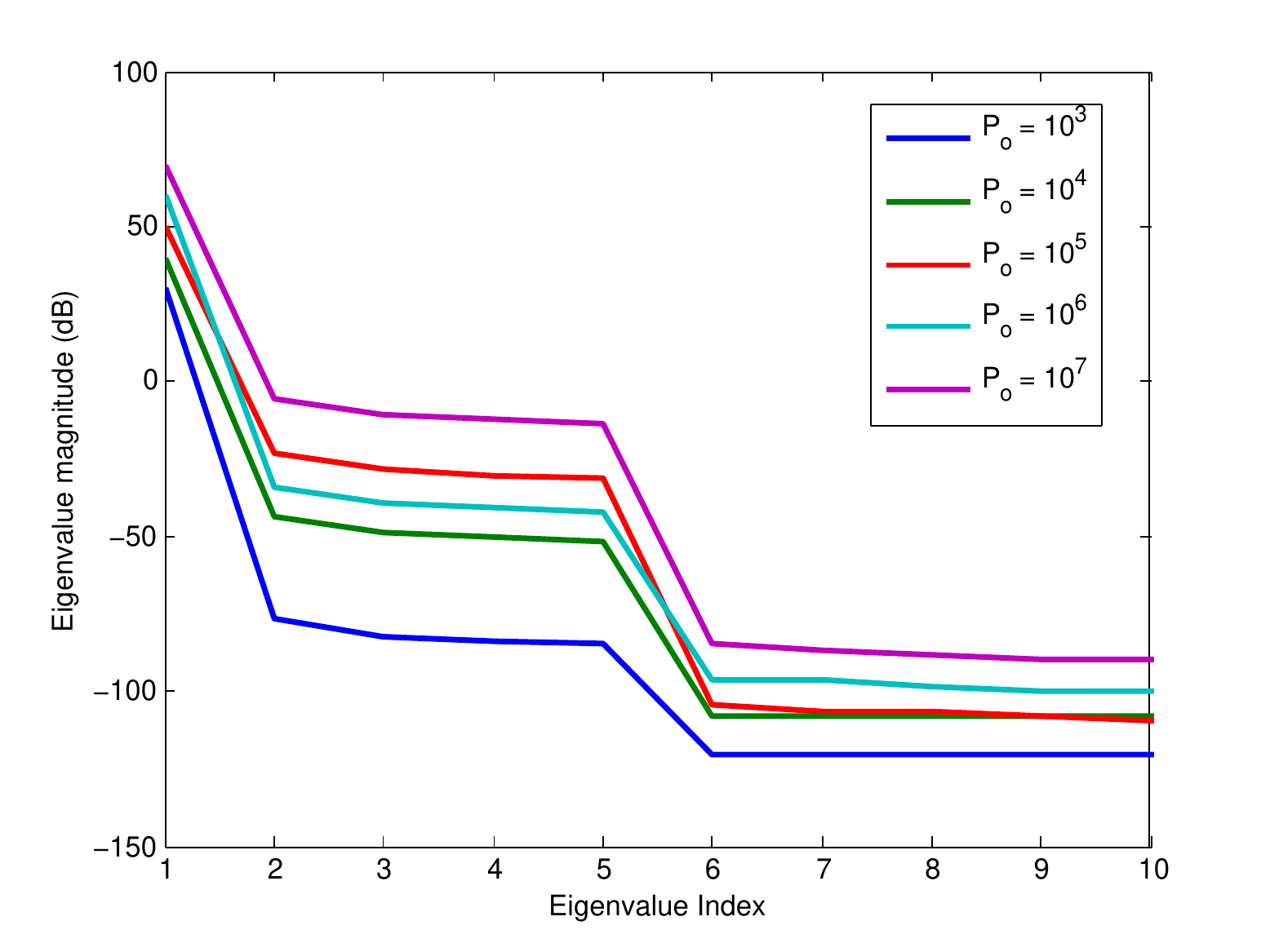}
	\caption{Closeup of the first $2N$ eigenvals. of $\Bbf$ using QuadSolver}
	\label{fig:PCS_Quad_ESpread_Zoom}
\end{figure}

First, we return to the eigenspread of the solution matrix $\Bbf$. Figures~\ref{fig:PCS_Quad_ESpread} \& \ref{fig:PCS_Quad_ESpread_Zoom} show the total eigenspread and its first $2N$ eigenvalues for QuadSolver. As $P_o$ increases, we see that the peak eigenvalue increases to maintain equality at the power bound; this also corresponds with an overall "lifting" of the eigenspectrum, which is most likely due to numerical precision limitations. In any case, the effective numerical rank of the solution is still one, while the overall effective rank of the solution is, at most, $N$. 

\begin{figure}[!ht]
	\centering
	\includegraphics[width=\textwidth]{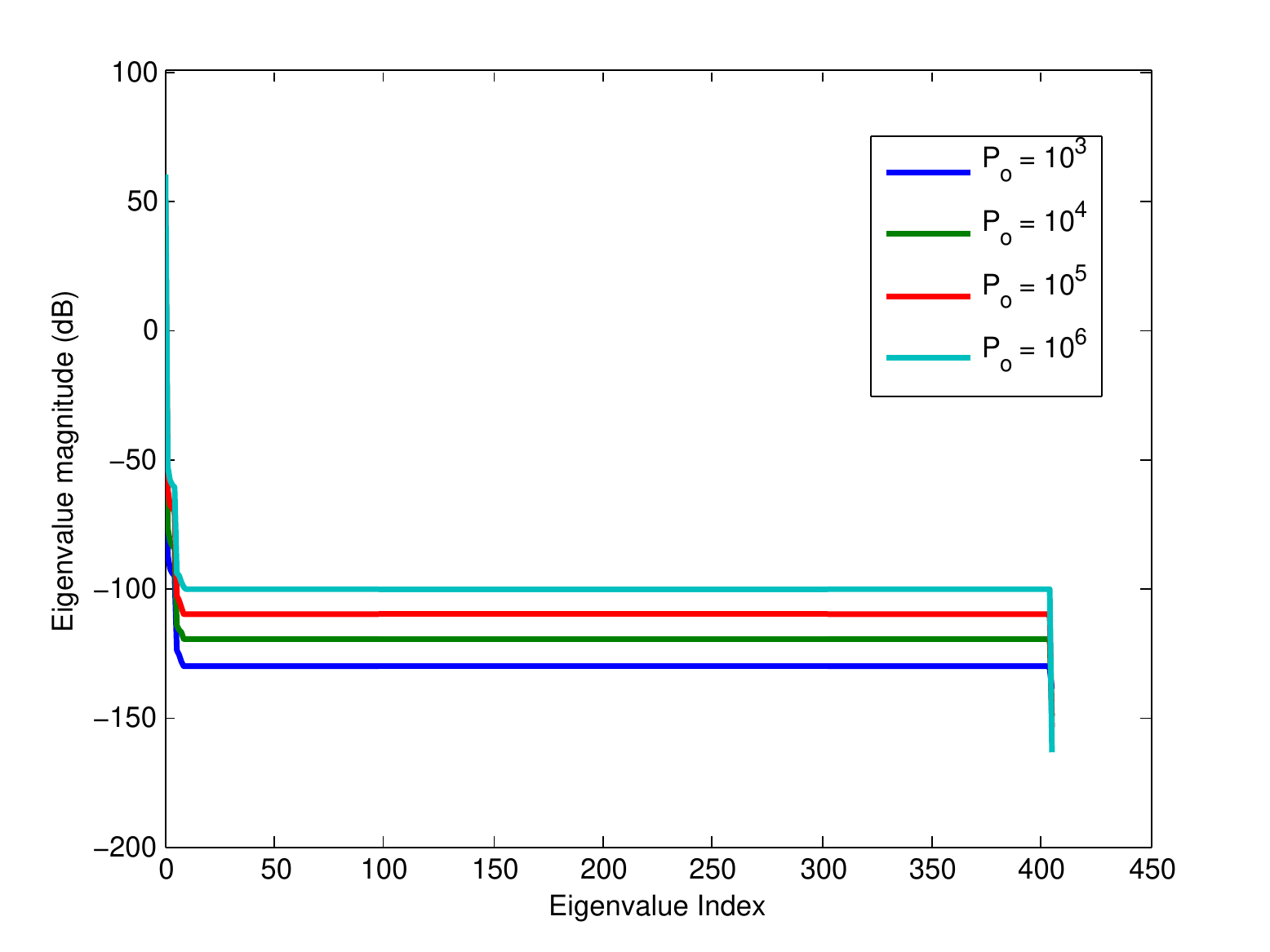}
	\caption{Eigenvalue spread of $\Bbf$ solution matrix using NormSolver}
	\label{fig:PCS_Norm_ESpread}
\end{figure}

\begin{figure}[!ht]
	\centering
	\includegraphics[width=\textwidth]{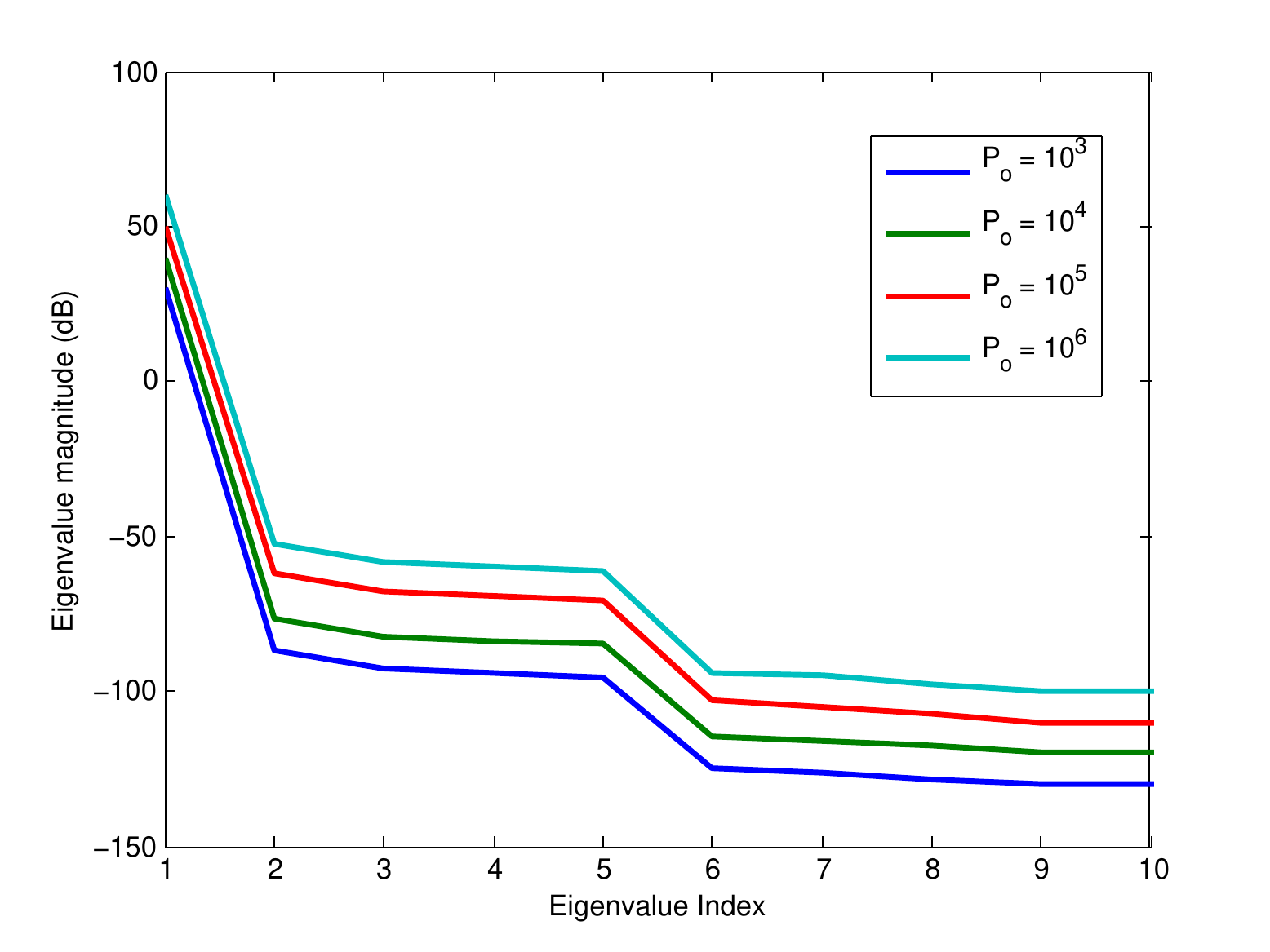}
	\caption{Closeup of the first $2N$ eigenvals. of $\Bbf$ using NormSolver}
	\label{fig:PCS_Norm_ESpread_Zoom}
\end{figure}

Similarly, the total eigenspread and first $2N$ eigenvalues for the NormSolver solution are provided in Figures~\ref{fig:PCS_Norm_ESpread} and \ref{fig:PCS_Norm_ESpread_Zoom}, respectively. We see similar behavior to the QuadSolver, but note the tighter spread in effectively non-zero eigenvalues with increasing $P_o$. This indicates that NormSolver is converging to the same solution at each $P_o$, scaled to match the power constraint. 

\begin{figure}[!ht]
	\centering
	\includegraphics[width=\textwidth]{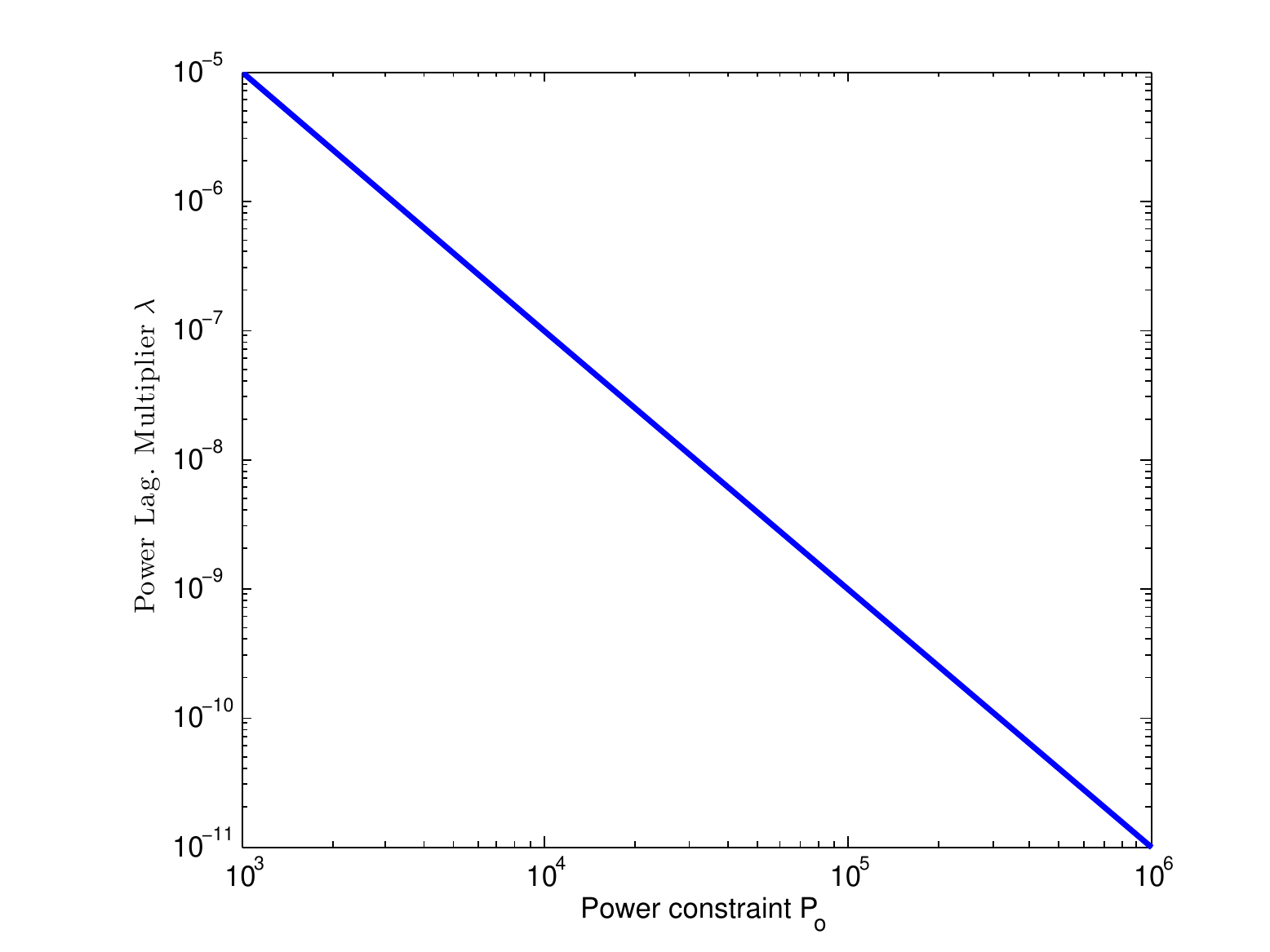}
	\caption{Optimal $\lambda$ as a function of the Capon constraint, NormSolver}
	\label{fig:PCS_Norm_Lambda}
\end{figure}

As above, we also explore the effect of the constraint on the Lagrange multiplier $\lambda$, shown in Figures~\ref{fig:PCS_Quad_Lamba} (for QuadSolver) and \ref{fig:PCS_Norm_Lambda} (for NormSolver). In both cases, the multiplier sharply declines as the constraint becomes looser, indicating that the overall filtered noise power remains nearly the same and is mostly a structural issue. 
\begin{figure}[!ht]
	\centering
	\includegraphics[width=\textwidth]{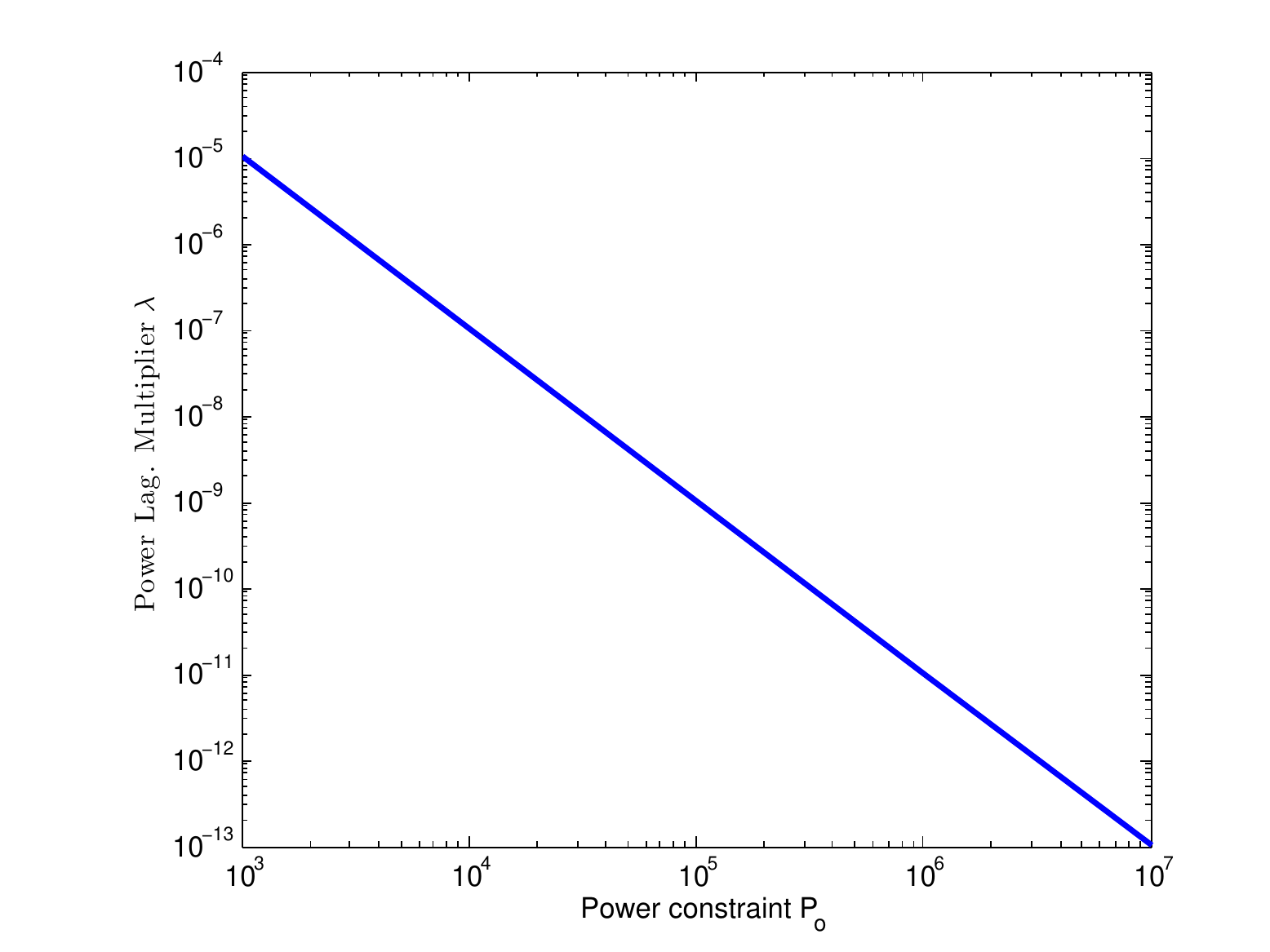}
	\caption{Optimal $\lambda$ as a function of the Capon constraint, QuadSolver}
	\label{fig:PCS_Quad_Lambda}
\end{figure}

\begin{figure}[!ht]
	\centering
	\includegraphics[width=\textwidth]{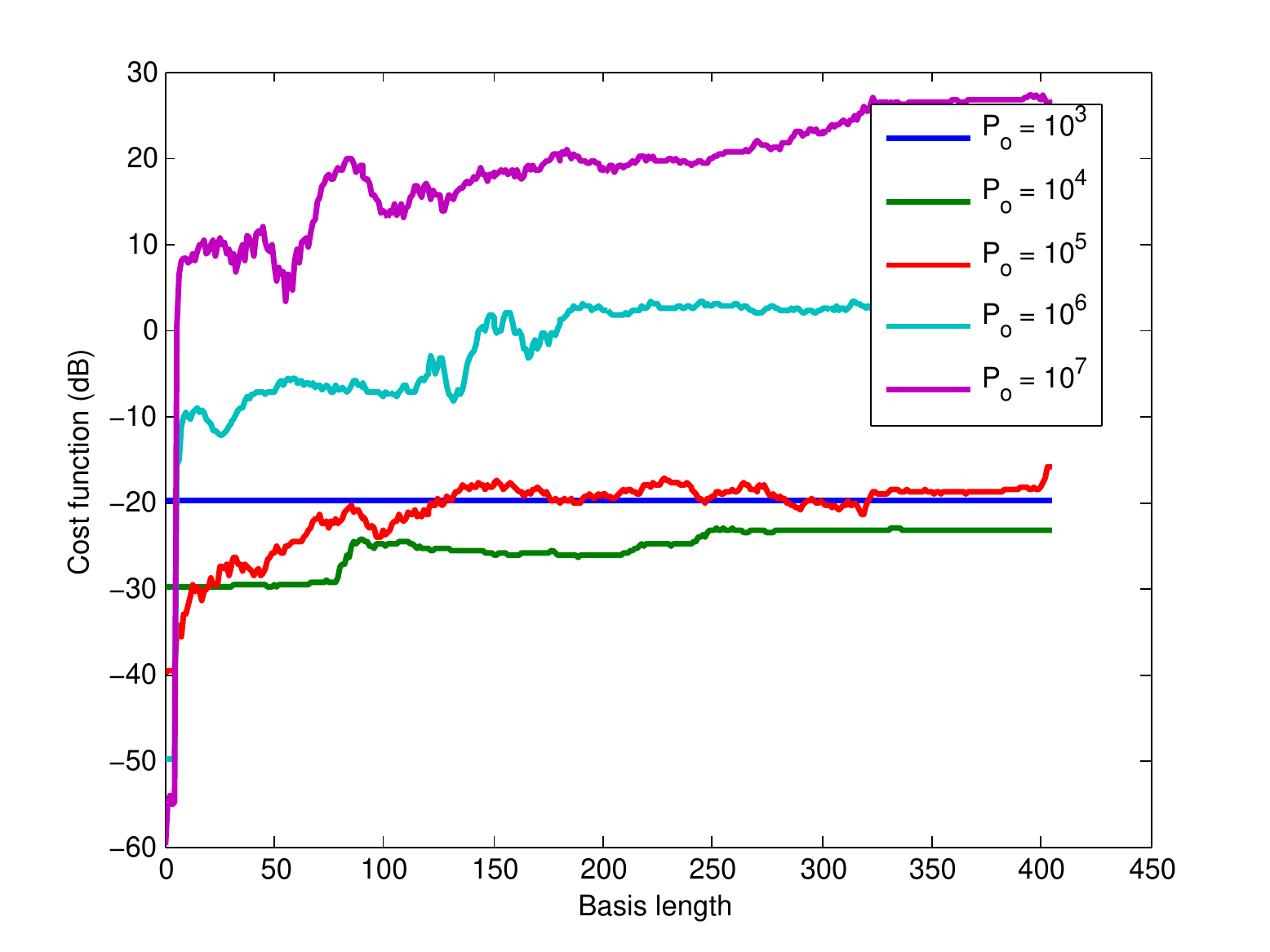}
	\caption{Original cost as function of basis length, QuadSolver}
	\label{fig:PCS_Quad_Costfunc}
\end{figure}

\begin{figure}[!ht]
	\centering
	\includegraphics[width=\textwidth]{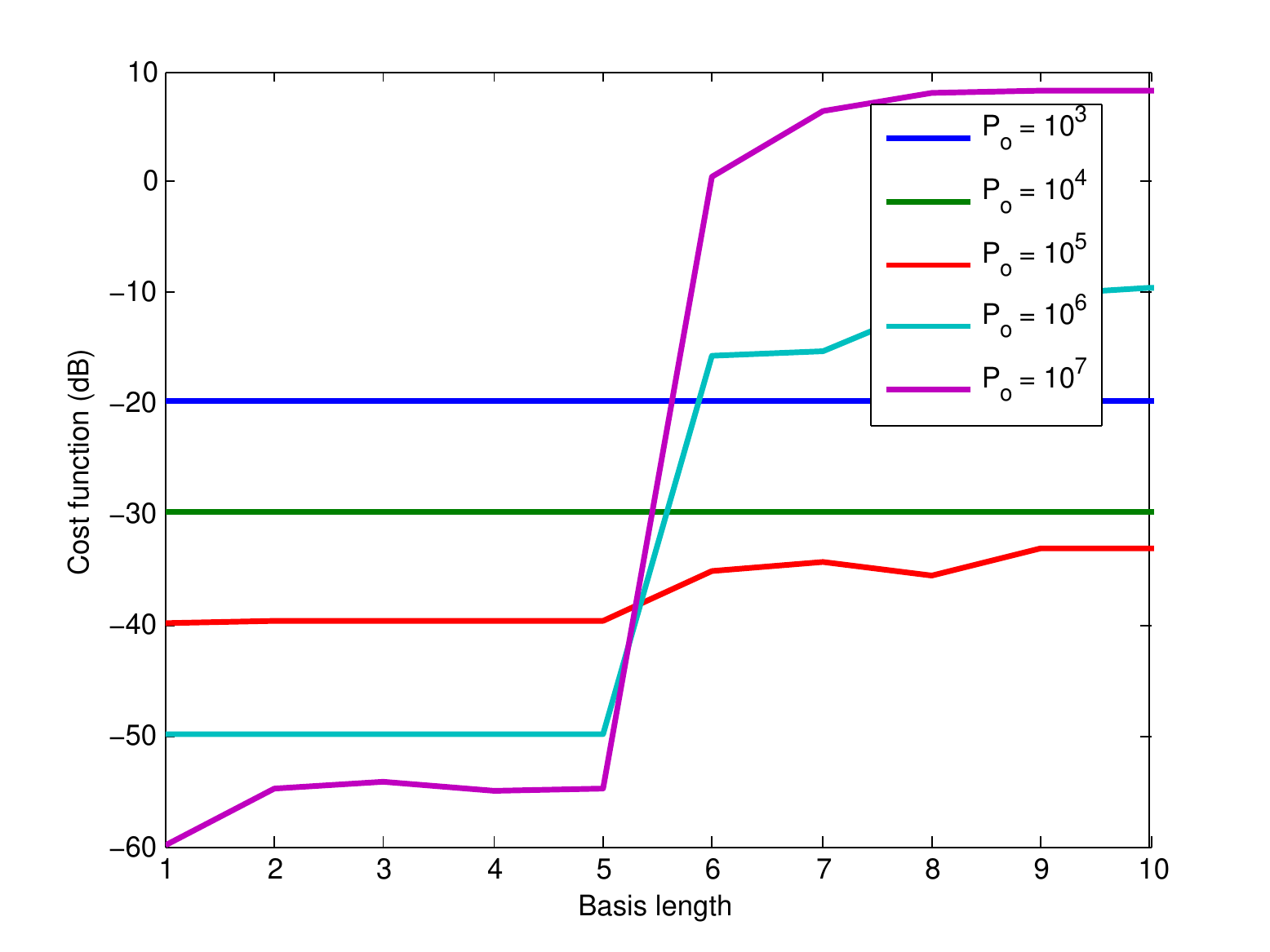}
	\caption{Original cost as function of basis length, first $2N$, QuadSolver}
	\label{fig:PCS_Quad_Costfunc_Zoom}
\end{figure}

We also reexamine the approximate rank-1 cost function described in the above section. For QuadSolver, we can see from Figures~\ref{fig:PCS_Quad_Costfunc} (the entire sweep of $J$ basis vectors) and \ref{fig:PCS_Quad_Costfunc_Zoom} (the first $2N$ basis vectors) that increasing the power lowers the overall cost when approximating using the first $N$ basis vectors. Outside these vectors, the cost dramatically increases for higher values of $P_o$. This indicates that low-rank approximation under weak power constraints is effectively unnecessary, while basis selection must be more careful in high-power scenarios. 

\begin{figure}[!ht]
	\centering
	\includegraphics[width=\textwidth]{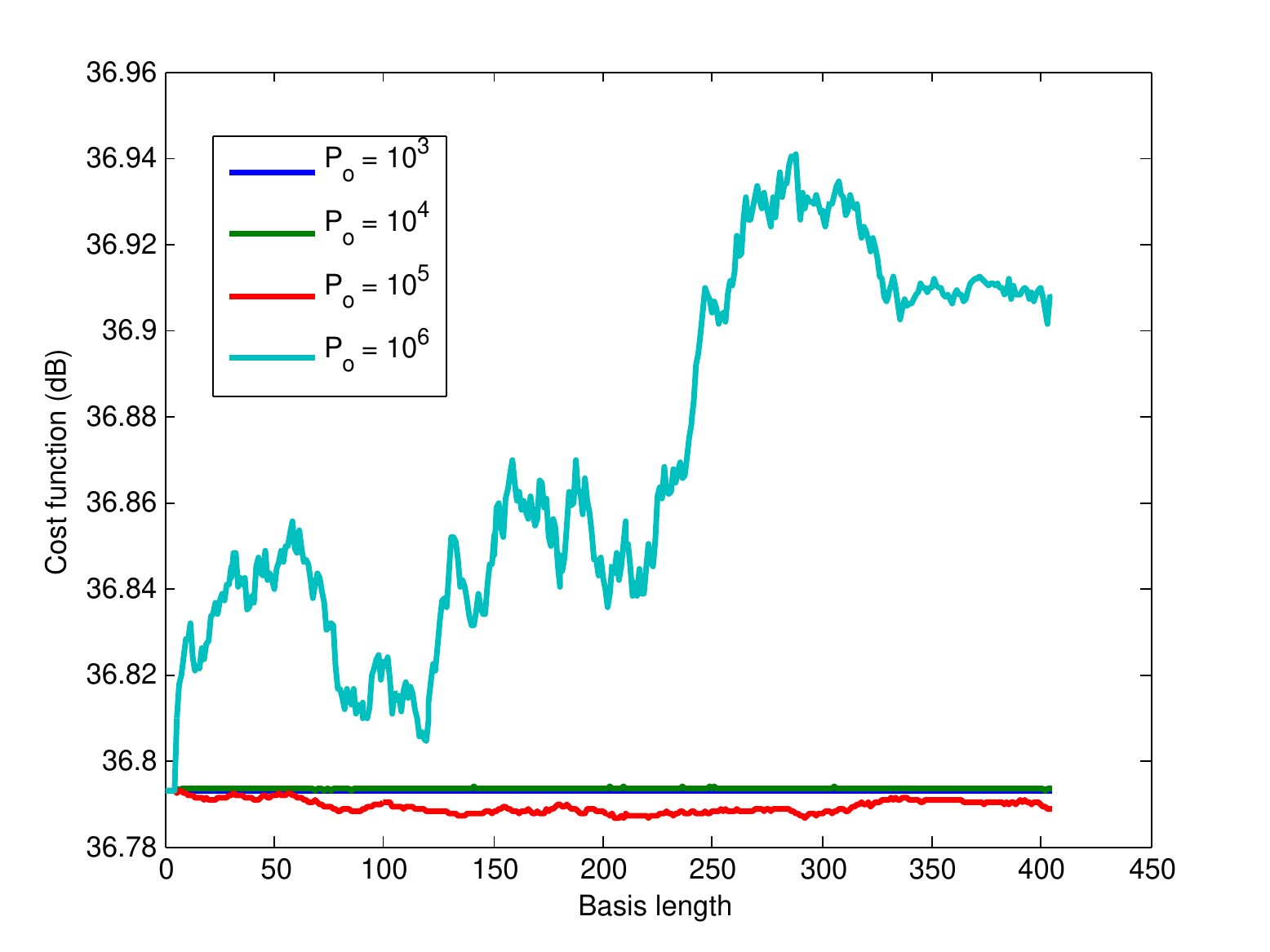}
	\caption{Original cost as function of basis length, NormSolver}
	\label{fig:PCS_Norm_Costfunc}
\end{figure}

\begin{figure}[!ht]
	\centering
	\includegraphics[width=\textwidth]{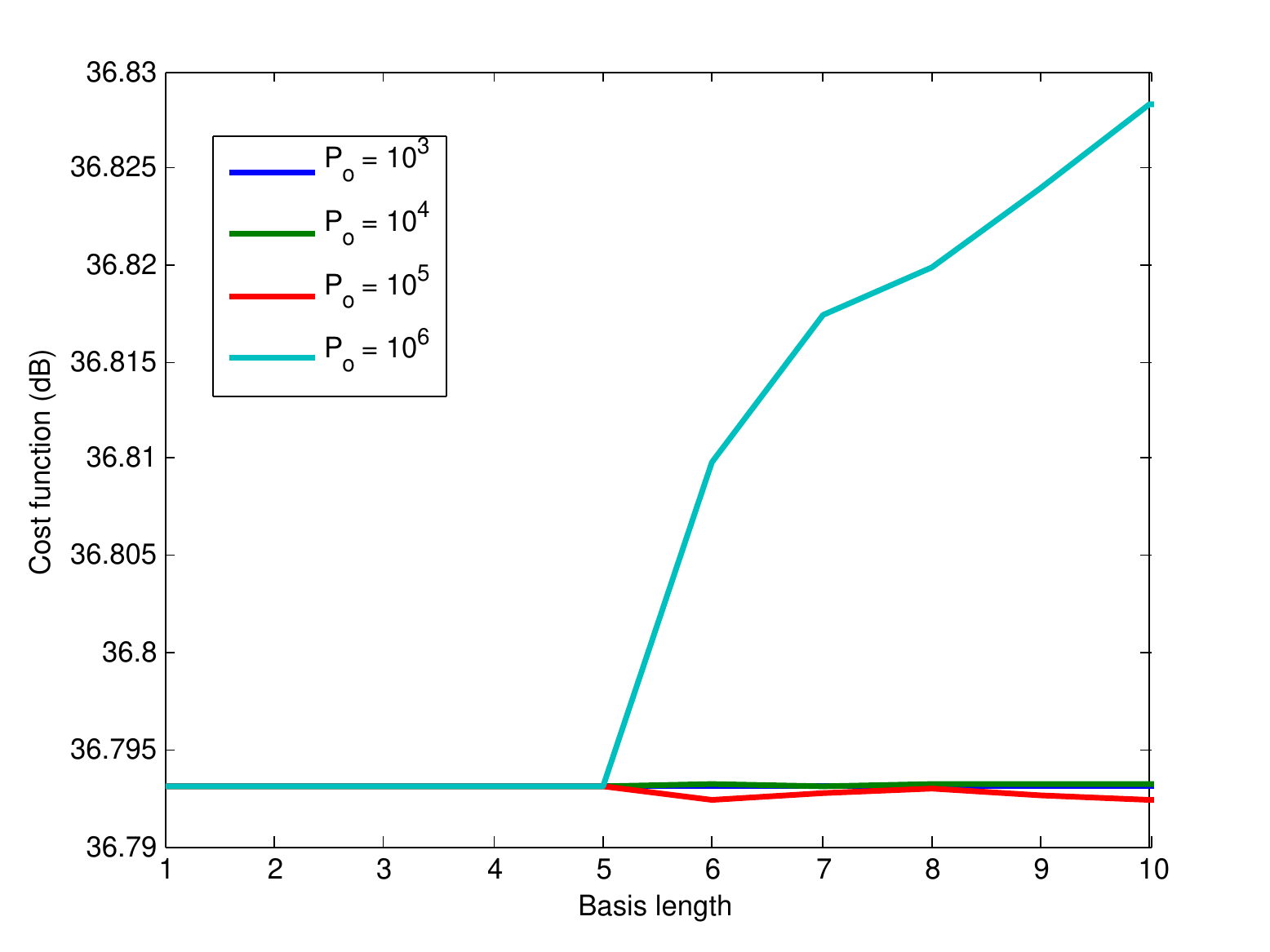}
	\caption{Original cost as function of basis length, first $2N$, NormSolver}
	\label{fig:PCS_Norm_Costfunc_Zoom}
\end{figure}

In contrast, NormSolver's solution remains functionally fixed in cost over the sweep (as seen in Figures~\ref{fig:PCS_Norm_Costfunc} and \ref{fig:PCS_Norm_Costfunc_Zoom}), which agrees with the behavior seen in \ref{fig:PCS_Norm_ESpread_Zoom} -- namely, the overall solution is minimally sensitive to the power constraint. This is most obvious when  $P_o = {10}^6$ , as even its ``dramatic'' increase is within a tenth of a decibel of its minimum.

\begin{figure}[!ht]
	\centering
	\includegraphics[width=\textwidth]{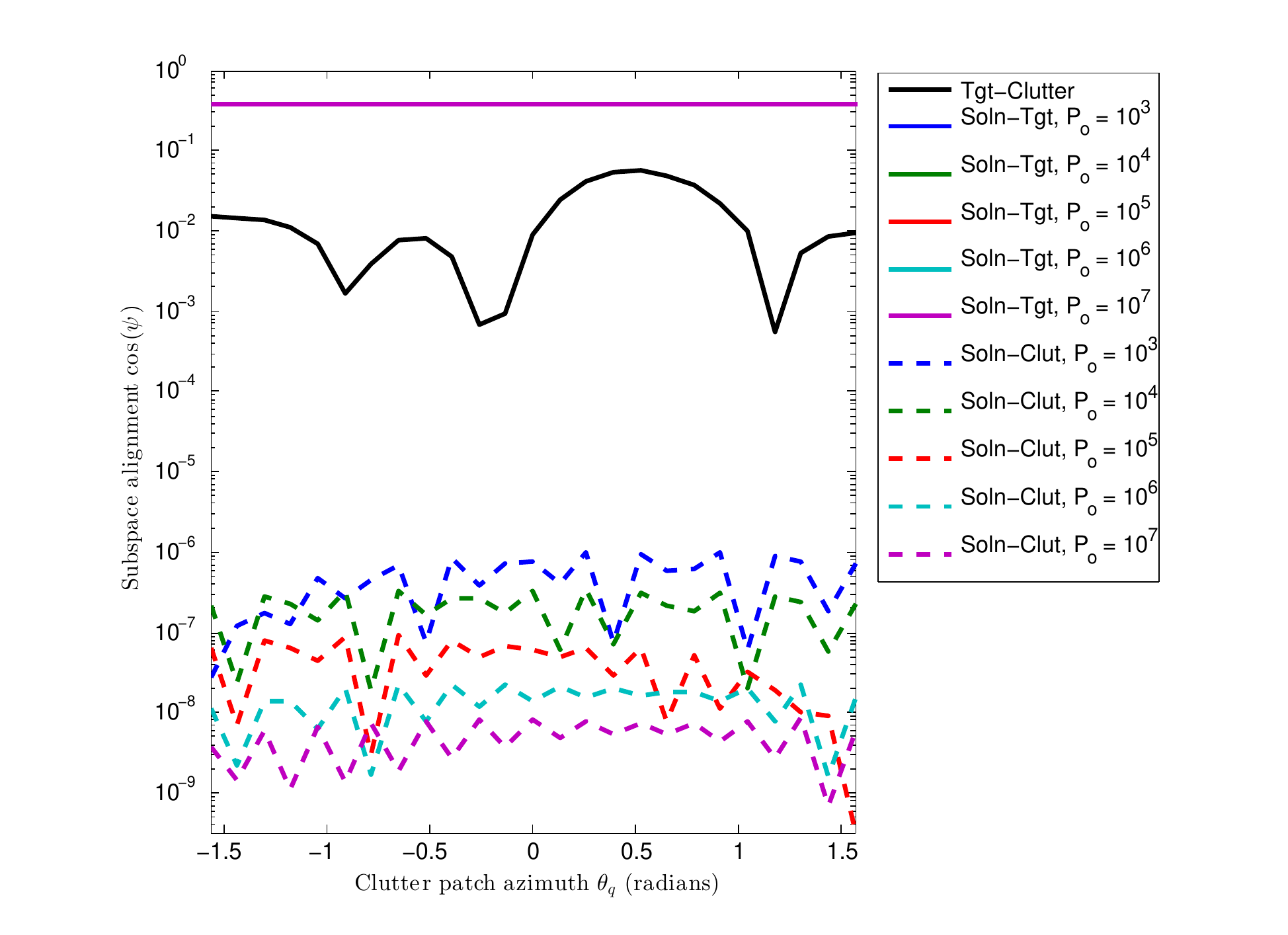}
	\caption{Subspace Cosines, QuadSolver}
	\label{fig:PCS_Quad_Subspace_Cosine}
\end{figure}

Finally, we investigate the subspace alignment as a function of the power constraint. As shown in Figure~\ref{fig:PCS_Quad_Subspace_Cosine}, QuadSolver's solution attempts to null the clutter spectrum entirely, with higher power corresponding to better nulling. That said, all solutions provide alignment figures less than ${10}^{-6}$, which is well below the target's alignment and any reasonable interpretation of ``aligned''.  Power does not seem to affect the solution-target alignment to a significant degree.

\begin{figure}[!ht]
	\centering
	\includegraphics[width=\textwidth]{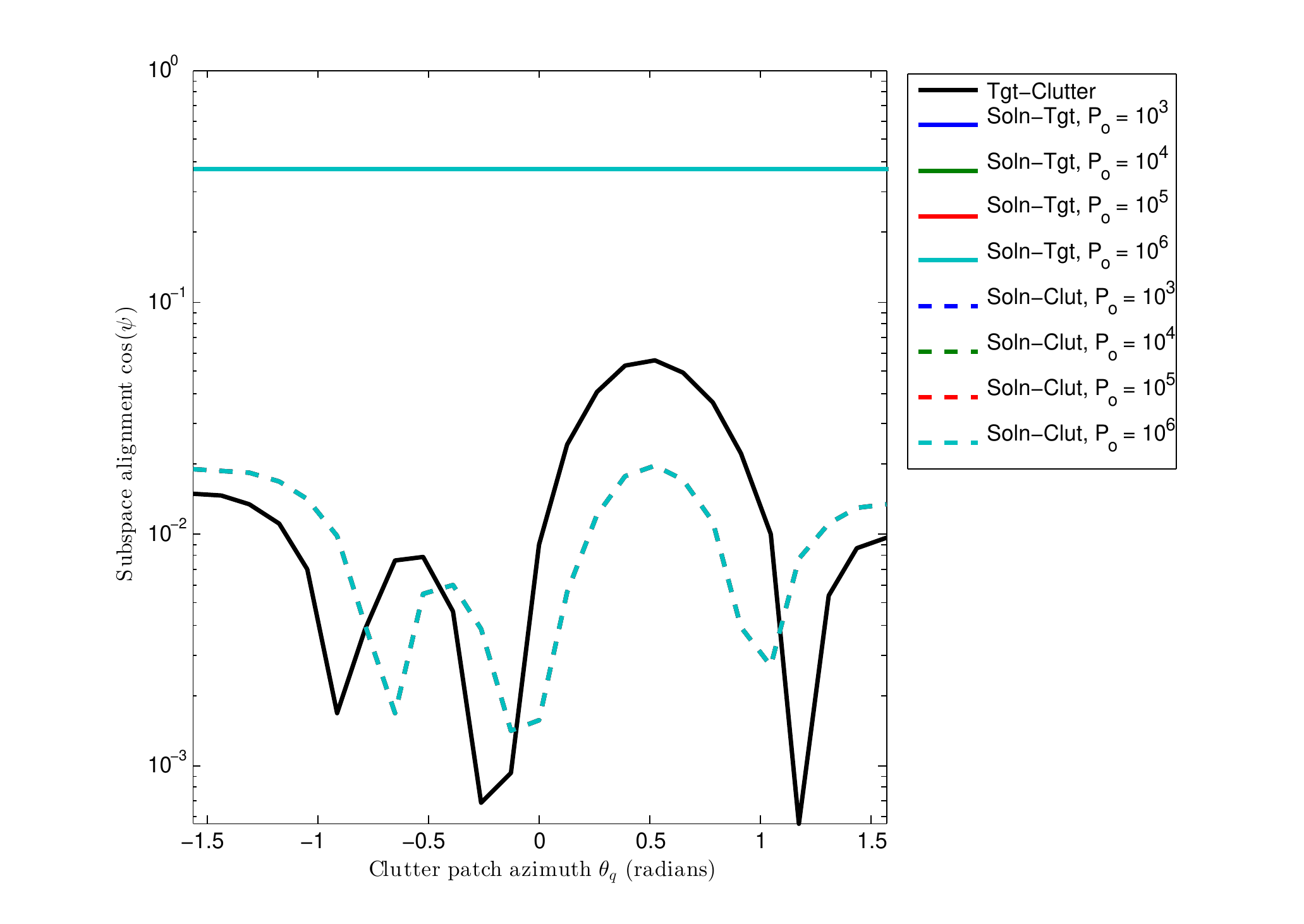}
	\caption{Subspace Cosines, NormSolver}
	\label{fig:PCS_Norm_Subspace_Cosine}
\end{figure}

Again, in contrast, NormSolver's solution, seen in Figure~\ref{fig:PCS_Norm_Subspace_Cosine}, attempts to align with the target's representation in the clutter and remains insensitive to the power constraint. This provides a final confirmation of the waterfilling interpretation above. 


\subsection{Interference Effects}
In this scenario, we consider the impact of interference on the relaxed solution and the effect each solver type has on this impact. Using the general scenario from above with both no interference and an interferer located at $(\theta_I, \phi_I) = (0.3941, 0.3)$ radians, we examine the impact on both subspace alignment and the overall adapted pattern. 

\begin{figure}[!ht]
	\centering
	\includegraphics[width=\textwidth]{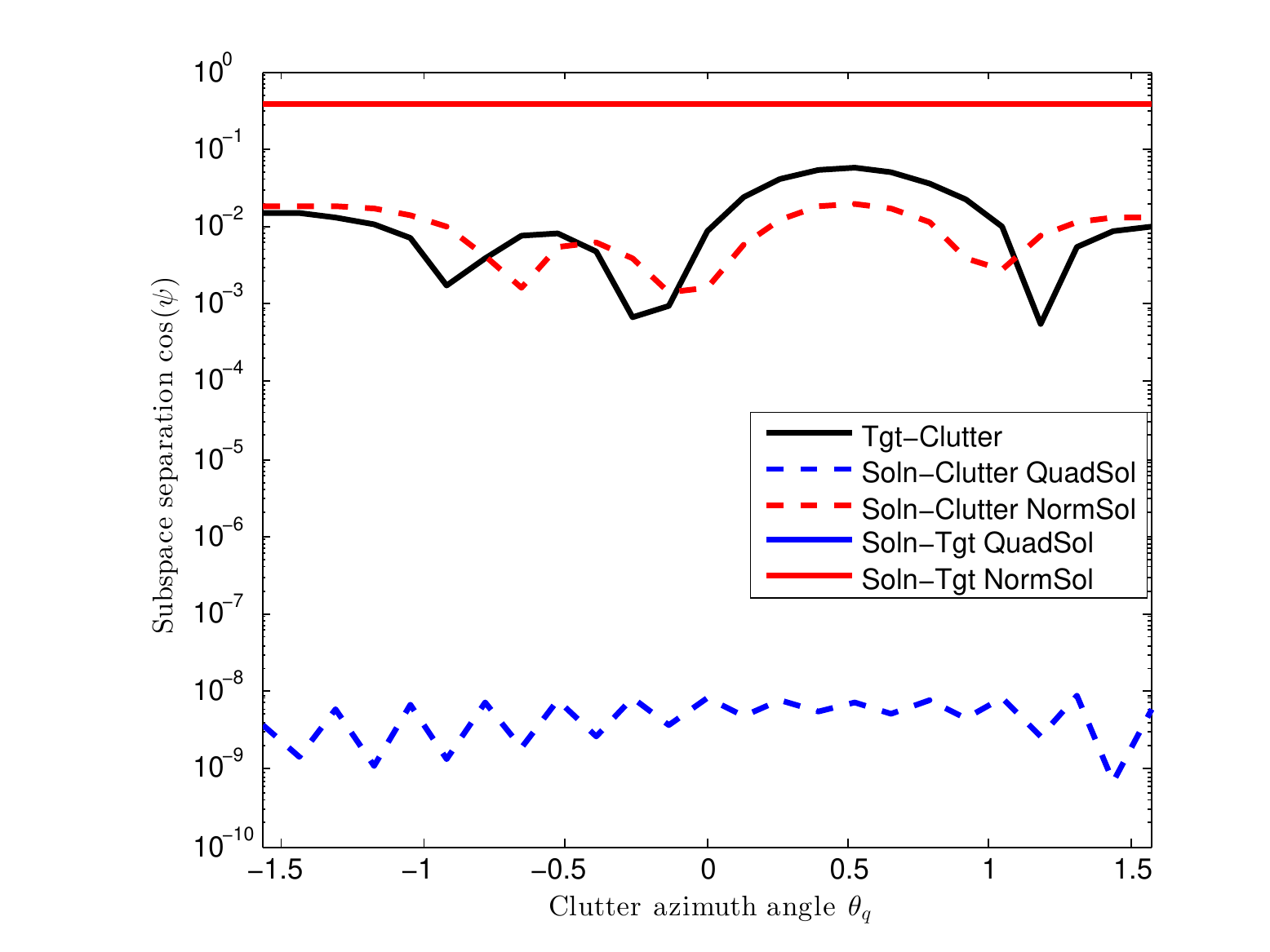}
	\caption{Subspace Alignment, No Interference}
	\label{fig:SubspaceCosineQNNoIntf}
\end{figure}

\begin{figure}[!ht]
	\centering
	\includegraphics[width=\textwidth]{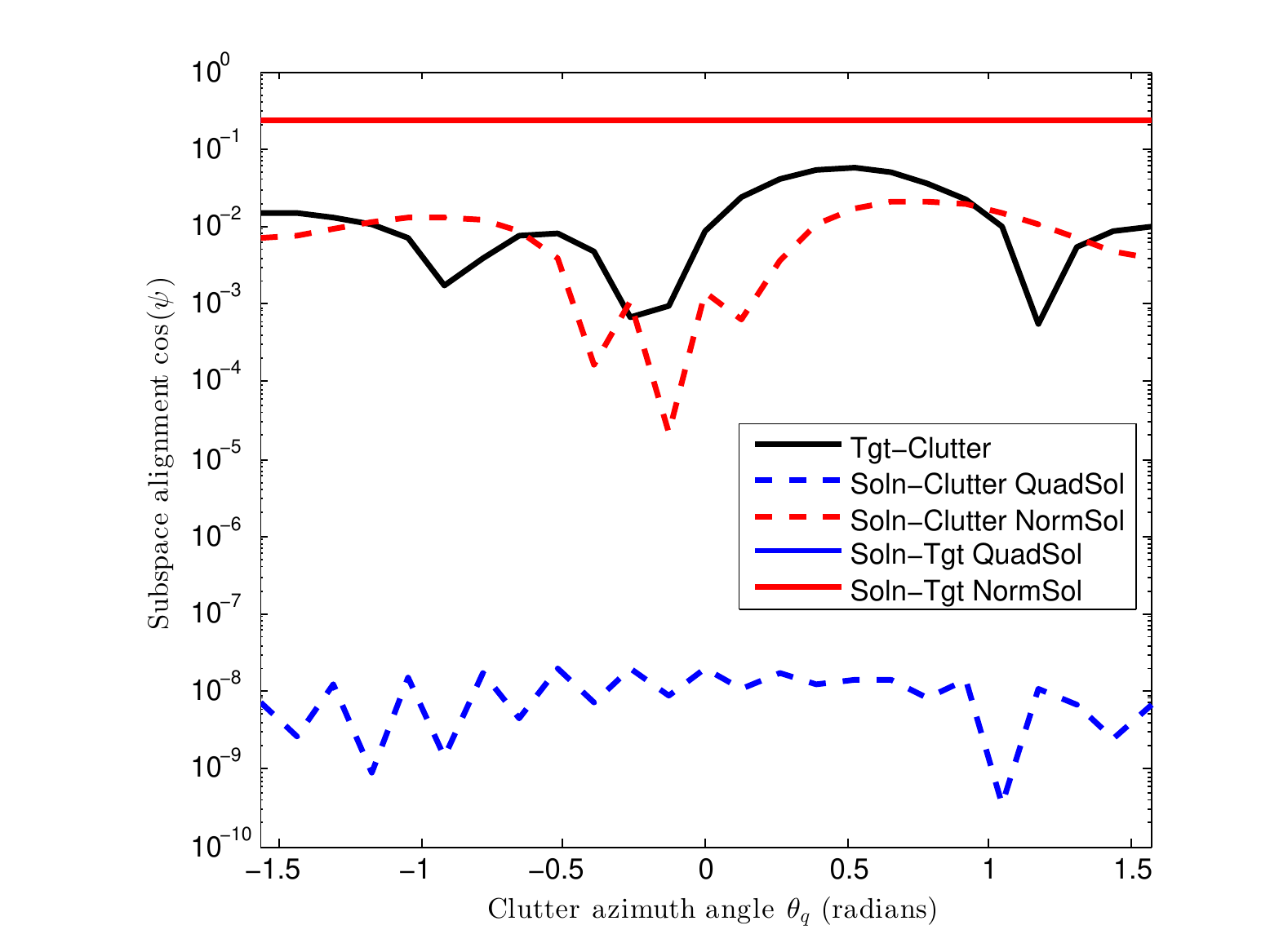}
	\caption{Subspace Alignment, Interferer at $(\theta, \phi) = (0.3941, 0.3)$ }
	\label{fig:SubspaceCosineQNIntf}
\end{figure}

First, we consider the subspace alignment as in Section~\ref{sec:KKTConsequences}. As above, the upper right corner of the solution matrix, $\Bws$, is compared against each clutter patch and the target. The subspace alignment cosine, $\cos(\psi)$, ranges from 0 to 1, with higher values indicating that the spans of the two matrices are more aligned. Figure~\ref{fig:SubspaceCosineQNNoIntf} shows the alignment with no interference, while Figure~\ref{fig:SubspaceCosineQNIntf} shows the alignment under the interference. In both cases, as in the previous sections, QuadSolver produces a solution that is effectively unaligned with the clutter, whereas NormSolver's solution attempts to follow the preexisting target-clutter alignment. The major difference under this metric is a clear broadening and shifting of the "mainlobe" of NormSolver's solution-clutter alignment spectrum near the direction of the interferer and beyond. QuadSolver's solution-clutter alignment appears generally unaffected, but we note that this does not mean the interferer has no impact, as we will soon see. 


Next, we consider the traditional adapted pattern for STAP, which plots the following function
\begin{equation}
\mathcal{P}(f_d,\theta) = |\wbf_o^{H}(\vbf(f_d)\kronecker\sbf_o\abf(\theta,\phi)|^{2}.
\end{equation}
That is, the adapted pattern for a given beamformer-signal pair $\wbf_o,\sbf_o$ is a function of the Doppler frequency $f_d$ and the azimuth $\theta$ at a given elevation $\phi$. We consider the same scenarios as above, with and without interference, and provide the overall adapted pattern as well as Doppler cuts at the target azimuth and azimuth cuts at the target Doppler frequency. 

\begin{figure}[!ht]
	\centering
	\includegraphics[width=\textwidth]{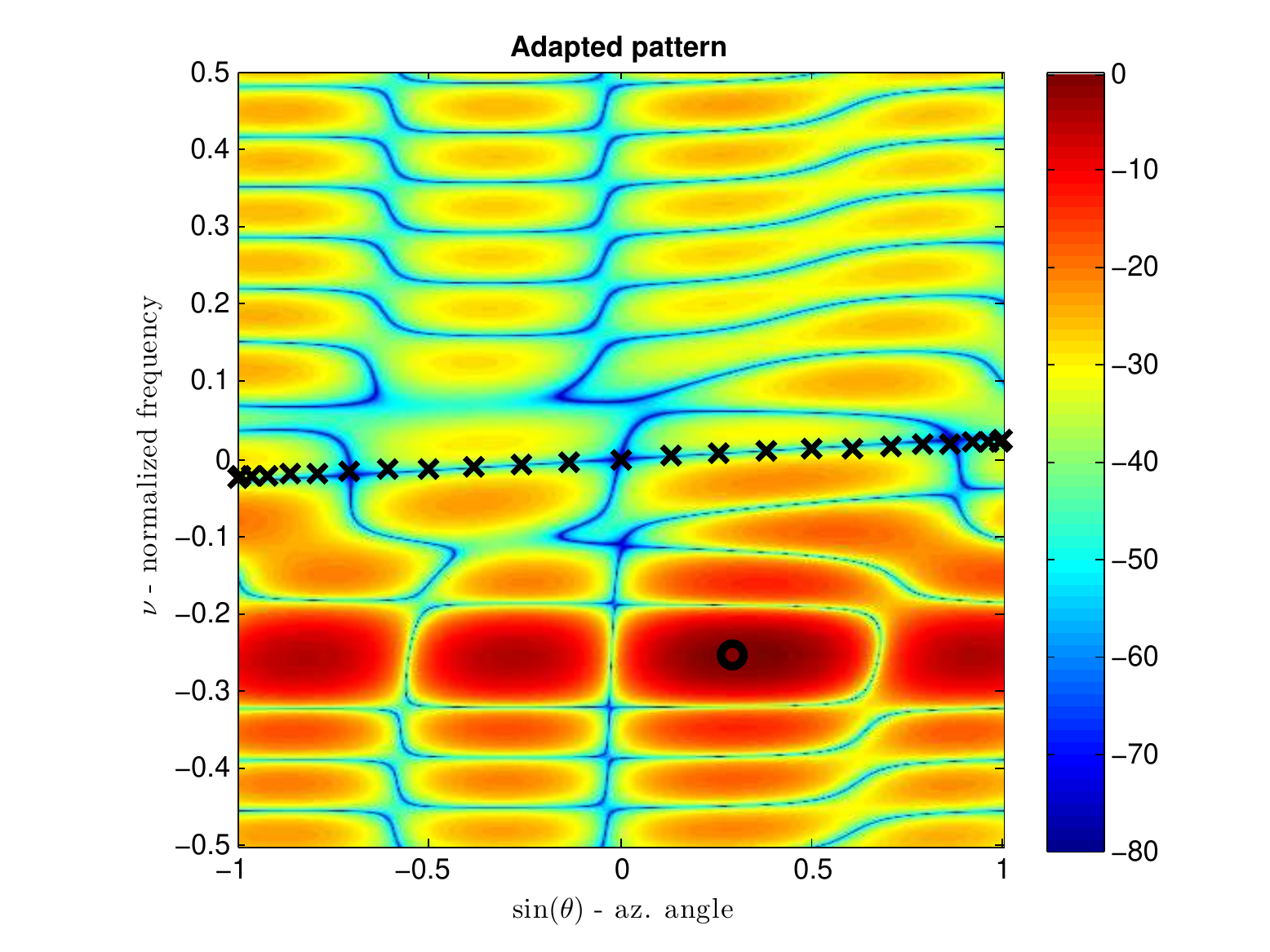}
	\caption{Adapted Pattern (dB scale), QuadSolver, No Interference. Target at $\bigcirc$, Clutter phase centers at $\times$. No Interference.}
	\label{fig:AdaptPatternQuadNoIntf}
\end{figure}

\begin{figure}[!ht]
	\centering
	\includegraphics[width=\textwidth]{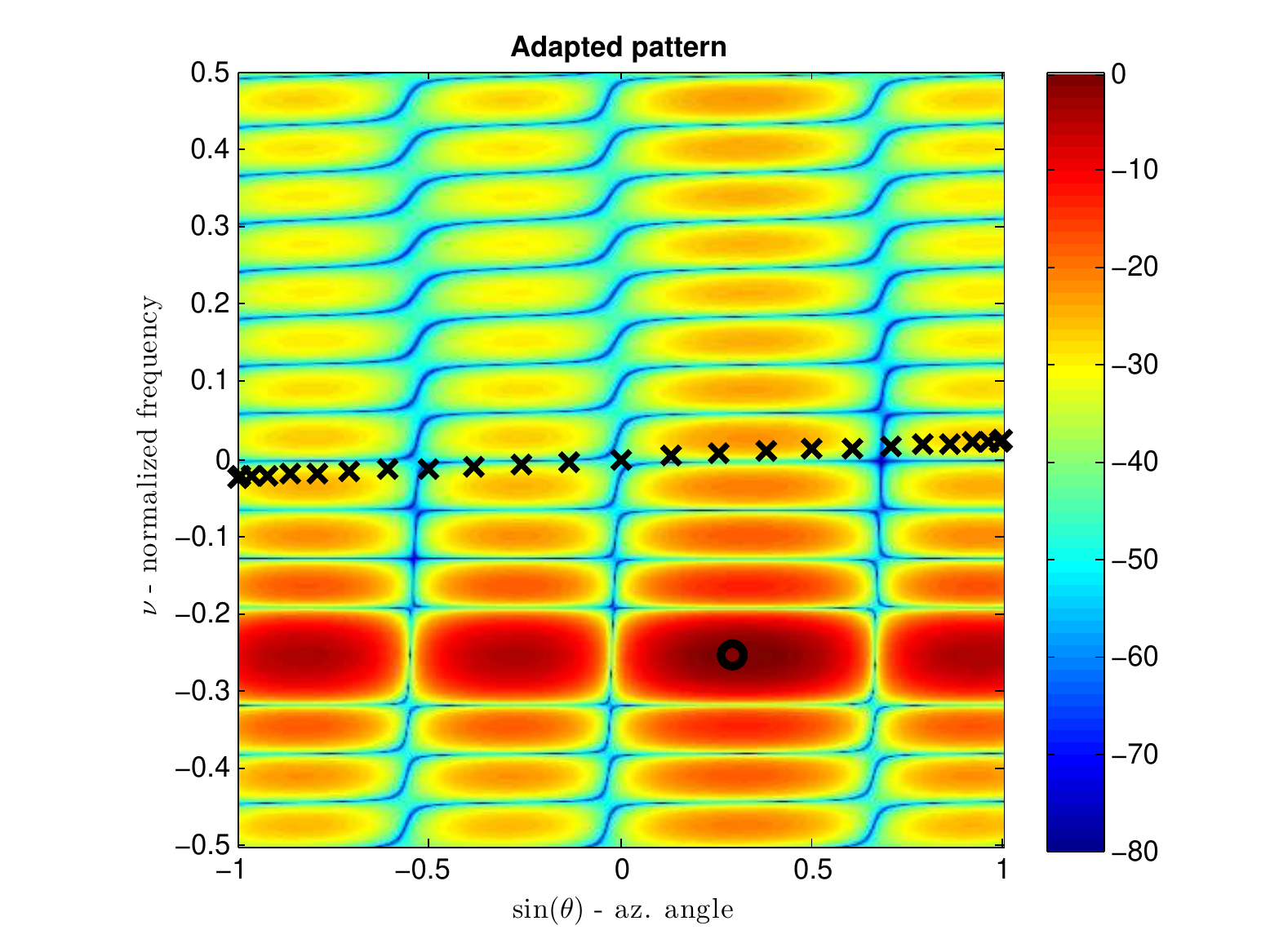}
	\caption{Adapted Pattern (dB scale), NormSolver, No Interference. Target at $\bigcirc$, Clutter phase centers at $\times$.}
	\label{fig:AdaptPatternNormNoIntf}
\end{figure}

\begin{figure}[!ht]
	\centering
	\includegraphics[width=\textwidth]{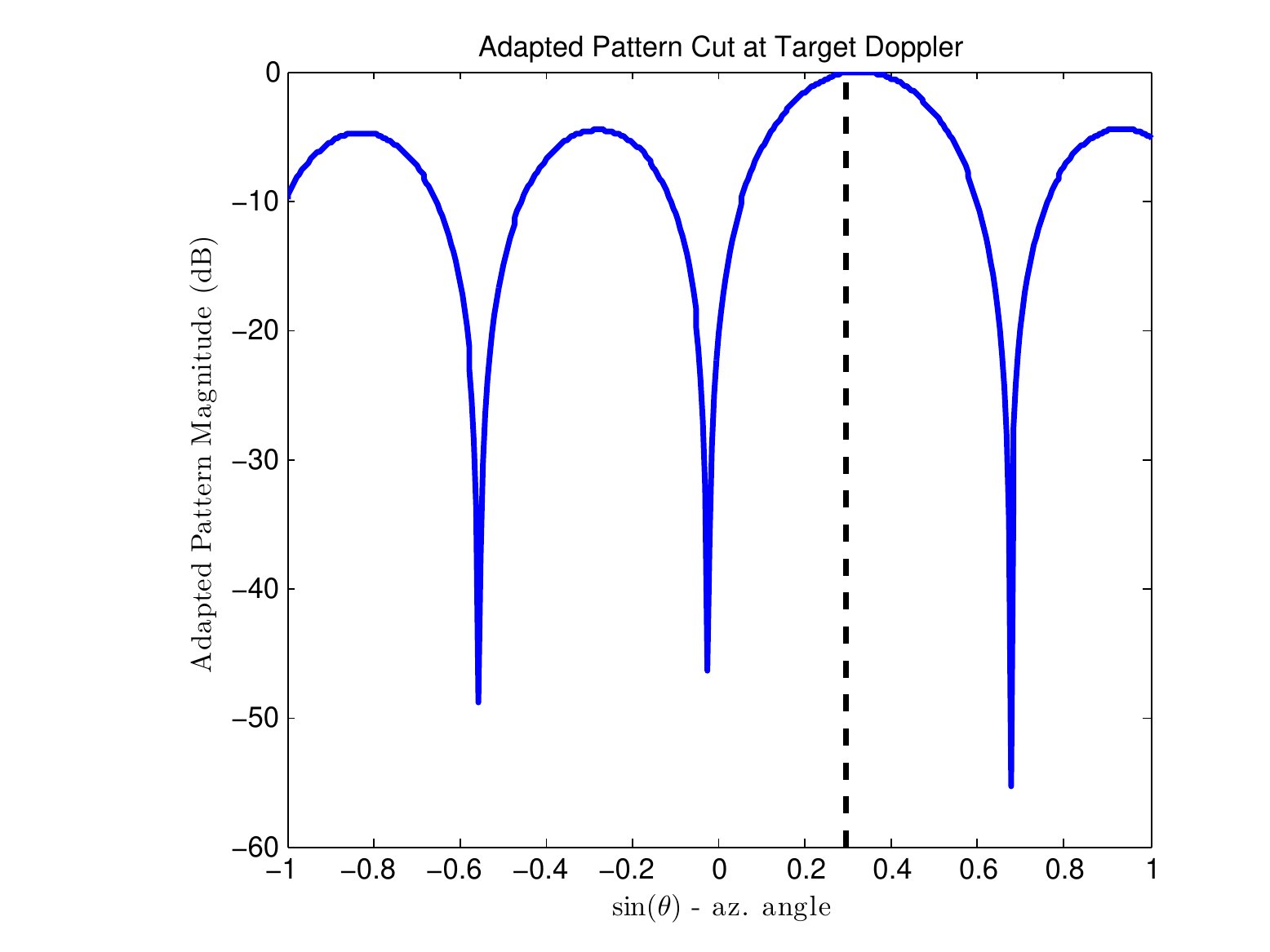}
	\caption{Adapted Pattern (dB scale), cut at target Doppler, QuadSolver.}
	\label{fig:AdaptPatternAzCutQuadNoIntf}
\end{figure}

\begin{figure}[!ht]
	\centering
	\includegraphics[width=\textwidth]{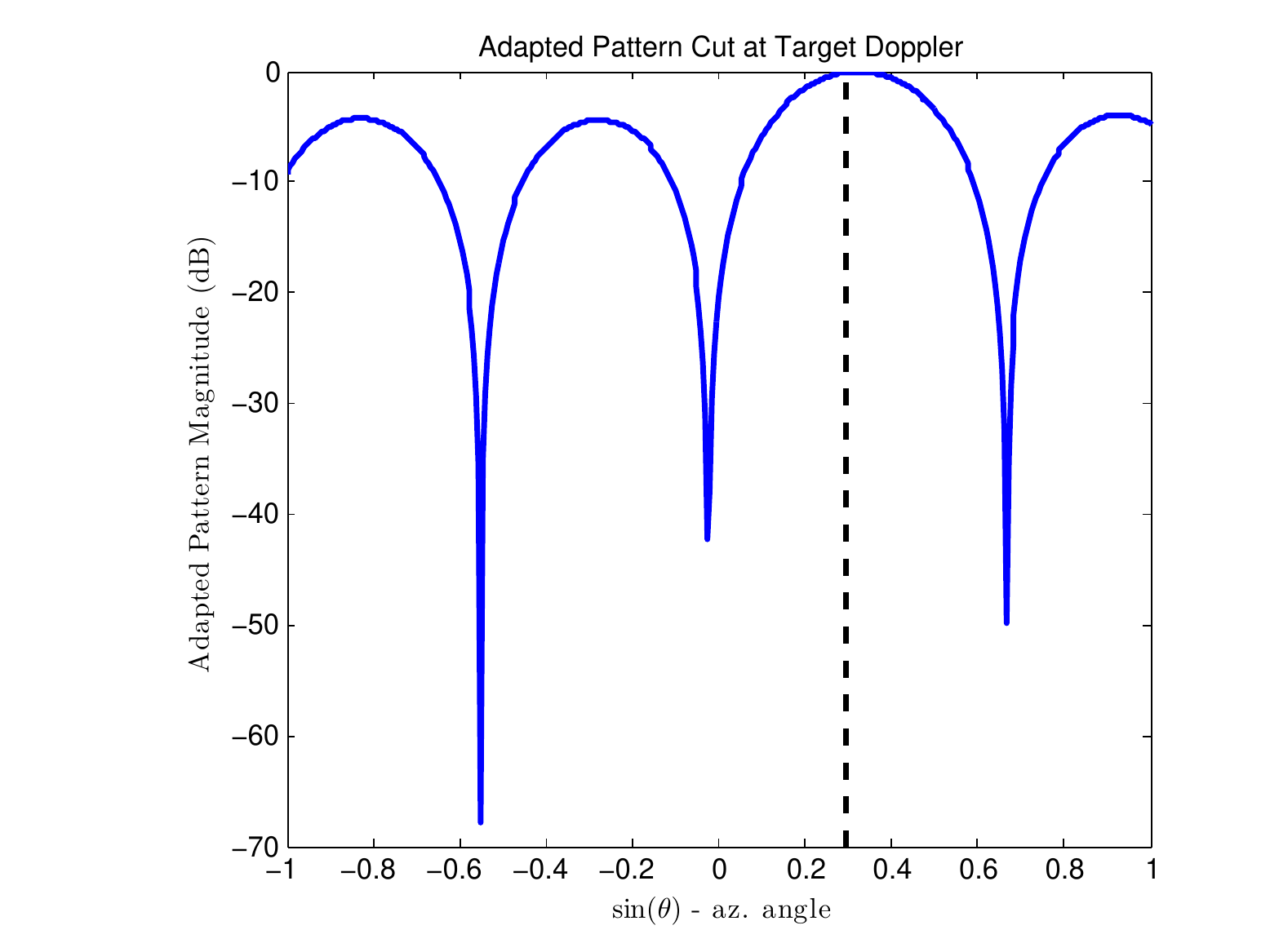}
	\caption{Adapted Pattern (dB scale), cut at target Doppler, NormSolver.}
	\label{fig:AdaptPatternAzCutNormNoIntf}
\end{figure}

\begin{figure}[!ht]
	\centering
	\includegraphics[width=\textwidth]{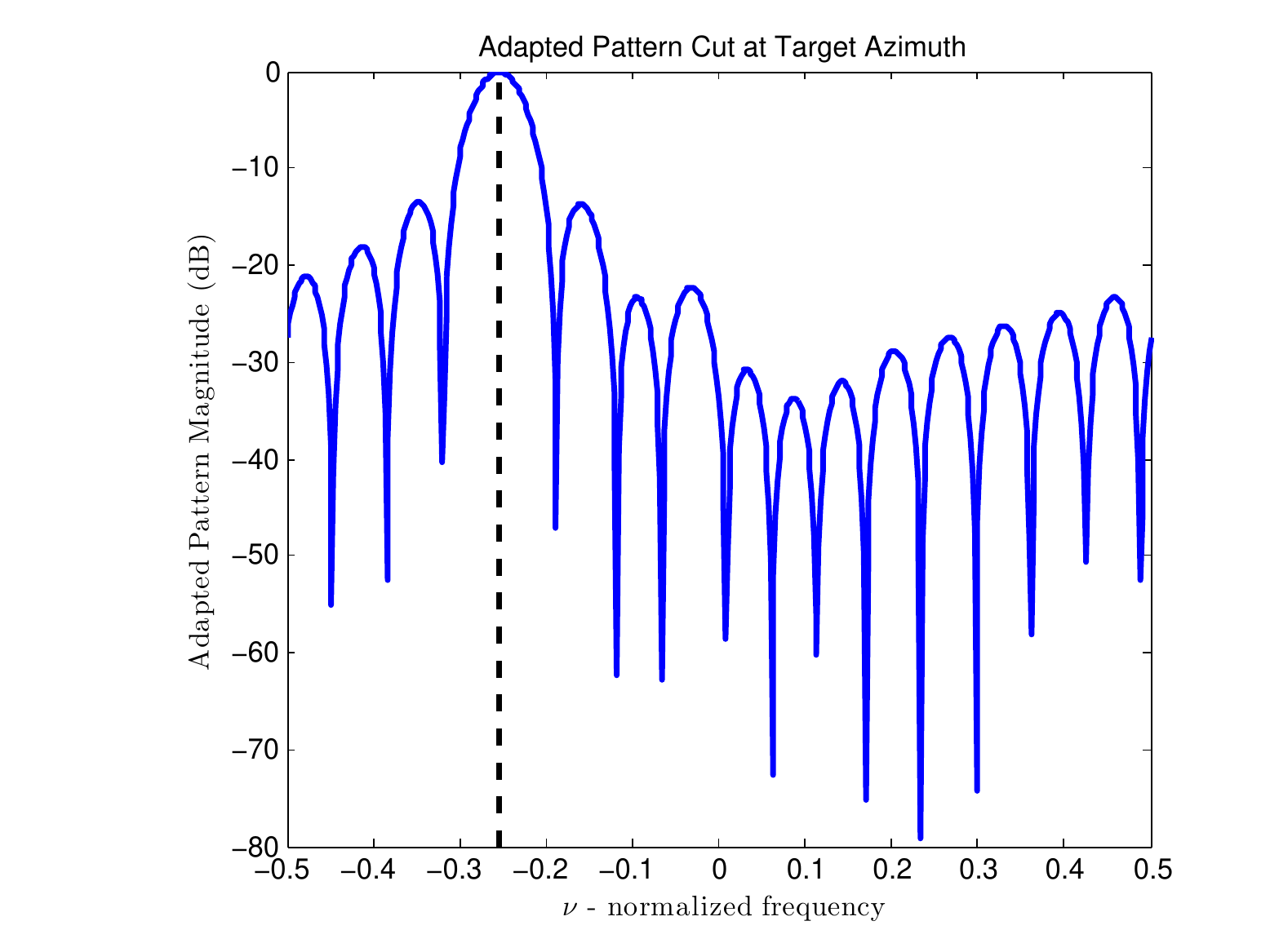}
	\caption{Adapted Pattern (dB scale), cut at target azimuth, QuadSolver.}
	\label{fig:AdaptPatternDoppCutQuadNoIntf}
\end{figure}

\begin{figure}[!ht]
	\centering
	\includegraphics[width=\textwidth]{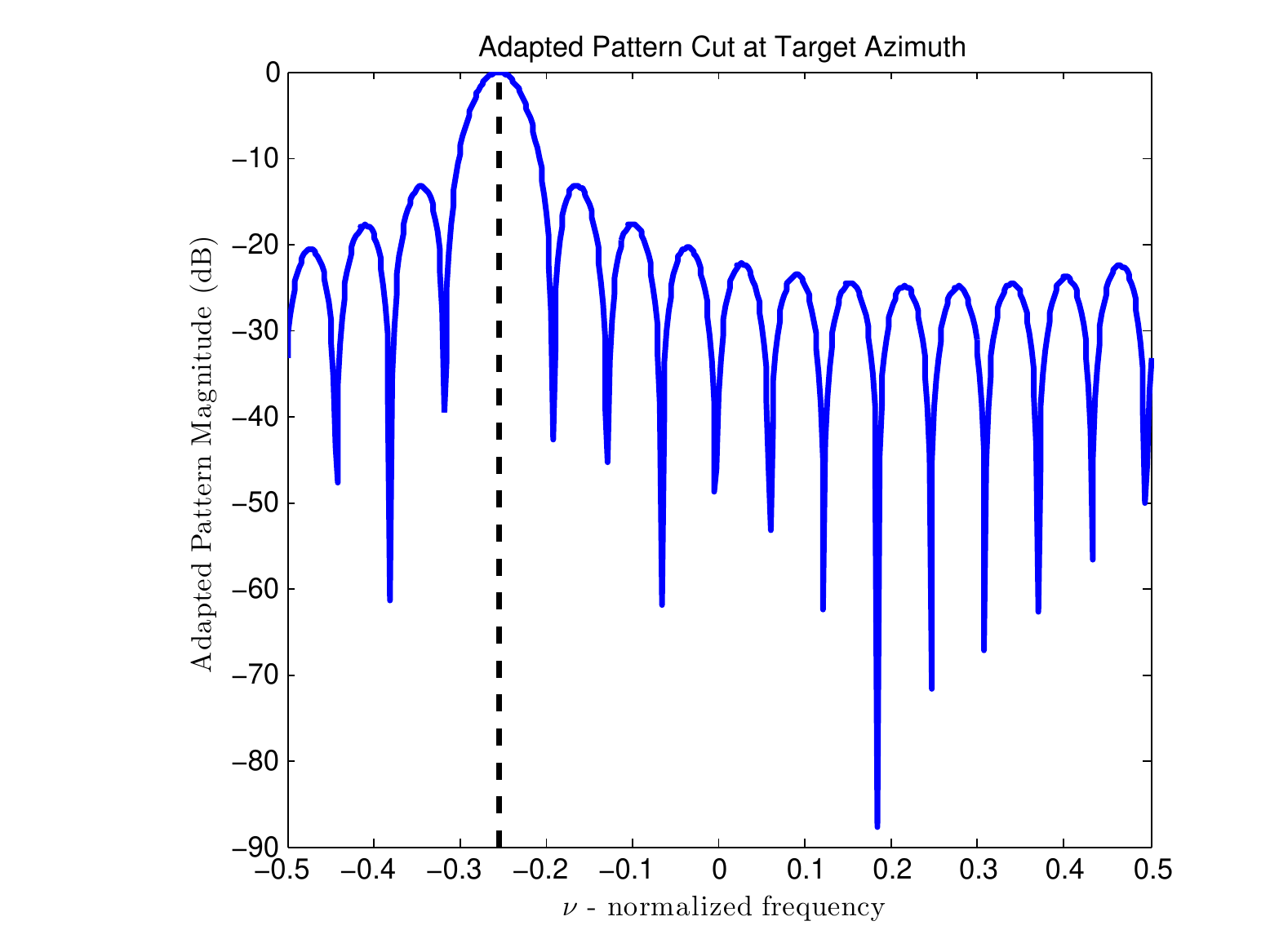}
	\caption{Adapted Pattern (dB scale), cut at target azimuth, NormSolver.}
	\label{fig:AdaptPatternDoppCutNormNoIntf}
\end{figure}

With no interference, the naive implementation performs relatively well, as illustrated in Figures~\ref{fig:AdaptPatternQuadNoIntf} (for QuadSolver) and \ref{fig:AdaptPatternNormNoIntf} (for NormSolver). The aforementioned clutter nulling peculiarities specific to each solver reappear, but the target is well localized in each case, with the peak of the pattern at its location. 
This behavior is confirmed by the cuts along the target Doppler (Figures~\ref{fig:AdaptPatternAzCutQuadNoIntf} and \ref{fig:AdaptPatternAzCutNormNoIntf}) and target azimuth (Figures~\ref{fig:AdaptPatternDoppCutQuadNoIntf} and \ref{fig:AdaptPatternDoppCutNormNoIntf}), with the target location in each denoted by the dashed line. While the azimuth plots have large sidelobes, this is mostly attributable to the limited number of antennas in this simulation ($M=5$) and not the optimization process.

\begin{figure}[!ht]
	\centering
	\includegraphics[width=\textwidth]{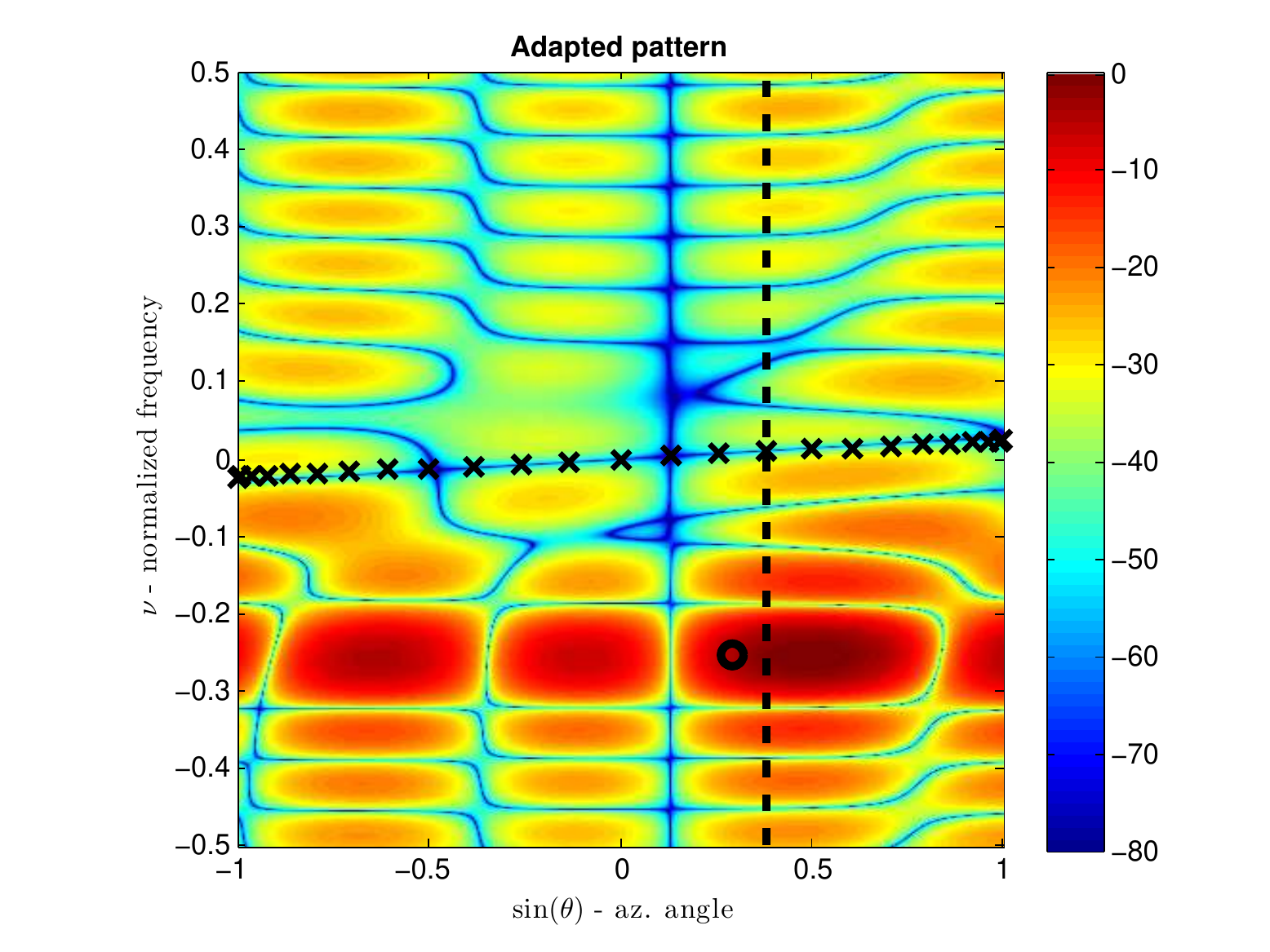}
	\caption{Adapted Pattern (dB relative to peak), QuadSolver. Target at $\bigcirc$, Clutter phase centers at $\times$, Interferer (dashed line) at $(\theta, \phi) = (0.3941, 0.3)$ radians}
	\label{fig:AdaptPatternQuadIntf}
\end{figure}

\begin{figure}[!ht]
	\centering
	\includegraphics[width=\textwidth]{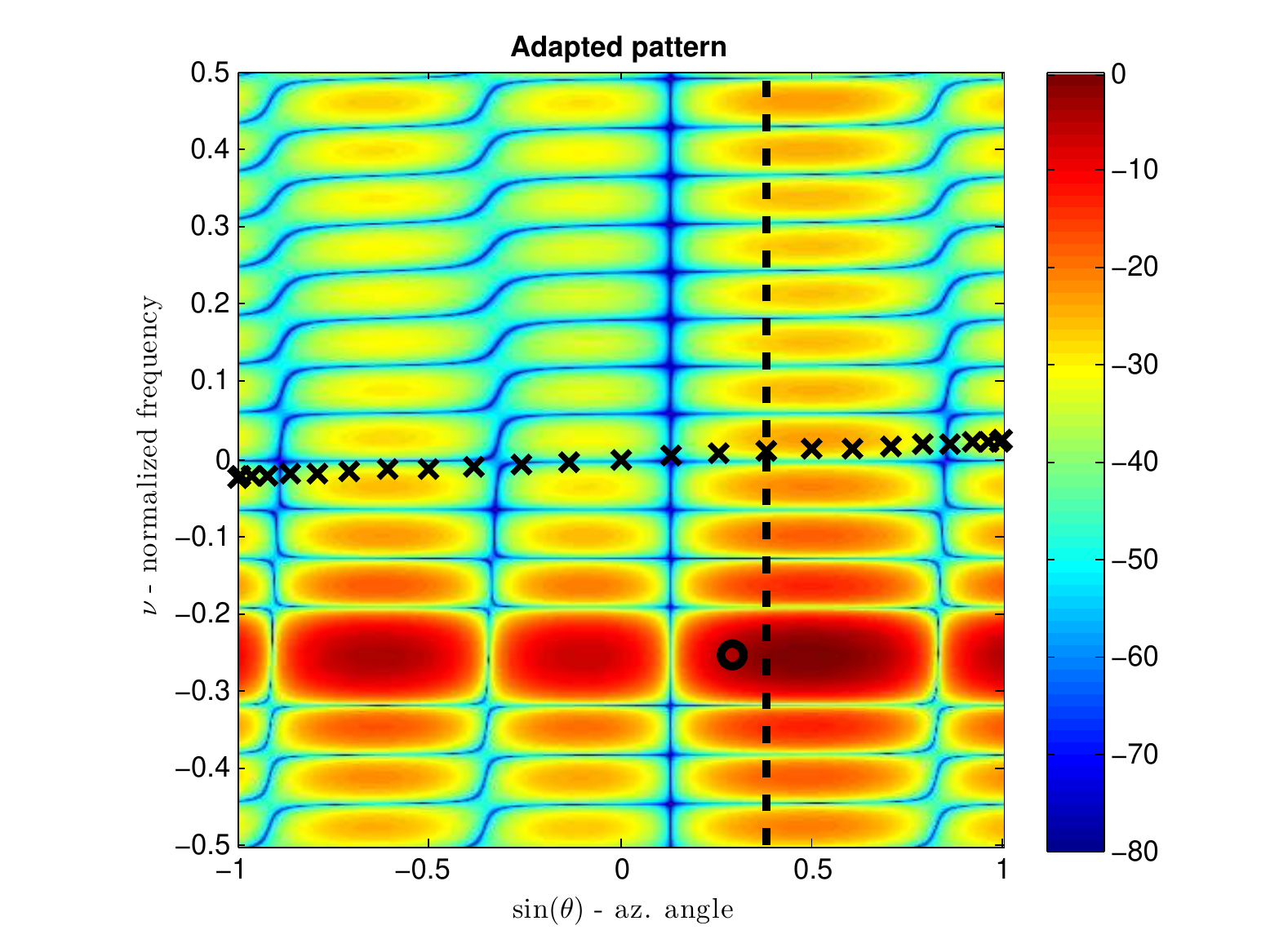}
	\caption{Adapted Pattern, NormSolver. Target at $\bigcirc$, Clutter phase centers at $\times$, Interferer (dashed line) at $(\theta, \phi) = (0.3941, 0.3)$ }
	\label{fig:AdaptPatternNormIntf}
\end{figure}

Now suppose we inject a broadband interferer "close" to the target -- as a reminder, the target is at $(\theta_t, \phi_t) = (0.3, \tfrac{\pi}{3})$ radians and the interferer is at $(\theta_I, \phi_I) = (0.3941, 0.3)$ radians. Figures~\ref{fig:AdaptPatternQuadIntf} and \ref{fig:AdaptPatternNormIntf} (for QuadSolver and NormSolver, respectively) show the adapted pattern under these conditions. As predicted by the subspace alignment in Figure~\ref{fig:SubspaceCosineQNIntf}, the peak in NormSolver's pattern is shifted away from the target quite significantly. What may be surprising, however, is that QuadSolver's adapted pattern shows the \emph{exact same error}, despite the subspace alignment appearing to suggest otherwise. There are a variety of factors that contribute to this, but the most salient is that the elevation angles of clutter patches are not identical to that of the interferer. Hence, the subspaces being nulled by QuadSolver do project into the adapted pattern as nulls, but are not identical to the subspace spanned by the interference. This lends credence to our proposition that the relaxed optimization process is a trans-recieve generalized whiten-and-match filter. In both cases, the resources available permit us to only \emph{whiten} the interference, especially because it is not signal-dependent. 

\begin{figure}[!ht]
	\centering
	\includegraphics[width=\textwidth]{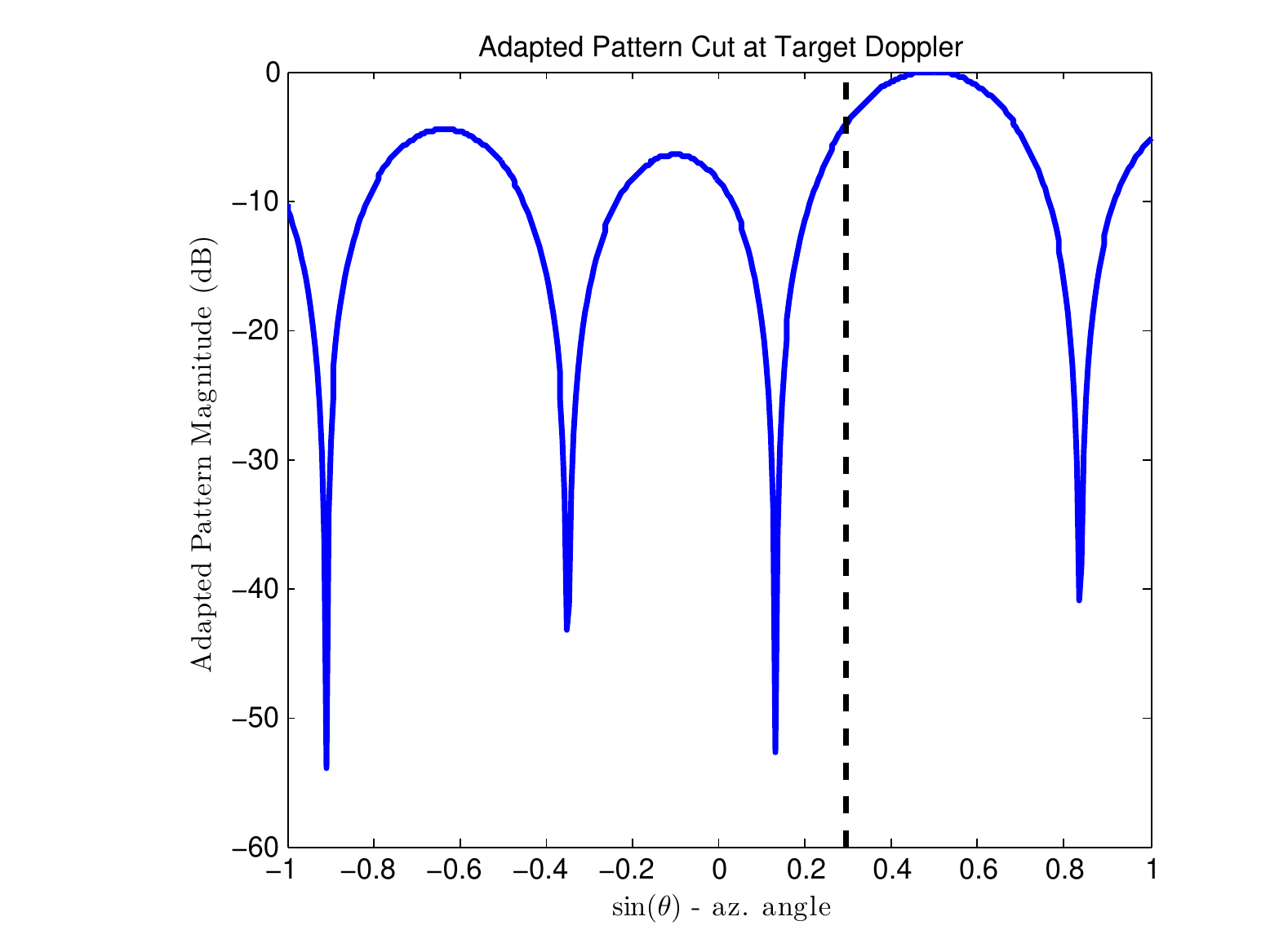}
	\caption{Adapted Pattern (dB scale), cut at target Doppler, QuadSolver. Interference.}
	\label{fig:AdaptPatternAzCutQuadIntf}
\end{figure}

\begin{figure}[!ht]
	\centering
	\includegraphics[width=\textwidth]{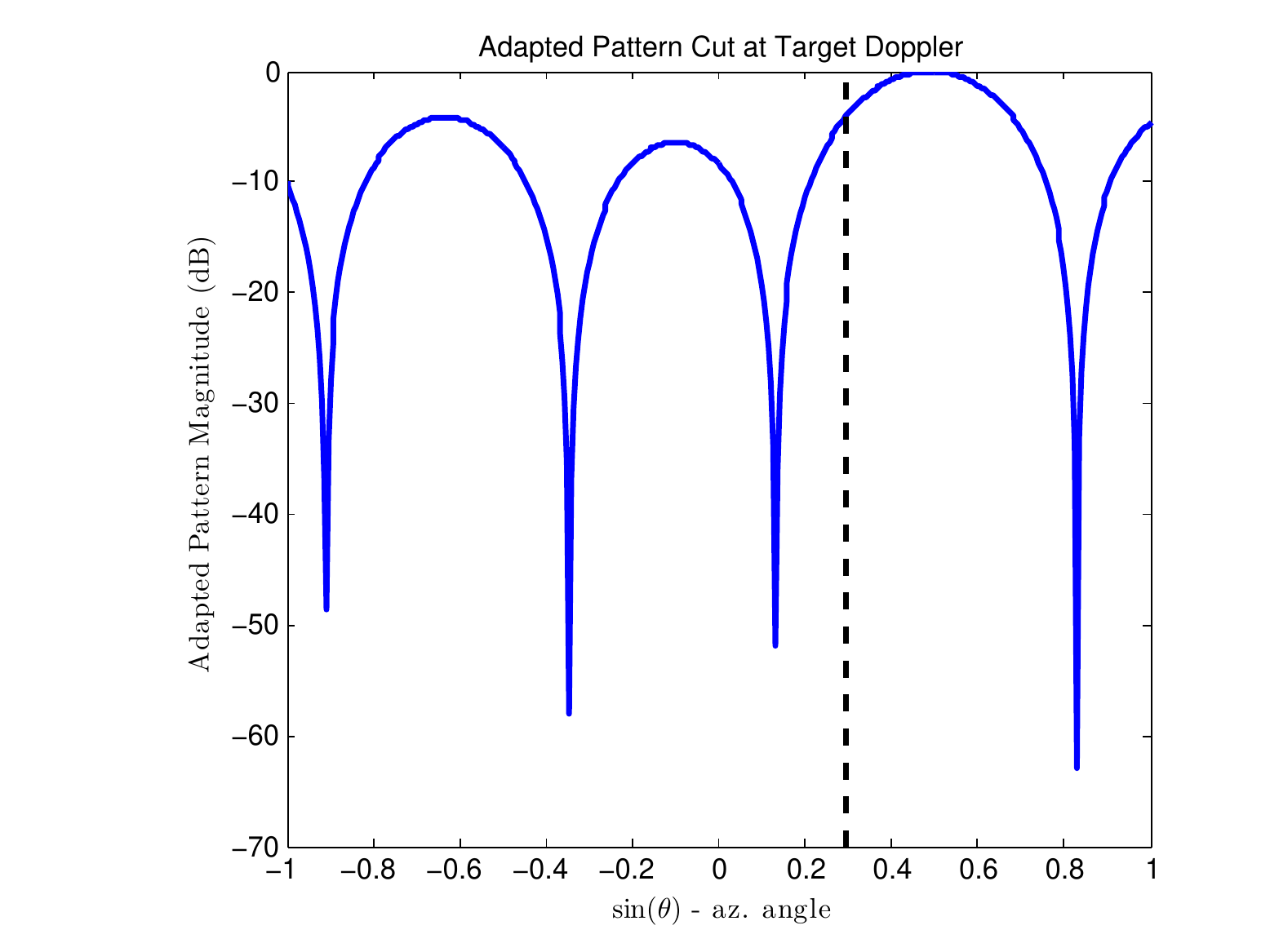}
	\caption{Adapted Pattern (dB scale), cut at target Doppler, NormSolver. Interference.}
	\label{fig:AdaptPatternAzCutNormIntf}
\end{figure}

\begin{figure}[!ht]
	\centering
	\includegraphics[width=\textwidth]{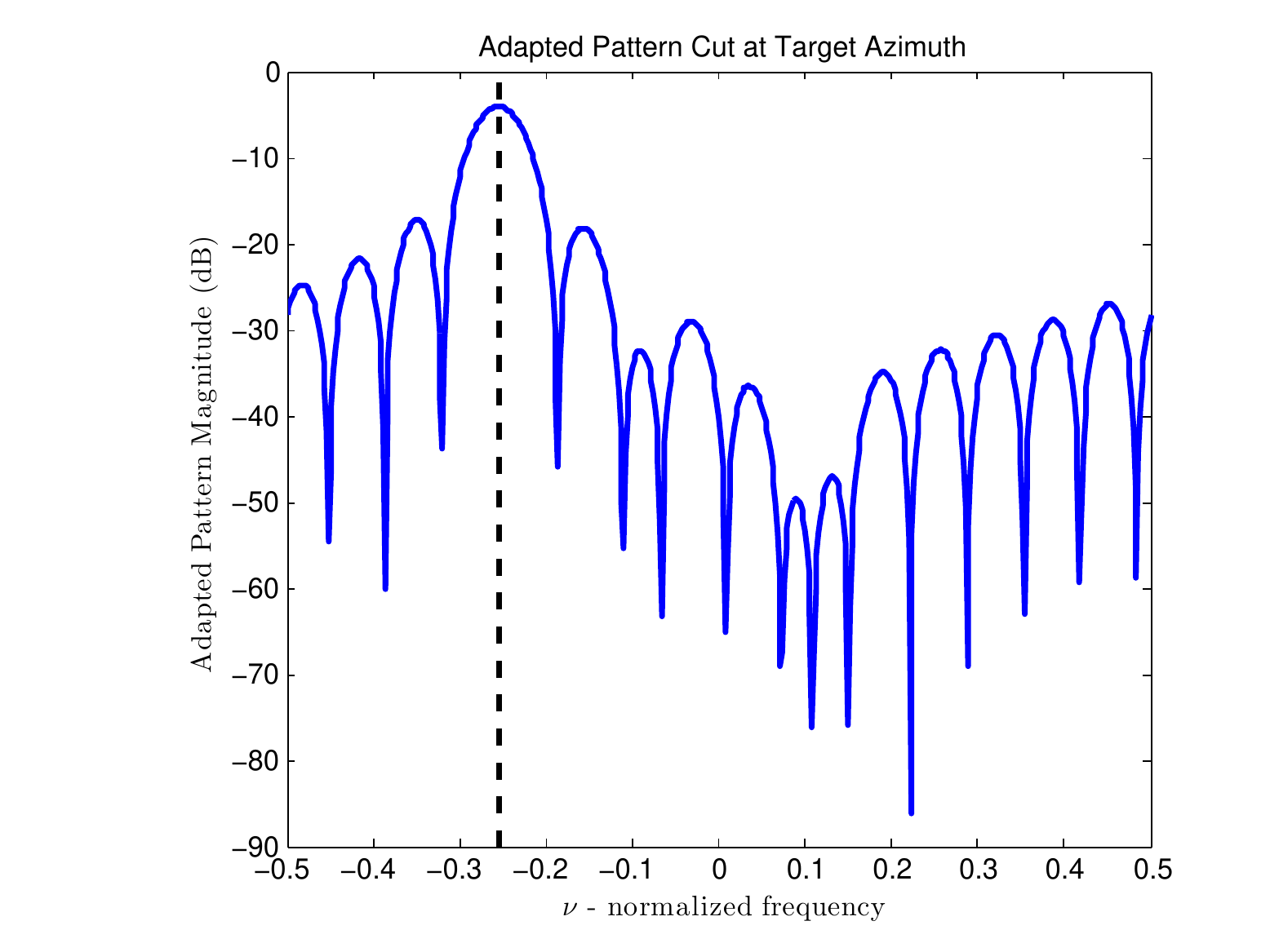}
	\caption{Adapted Pattern (dB scale), cut at target azimuth, QuadSolver. Interference.}
	\label{fig:AdaptPatternDoppCutQuadIntf}
\end{figure}

\begin{figure}[!ht]
	\centering
	\includegraphics[width=\textwidth]{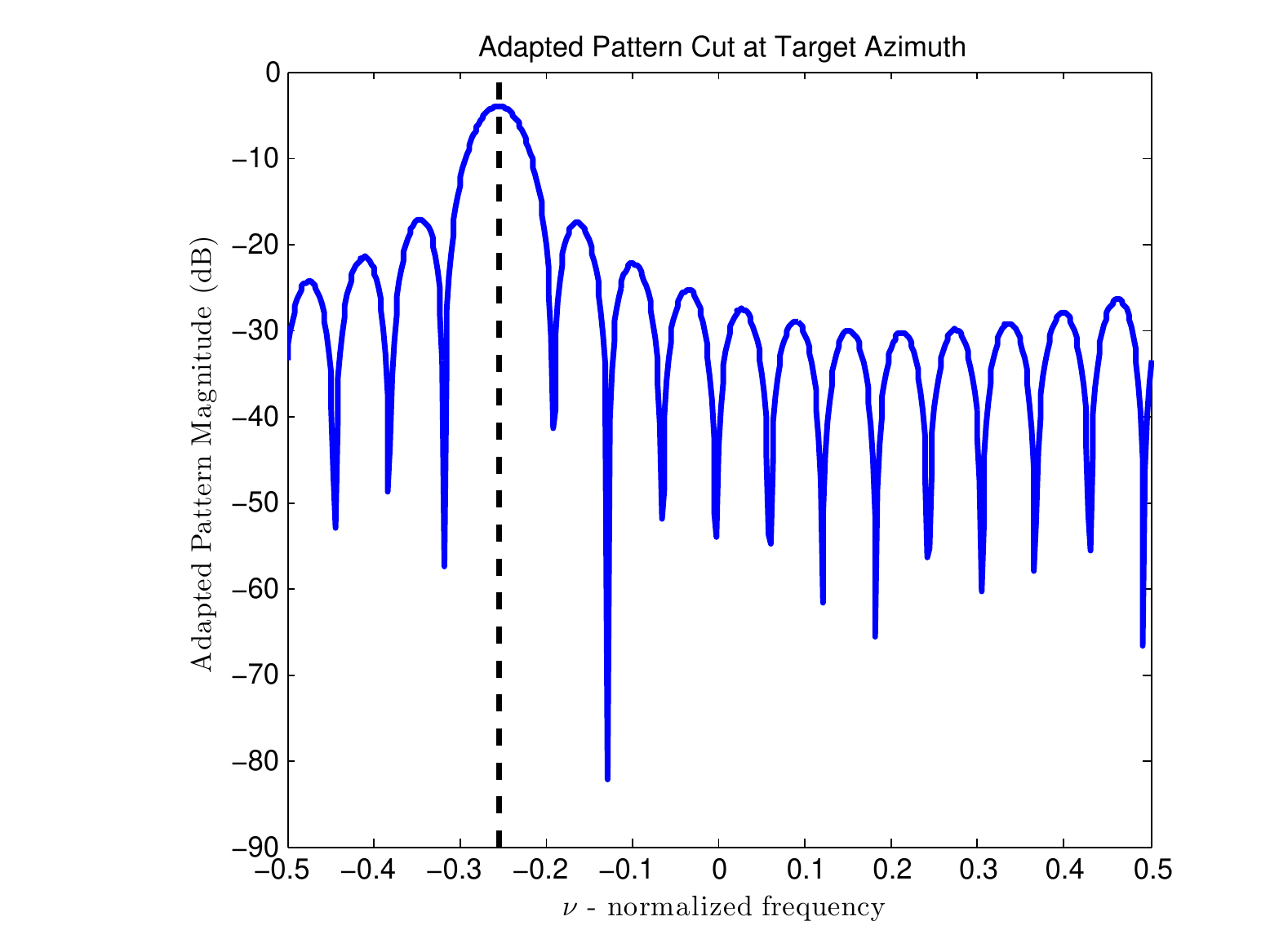}
	\caption{Adapted Pattern (dB scale), cut at target azimuth, NormSolver. Interference.}
	\label{fig:AdaptPatternDoppCutNormIntf}
\end{figure}

For completeness, we also show the adapted pattern cuts under the interference scenario. Again, in all figures, the dashed line represents the target location.  Figures~\ref{fig:AdaptPatternAzCutQuadIntf} and \ref{fig:AdaptPatternAzCutNormIntf} show the cut along the target Doppler. The significant bias seen above is more clear here, which would result in a rather large angle estimation error. 
The cuts along the target azimuth, shown in Figures~\ref{fig:AdaptPatternDoppCutQuadIntf} and \ref{fig:AdaptPatternDoppCutNormIntf}, still appropriately localize the target in some sense. However, note that in both cases, the peak gain at the target Doppler has dropped about 5 dB from the gain in Figures~\ref{fig:AdaptPatternDoppCutQuadNoIntf} and \ref{fig:AdaptPatternDoppCutNormNoIntf}. Thus, there is a clear loss in both dimensions due to the interferer. 

We note that this interference impact (in all cases) is lessened by increasing $M$ in our simulations. This is not surprising, since more antenna elements over the same aperture increases the degrees of freedom available to null interference and localize the target.  

\section{Conclusions}\label{sec:conclusion}
In this report, we reconsidered the problem of \cite{SetlurRangaswamy2016} under the relaxed biquadratic program framework. We demonstrated definitively that the problem is non-convex, then showed the relaxation process. We then showed that the KKTs require power-bounded solutions when the noise-and-interference matrix is full rank, and that such a solution admits a waterfilling interpretation. Simulations demonstrated that numerical solvers can provide divergent solution paths, depending on scaling -- one closer to the traditional clutter-nulling process, the other relying on a matched target-in-clutter response. Future work will attempt to generalize the findings made in this report and demonstrate its utility for other radar system models that admit a channel response representation.
\bibliographystyle{IEEEtran} 
\bibliography{BQP_report}
\end{document}